\documentclass[oneside]{elsart}

\usepackage{epsfig}
\usepackage{rotate}
\usepackage{amsmath}
\usepackage{amssymb}
\usepackage{array}
\usepackage[dvips]{color}
\usepackage{wrapfig}
\usepackage{bm}

\newcommand{\vo}{\mbox{$V^0$}}
\newcommand{\lam}{\mbox{$\rm \Lambda$}}

\newcommand{\alam}{\mbox{$\rm \bar \Lambda$}}

\newcommand{\ko}{\mbox{$K^0_S$}}
\newcommand{\kodecay}{\mbox{$K^0_S \to \pi^+ \pi^-$}}

\newcommand{\numuCC}{$\nu_\mu\ \mbox{CC }$}
\newcommand{\nuNC}{$\nu\ \mbox{NC }$}
\newcommand{\Kstar}{\mbox{$K^\star$}}

\newcommand{\rhoOO}{\mbox{$\rho_{00}$}}

\begin{document}
\begin{frontmatter}
\title{\boldmath 
Production properties of K$^\star (892)^\pm$ \\ 
vector mesons and their spin alignment \\
as measured in the NOMAD experiment}

\centerline{\bf NOMAD Collaboration}
\vskip 0.5cm
\author[Dubna]             {A.~Chukanov\corauthref{corr}},
\corauth[corr]                {Corresponding author.}
\ead                          {chukanov@nusun.jinr.ru}
\author[Dubna]             {D.~Naumov},
\author[Dubna,Paris]       {B.~Popov},
\author[Paris]             {P.~Astier},
\author[CERN]              {D.~Autiero},
\author[Saclay]            {A.~Baldisseri},
\author[Padova]            {M.~Baldo-Ceolin},
\author[Paris]             {M.~Banner},
\author[LAPP]              {G.~Bassompierre},
\author[Lausanne]          {K.~Benslama},
\author[Saclay]            {N.~Besson},
% ...

\address[LAPP]           {LAPP, Annecy, France}
% ...
\address[Dubna]          {JINR, Dubna, Russia}
% ...
\address[CERN]           {CERN, Geneva, Switzerland}
\address[Lausanne]       {University of Lausanne, Lausanne, Switzerland}
% ...
\address[Padova]         {Univ. of Padova and INFN, Padova, Italy}
\address[Paris]          {LPNHE, Univ. of Paris VI and VII, Paris, France}
% ...
\address[Saclay]         {DAPNIA, CEA Saclay, France}
% ...

\thanks[Scuola]          {Now at Scuola Normale Superiore, Pisa, Italy}
\thanks[Perugia]         {Now at Univ. of Perugia and INFN, Perugia, Italy}

\clearpage
\begin{abstract}
First measurements of $K^\star (892)^\pm$ mesons production properties and 
their spin alignment in $\nu_\mu$ charged current (CC) and neutral current (NC)
interactions are presented. The analysis of the full data sample of the NOMAD 
experiment is performed in different kinematic regions. For $\Kstar^+$ and 
$\Kstar^-$ mesons produced in $\nu_\mu$ CC interactions and decaying into 
$K^0 \pi^\pm$ we have found the following yields per event: 
$(2.6 \pm 0.2\, (stat.) \pm 0.2\, (syst.))\%$ and 
$(1.6 \pm 0.1\, (stat.) \pm 0.1\, (syst.))\%$ respectively, while for the 
$\Kstar^+$ and $\Kstar^-$ mesons produced in \nuNC interactions the 
corresponding yields per event are: 
$(2.5 \pm 0.3\, (stat.) \pm 0.3\, (syst.))\%$ and 
$(1.0 \pm 0.3\, (stat.)\pm 0.2\, (syst.))\%$. The results obtained for the 
$\rho_{00}$ parameter, $0.40 \pm 0.06\, (stat) \pm 0.03\, (syst)$ and
$0.28 \pm 0.07\, (stat) \pm 0.03\, (syst)$ for $\Kstar(892)^+$ and 
$\Kstar(892)^-$ produced in $\nu_\mu$ CC interactions, are compared to 
theoretical predictions tuned on LEP measurements in $e^+e^-$ annihilation at 
the $Z^0$ pole. For $\Kstar(892)^+$ mesons produced in \nuNC interactions the 
measured $\rho_{00}$ parameter is $0.66 \pm 0.10\, (stat) \pm 0.05\, (syst)$.
\end{abstract}
\end{frontmatter}
\section{Introduction}

Following the analyses of strange particles in neutrino interactions
reported earlier~\cite{nomad-strange-cc,lam_polar,alam_polar,nc_event} 
we present a study of the production properties of $K^\star (892)^\pm$ vector 
mesons observed through the $K^0_S \pi^\pm$ decay modes. The full data sample 
of the NOMAD experiment divided into subsamples of neutrino charged current 
(CC) and neutral current (NC) interactions is used for this analysis.

For the first time in neutrino experiments the acquired statistics of 
the $K^\star (892)^\pm$ mesons allows the measurement of the absolute and
relative yields, the determination of their dependence on relevant kinematic 
quantities as well as the extraction of the spin alignment of these vector 
mesons. 

\subsection{$K^\star (892)^\pm$ production in neutrino interactions}

\begin{figure}[htb]
\begin{center}
\mbox{\epsfig{file=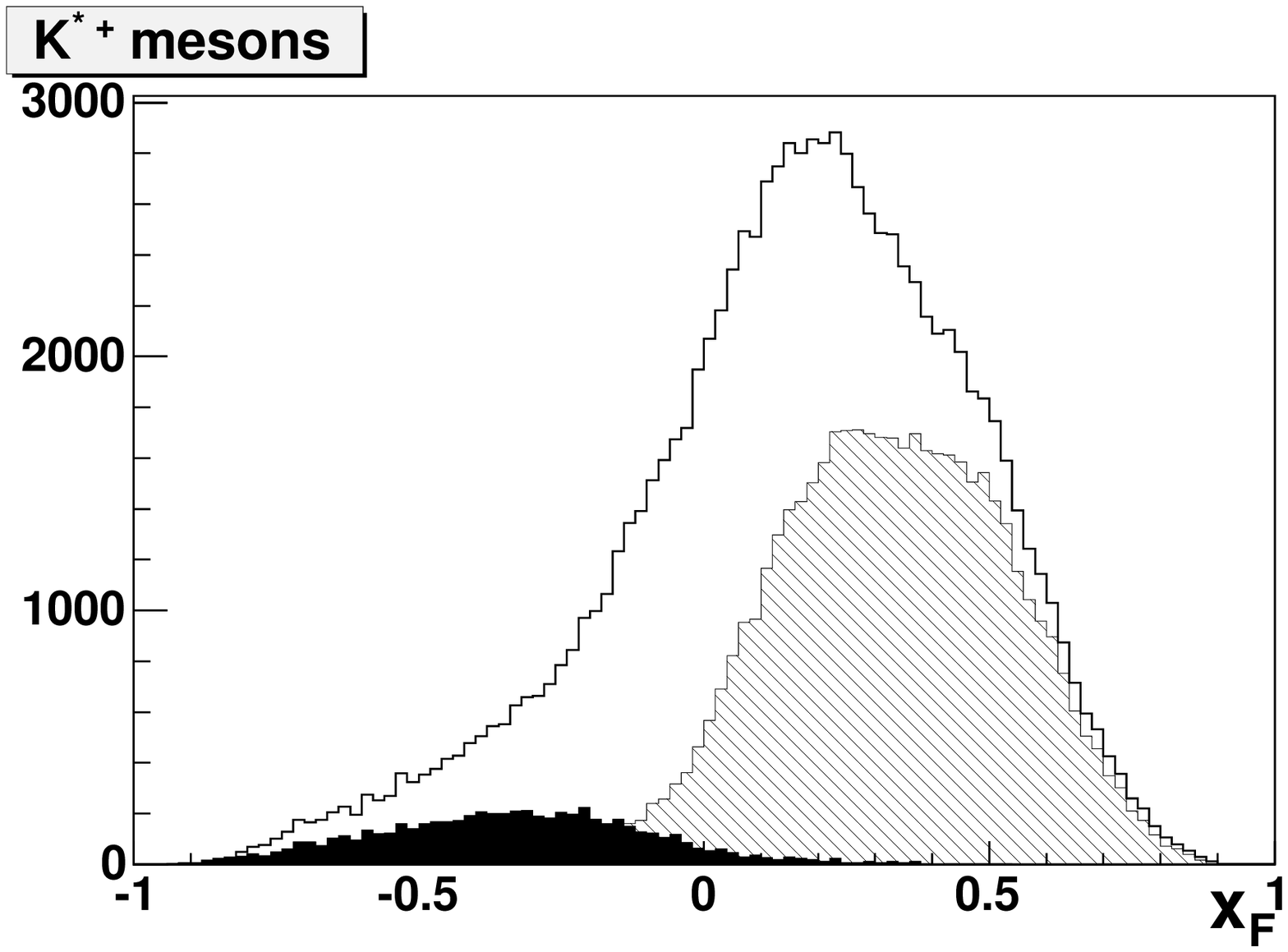,width=0.45\linewidth,height=0.25\linewidth}}
\mbox{\epsfig{file=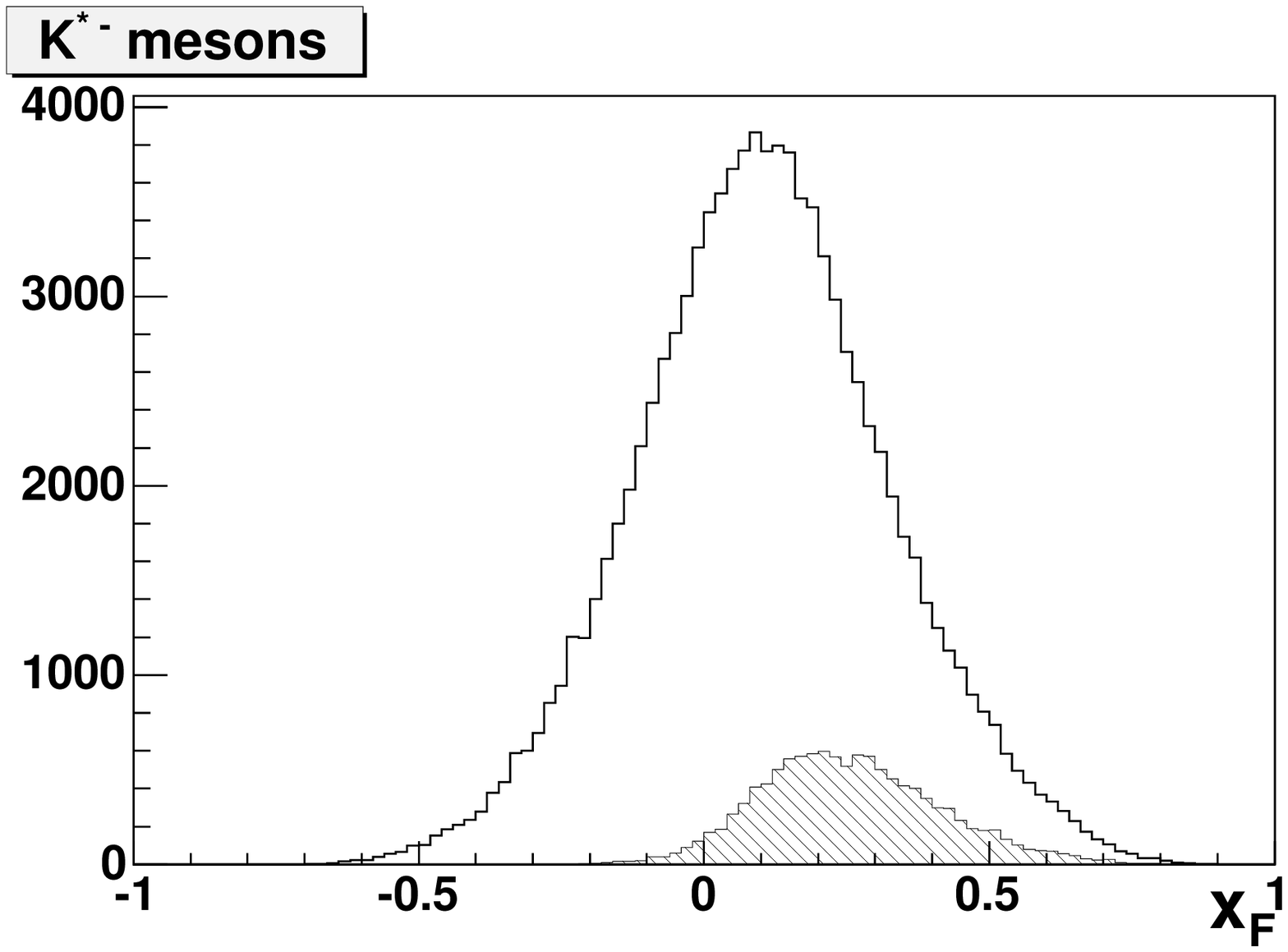,width=0.45\linewidth,height=0.25\linewidth}}
\protect\caption{\label{fig:xF_distr} \it 
\numuCC interactions in the NOMAD experiment: LUND model predictions for the 
$x_F$ distribution of all $\Kstar^+$ (left) and $\Kstar^-$ (right) mesons, 
showing those originating from diquark fragmentation (filled area) and those 
originating from quark fragmentation (hatched area). The remainder of 
$\Kstar^{\pm}$ in the unshaded area come from string fragmentation.}
\end{center}
\end{figure}

In \numuCC interactions with nucleons, in a dominant number of cases the 
produced $u$-quark is in a 100\% left-polarized state, and it can eventually 
fragment into a $\Kstar^+(u\bar s)$ vector meson ($J^P=1^-$): 
$\nu_\mu N \to \mu^-\Kstar^+X$. The $\Kstar^+$ mesons containing the leading 
$u$-quark populate the current fragmentation region (the $x_F>0$ 
region\footnote{$x_F$ is defined as $x_F \equiv 2p^*_l/\hat{W}$, $p^*_l$ being 
the momentum of the vector meson along the $W$ boson direction and $\hat{W}$ 
the hadronic energy, both calculated in the hadronic centre of mass system.}). 
However, according to the LUND model~\cite{LUND} predictions this population 
is of the same order of magnitude as the fraction of $\Kstar^+$ mesons 
produced more centrally in the string fragmentation process, as seen in 
Fig.~\ref{fig:xF_distr}~(left). It is therefore interesting to study the spin 
alignment in different kinematic regions, not only to relate a possible 
effect to a well defined initial state (at large $x_F$), but also to improve 
our knowledge of the spin transfer in the string fragmentation process. This 
is the main production mechanism for the $\Kstar^-$ mesons as seen in 
Fig.~\ref{fig:xF_distr}~(right).

Let us stress that neutrino NC interactions are different from CC 
interactions at the quark level: for example, a leading down quark, a leading 
up quark and even a leading strange quark can be produced in \nuNC 
interactions.

\begin{table}[htb]
\begin{center}
\caption{\it The spin density matrix element $\rho_{00}$ as measured for 
different vector mesons.}
\vspace*{0.3cm}
\label{tab:exp_res}
\begin{tabular}{ccc}
\hline
Meson         & Results (Experiment)   & Comments      \\ \hline
$\rho^{\pm}$  & $0.373\pm0.052$ (OPAL) & $0.3<x_E<0.6$ \\ \hline
$\rho^0$      & $0.43\pm0.05$ (DELPHI) & $x_E>0.4$     \\ \hline
$\omega$      & $0.142\pm0.114$ (OPAL) & $0.3<x_E<0.6$ \\ \hline
$K^{\star 0}$ & 
    \begin{tabular}{c}
     $0.46\pm0.08$ (DELPHI) \\ 
     $0.66 \pm 0.11$ (OPAL) \\
    \end{tabular} &
    \begin{tabular}{c}
     $x_E>0.4$ \\
     $x_E>0.7$ \\
    \end{tabular} \\ \hline
$\phi  $      &
    \begin{tabular}{c}
     $0.54\pm0.06\pm0.05$ (OPAL)\\
     $0.55\pm0.10$ (DELPHI) \\
    \end{tabular}
              & $x_E>0.7$ \\ \hline
$D^{*\pm}$    & $0.40\pm0.02\pm0.01$ (OPAL) & $x_E > 0.5$ \\ \hline
$B^*$         &
   \begin{tabular}{c}
    $0.32\pm0.04\pm0.03$ (DELPHI)\\
    $0.33\pm0.06\pm0.05$ (ALEPH)\\
    $0.36\pm0.06\pm0.07$ (OPAL) \\
   \end{tabular}
              & $0 < x_E < 1$ \\ \hline
$K^{\star +}$ & $0.424\pm 0.011$ (EXCHARM) & in the transversity \\
$K^{\star -}$ & $0.393\pm 0.025$ (EXCHARM) & frame of $K^\star$ \\ \hline
$\rho^0$ & 
   \begin{tabular}{c}
    $0.65\pm0.18\pm0.10$ (BEBC, $\bar \nu$Ne) \\
    $0.41\pm0.13\pm0.07$ (BEBC, $\nu$Ne) \\
   \end{tabular} &
   \begin{tabular}{c}
    $x_F>0,\ z>0.4$ \\
   \end{tabular} \\
\hline
\end{tabular}
\end{center}
\end{table}

\subsection{Spin-related production properties of vector mesons}

The production and decay properties of mesons carrying spin are described
in terms of the spin density matrix $\rho_{mm'}$, where $m$ and $m'$ label the
spin components along the quantization axis. The Hermitian $3\times 3$ matrix 
$\rho$ with unit trace is built as a direct product of the quark and 
antiquark spin states. This matrix is usually defined in the helicity basis.
The diagonal elements $\rho_{00},\ \rho_{11}$ and $\rho_{1;-1}$ describe the 
relative intensities of the $0,\ +1$ and $-1$ spin states of the particle. 
It is common to refer to the situation with $\rho_{00} = 1/3$ as to the no 
spin alignment case, regardless of the values of $\rho_{11}$ and 
$\rho_{1;-1}$. Note that the spin alignment is not equivalent to the 
polarization in the conventional sense. For example, 
$\rho_{11}=\rho_{1;-1}=1/2$ and $\rho_{00}=0$ is {\em unpolarized} but 
{\em spin aligned}. Further details about the spin density matrix can be found 
in~\cite{alignMore}.

The elements $\rho_{11}$ and $\rho_{1;-1}$ cannot be measured separately 
since vector mesons decay via strong interactions and therefore conserve 
parity. Thus, the spin alignment of a particle can only be studied through 
the diagonal element $\rho_{00}$. For $J^P = 1^-$ states it can be 
experimentally measured using the angular distribution of the meson decay 
products~\cite{donoghue78}:
\begin{equation}
\label{rho00_distr}
W(\theta)=\frac{3}{4}[(1-\rho_{00})+(3\rho_{00}-1)\cos^2\theta],
\end{equation}
where $\theta$ is the angle between the direction of one of the decay 
products and the direction of the vector meson ($z$-axis) in its rest frame. 
The following are interpretations of some special cases for the $\rho_{00}$ 
parameter: 

\begin{description}
\item[\mbox{$\rho_{00} = \frac{1}{3}$ --}]
     no spin alignment; the probability of projections $+1,\ -1$ and 0 of the 
     meson spin onto the $z$ axis are equal (in this case there is no 
     dependence of Eq.~(\ref{rho00_distr}) on $\cos\theta$);
\item[\mbox{$\rho_{00} = 0$ --}]
     spin alignment; only the $+1$ and $-1$ projections are possible; 
\item[\mbox{$\rho_{00} = 1$ --}] 
     spin alignment; only the 0 projection is possible.
\end{description}

\subsection{Review of experimental results on $\rho_{00}$}

The spin alignment of vector mesons was measured previously mainly in the LEP 
experiments in $e^+e^-$ annihilation at the $Z^0$ pole. A summary of available
experimental results is given in Table~\ref{tab:exp_res}. Spin alignment for 
the $\rho^0,\ \omega,\ K^{\star 0}$ and $D^{\star \pm}$ vector mesons was 
observed at high $x_E$, where $x_E$ is the ratio of the meson energy to the 
beam energy, and there was no spin alignment found for the $\rho^\pm$ and 
$B^\star$ mesons~\cite{ALEPH,DELPHI,OPAL}. The EXCHARM collaboration observed 
spin alignment of $\Kstar^\pm$ mesons in neutron-carbon interactions in the 
transversity frame (the $z$-axis was defined to be normal to the production 
plane) of the $\Kstar^\pm$ at rest~\cite{EXCHARM}.

In a previous neutrino experiment the $\rho_{00}$ parameter of the $\rho^0$ 
vector meson was measured by the BEBC WA59 collaboration~\cite{BEBC_align}. 
Large uncertainties do not allow to draw any conclusion about the spin 
alignment.

Note that a 3$\sigma$ statistical significance for spin alignment is achieved 
only for $\mbox{$D^\star$}^\pm$ mesons by the OPAL collaboration~\cite{OPAL} 
and for $\Kstar^+$ mesons by the EXCHARM collaboration~\cite{EXCHARM}.

\subsection{Theoretical predictions for the \rhoOO \ parameter}

\begin{figure}[htb]
\begin{center}
\epsfig{file=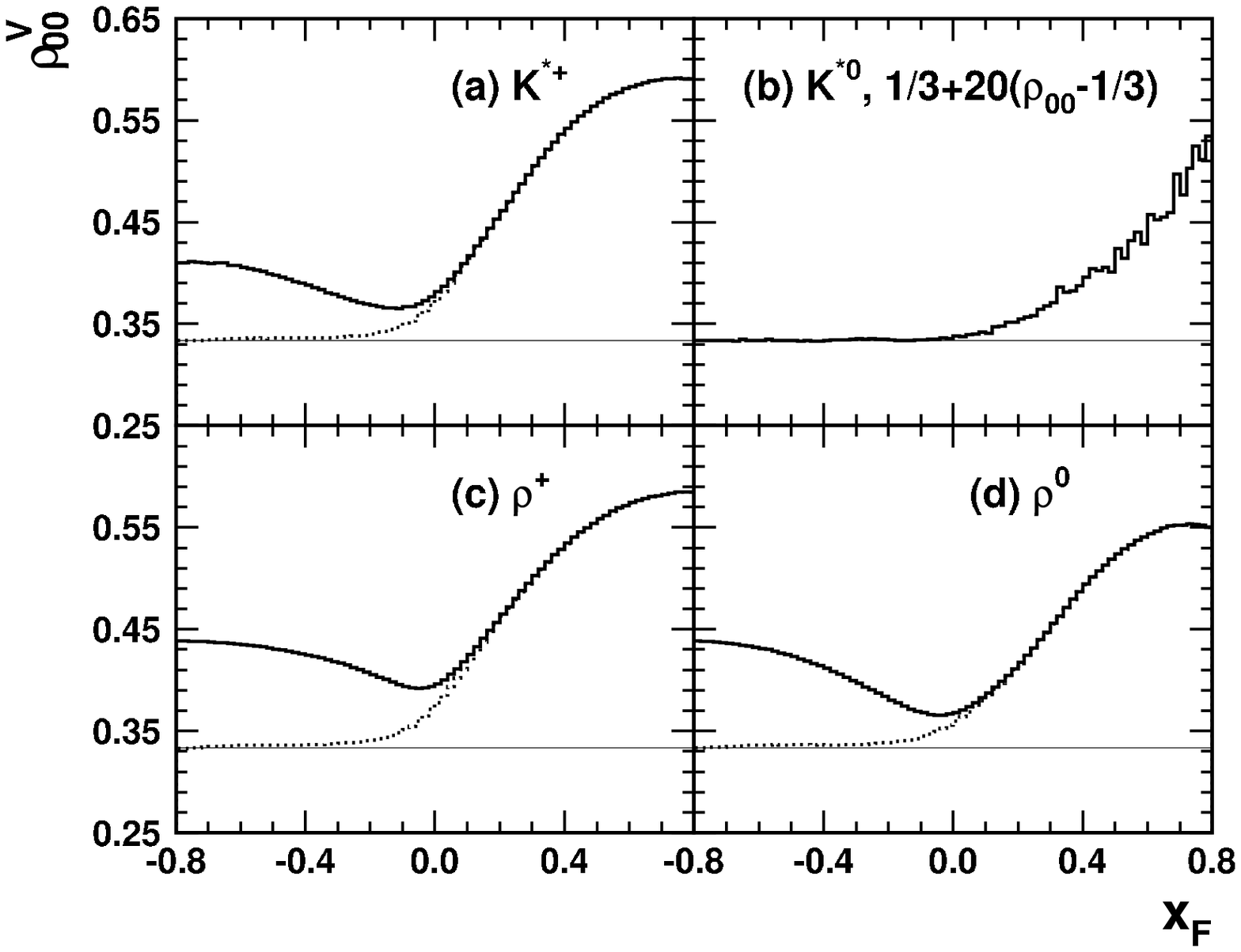,width=0.6\linewidth,height=0.4\linewidth}
\protect\caption{\label{fig:rho_predicted} \it The $\rho_{00}^V$ parameter in 
$\nu_\mu p \to \mu^-VX$ at $E_\nu=43.8$ GeV. The solid line represents the 
results where the contribution of target fragmentation is taken into account, 
while the dotted line represents the results where only the contribution of 
the current fragmentation is included. The horizontal line shows the no spin 
alignment case, $\rho_{00} = 1/3$.}
\end{center}
\end{figure}

Different theoretical approaches (see for 
example~\cite{Donoghue:1978yb,EfremovTeryaev}) have been developed for the 
prediction of the $\rho_{00}$ parameter. All these models give a value of 
$1/3$ for the $\rho_{00}$ parameter in the case of no spin alignment, but 
differ significantly in the treatment of the spin correlation mechanisms 
during the fragmentation process. For example, a model~\cite{Liang,Liang1} 
has been built to describe the results obtained in the LEP experiments 
(see previous subsection). This model can be used to predict the \rhoOO \ 
parameter dependence on $x_F$ in other vector meson production processes.
In particular, this model predicts spin alignment of vector mesons 
produced in neutrino interactions in the NOMAD energy region~\cite{Liang1}. 
These predictions for $\rho_{00}^V$ in $\nu_\mu p \to \mu^-VX$ at incoming 
neutrino energy $E_\nu=43.8$~GeV both in the current and target fragmentation 
regions are shown in Fig.~\ref{fig:rho_predicted}.

\section{Experimental Procedure} \label{sec:procedure}
\subsection{The NOMAD experiment} \label{sec:nomad}
 
The main goal of the NOMAD experiment~\cite{NOMAD_NIM} was the search for
$\nu_\mu \rightarrow \nu_\tau$ oscillations in a wide-band neutrino beam from 
the CERN SPS. This search used kinematic criteria to identify $\nu_\tau$ CC 
interactions~\cite{NOMAD_OSC} and required a very good quality of event 
reconstruction, in particular the ability to reconstruct individual particles.
This has indeed been achieved by the NOMAD detector, and moreover, the large 
data sample collected during four years of data taking (1995-1998) has 
allowed for detailed studies of neutrino interactions. The full data sample, 
corresponding to about $1.3\times10^6$ $\nu_\mu$ CC interactions in the 
detector fiducial volume, is used in the present analysis. A complete
reprocessing of the whole NOMAD data sample has been performed using improved 
reconstruction algorithms with respect to those used for the previous NOMAD 
publications related to the studies of strange 
particles~\cite{nomad-strange-cc,lam_polar,alam_polar}. In particular, the 
cut on the density of hits in the drift chambers has been removed (see 
discussion in~\cite{NOMAD_NUE}). The data are compared to the results of a 
Monte Carlo (MC) simulation based on modified versions of 
LEPTO~6.1~\cite{LEPTO} and JETSET~7.4~\cite{JETSET} generators for neutrino 
interactions (with $Q^2$ and $W^2$ cutoff parameters removed, where $Q$ is 
the four-momentum transferred from the incoming neutrino to the target 
nucleon) and on a GEANT~\cite{GEANT} based program for the detector response. 
The relevant JETSET parameters have been tuned in order to reproduce the 
yields of strange particles measured in $\nu_\mu$ CC interactions in 
NOMAD~\cite{nomad-strange-cc}. A detailed description of the tuning of the MC 
simulation program will be the subject of a forthcoming publication. 
To define the parton content of the nucleon for the cross-section calculation 
we use the fixed-flavour parameterization~\cite{Alekhin} in the NNLO 
approximation. We do not include the parton shower treatment from JETSET. 
The reinteractions of hadrons with surrounding nucleons in target nuclei are 
described within the DPMJET~\cite{DPMJET} package. For the analysis reported 
below we used a MC sample consisting of about 3 million $\nu_\mu$ CC events 
and 2.6 million $\nu_\mu$ NC events. The MC assumes no spin alignment 
($\rho_{00}=1/3$) for \Kstar \ mesons.

\subsection{Signal extraction}\label{sec:kstar_signal}

The selection procedure for the \numuCC and \nuNC event samples has been 
described in~\cite{lam_polar,nc_event} and is used in the current analysis 
along with the additional cut on the total visible hadronic energy: 
$E_{jet}>3 \mbox{ GeV}$. For the $\nu_\mu$ CC sample a further cut, 
$Q^2 > 0.8 \mbox{ GeV}^2$, is applied. We identified $8 \times 10^5$ 
\numuCC events with efficiency $\epsilon_{\nu_\mu CC} = (77.16 \pm 0.03) \%$ 
and $2.3 \times 10^5$ \nuNC events with efficiency 
$\epsilon_{\nu NC} = (67.94 \pm 0.03) \%$. The efficiencies are computed with the help of the MC and are defined 
as ratios of the number of events reconstructed and identified as $\nu_\mu$  CC (NC) to the number of
simulated $\nu_\mu$ CC (NC) events. The errors include only statistical uncertainties. The contamination
of NC events in the CC event sample is estimated to be less than $0.1\%$, while the CC contamination in the NC 
sample is estimated to be about 8\% (see~\cite{nc_event} for details).

The procedure for the $K_S^0$ and $\Kstar$ signal extraction was described 
in~\cite{lam_polar,nomad-strange-cc}. Here we present only those details 
relevant to the yield and spin alignment measurements.

\begin{figure}[htb]
\begin{center}
\epsfig{file=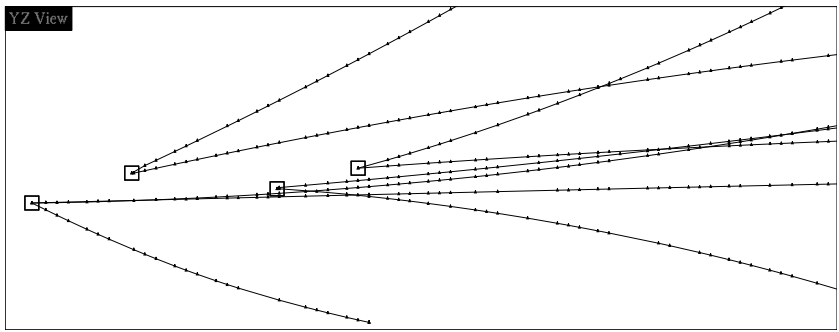,angle=-90,width=0.8\linewidth}
\protect\caption{\it A reconstructed data event containing 3 $\vo$ 
vertices identified as $\ko$ decays by the identification procedure. 
The scale on this plot is given by the size of the vertex boxes 
($3\times3$ cm$^2$).}
\label{fig:K0s_decays}
\end{center}
\end{figure}

The NOMAD experiment has observed an unprecedented number of neutral strange 
particle decays in a neutrino experiment~\cite{nomad-strange-cc}. These 
decays appear in the detector as a $\vo$-like vertex: two tracks of opposite 
charge emerging from a common vertex separated from the primary neutrino 
interaction vertex (see Fig.~\ref{fig:K0s_decays}).

Since the NOMAD detector has limited possibilities to distinguish 
(anti)protons from pions in the momentum range relevant for this analysis, 
our $\vo$ identification procedure relied on the kinematic properties of a 
$\vo$ decay to reject \lam \ and \alam. In Table~\ref{tab:k0_ident} we 
summarize the numbers of identified \kodecay \ decays in \numuCC and
\nuNC interactions as well as their identification and reconstruction 
efficiencies and purities evaluated with the help of the MC.

\begin{table}[htb]
\begin{center}
\caption{\it Number of events, purity and efficiency of identified \kodecay \ 
decays in \numuCC and \nuNC interactions in the data (efficiency includes the 
reconstruction and identification efficiencies of neutrino interactions).}
\vspace*{0.3cm}
\begin{tabular}{cccc}
\hline
Sample & $N_{K_S^0}$ & $P_{K_S^0}$(\%) & $\epsilon_{K_S^0}$(\%) \\
\hline
\numuCC & 14280 & $97.1 \pm 0.1$ & $23.9 \pm 0.1$ \\
\nuNC   & 3718  & $96.8 \pm 0.1$ & $17.7 \pm 0.1$ \\
\hline
\end{tabular}
\label{tab:k0_ident}
\end{center}
\end{table}

For the $\Kstar$ signal extraction we built an invariant mass distribution 
of any $\ko+\mbox{\it charged track}$ system and fit it using the following 
relativistic Breit-Wigner function~\cite{Jackson}:
\begin{equation}
\label{eq:BW}
BW(m) = \frac{\Gamma}{(m^2 - M_0^2)^2 + M_0^2\Gamma^2} \left( \frac{m}{q}
\right),
\end{equation}
$$
\mbox{with} \quad \Gamma = \Gamma_0 \left( \frac{q}{q_0} \right)^{2l+1} 
\frac{M_0}{m}
$$
where $M_0$, $\Gamma_0$ are the resonance mass and width, respectively, 
$q$ is the momentum of the decay product in the resonance rest frame ($q_0$ 
corresponds to $M_0$), and $l=1$. We have chosen the following background (BG) 
parametrization:
\begin{equation}
BG = a_1 \Delta^{a_2} e^{-(a_3\Delta + a_4\Delta^2)},
\end{equation}
where $\Delta = m-M_{th}$, $M_{th}$ being the threshold mass 
($m_{K_S^0}+m_{\pi}$). The number of the \Kstar \ mesons extracted from the 
fit to the data (DATA) are given in Tables~\ref{tab:Kstar_stat_cc} 
and~\ref{tab:Kstar_stat_nc}.

\begin{figure}[htb]
\begin{center}
\begin{tabular}{cc}
\mbox{\epsfig{file=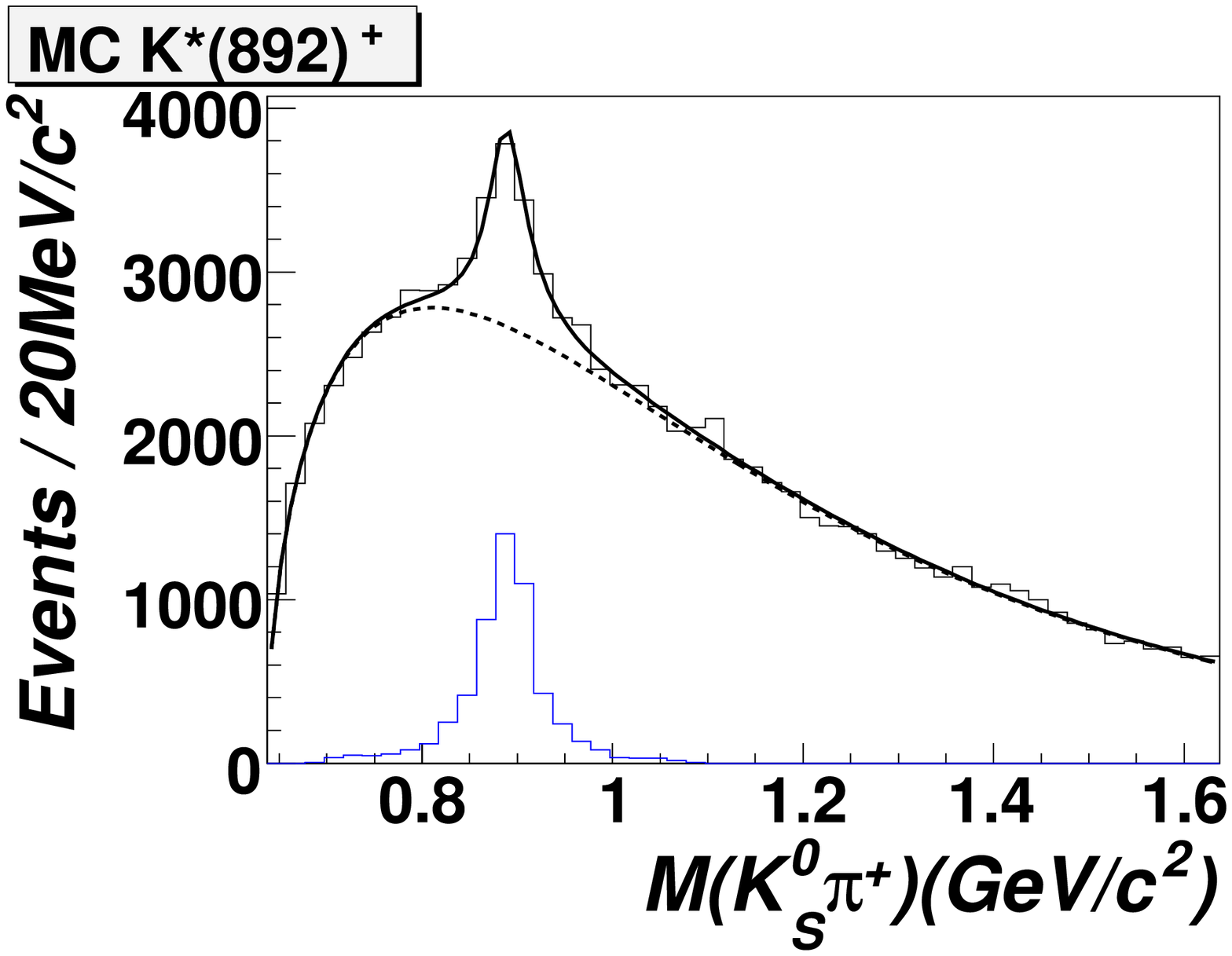,width=0.45\linewidth}} &
\mbox{\epsfig{file=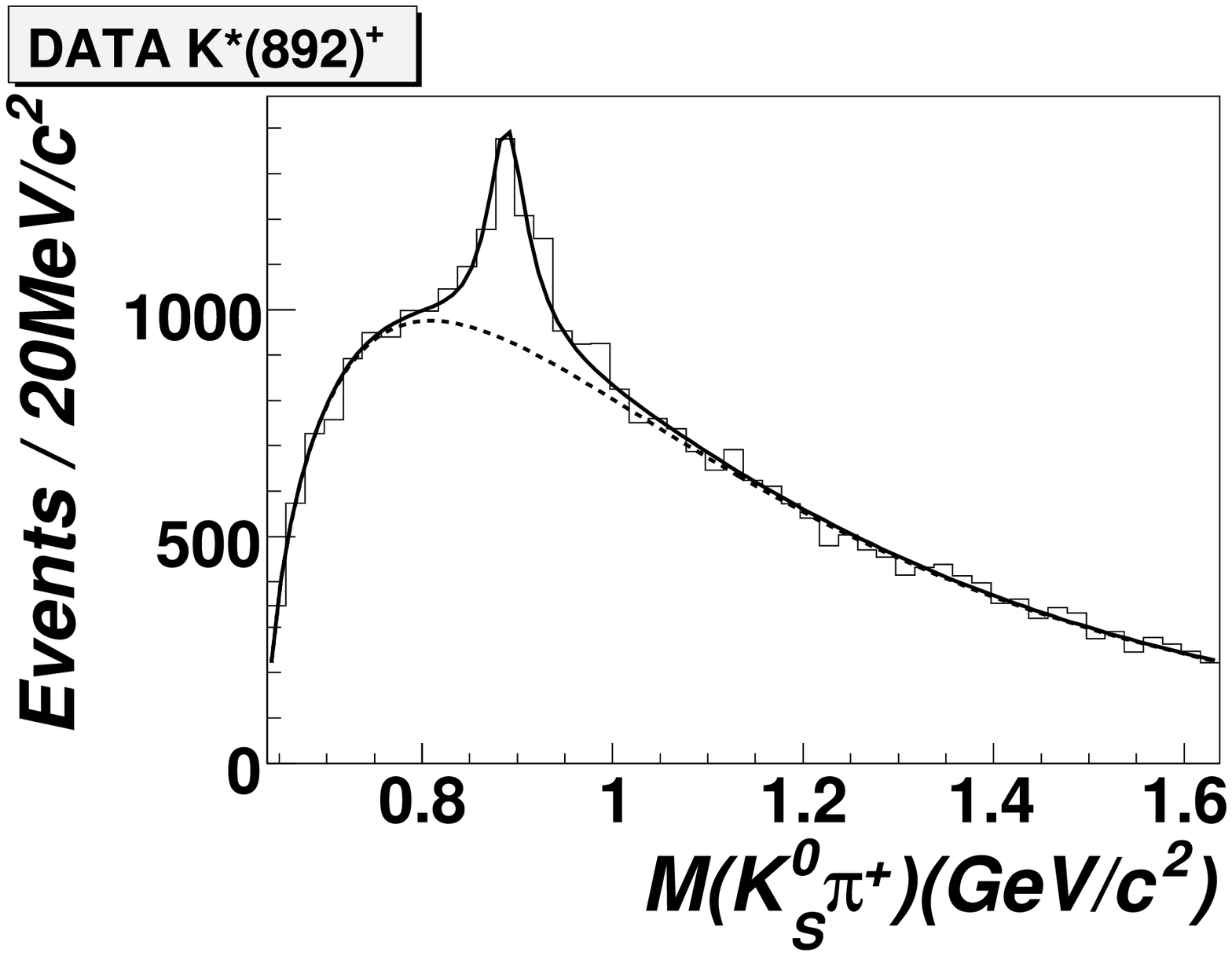,width=0.45\linewidth}}\\
\mbox{\epsfig{file=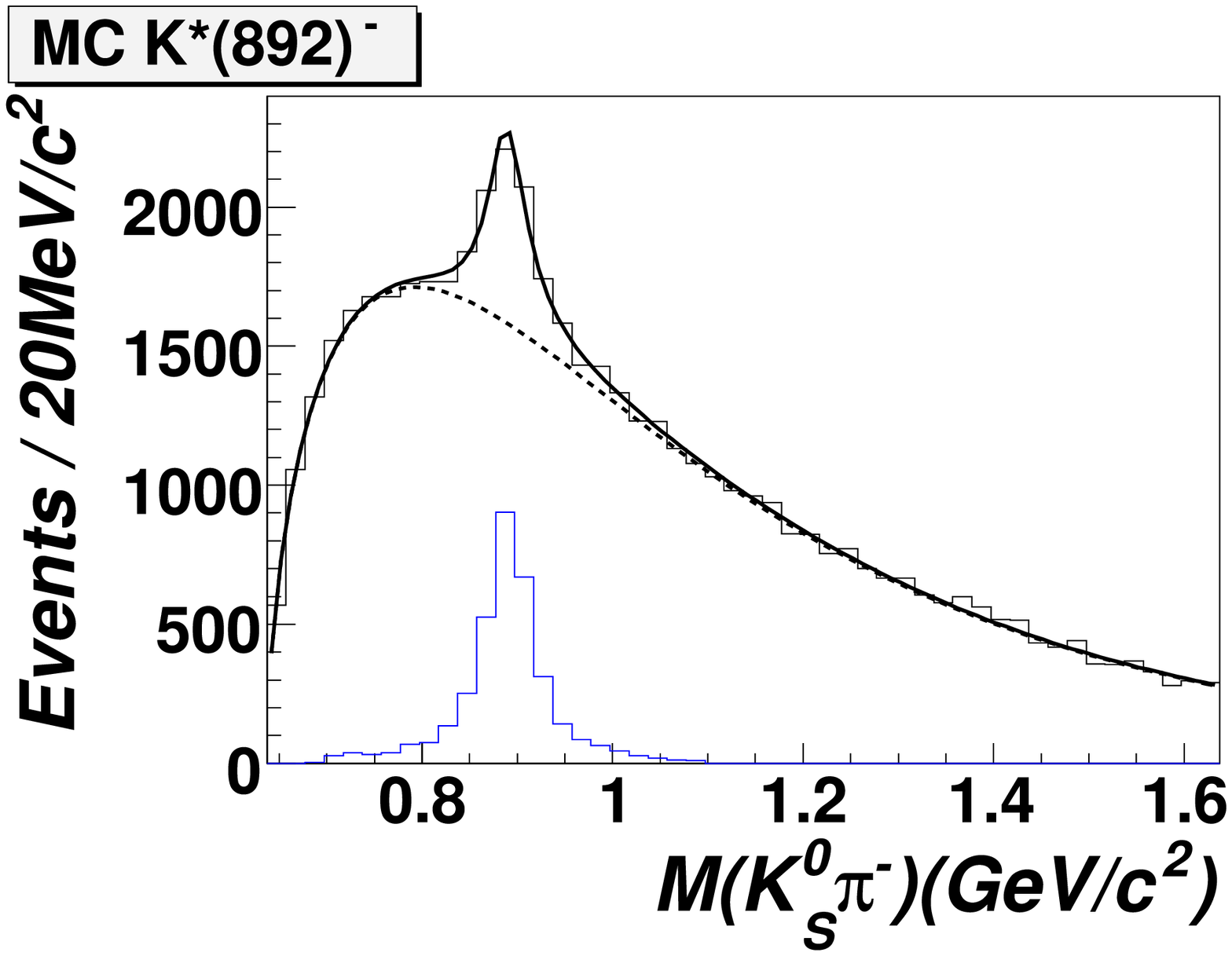,width=0.45\linewidth}} &
\mbox{\epsfig{file=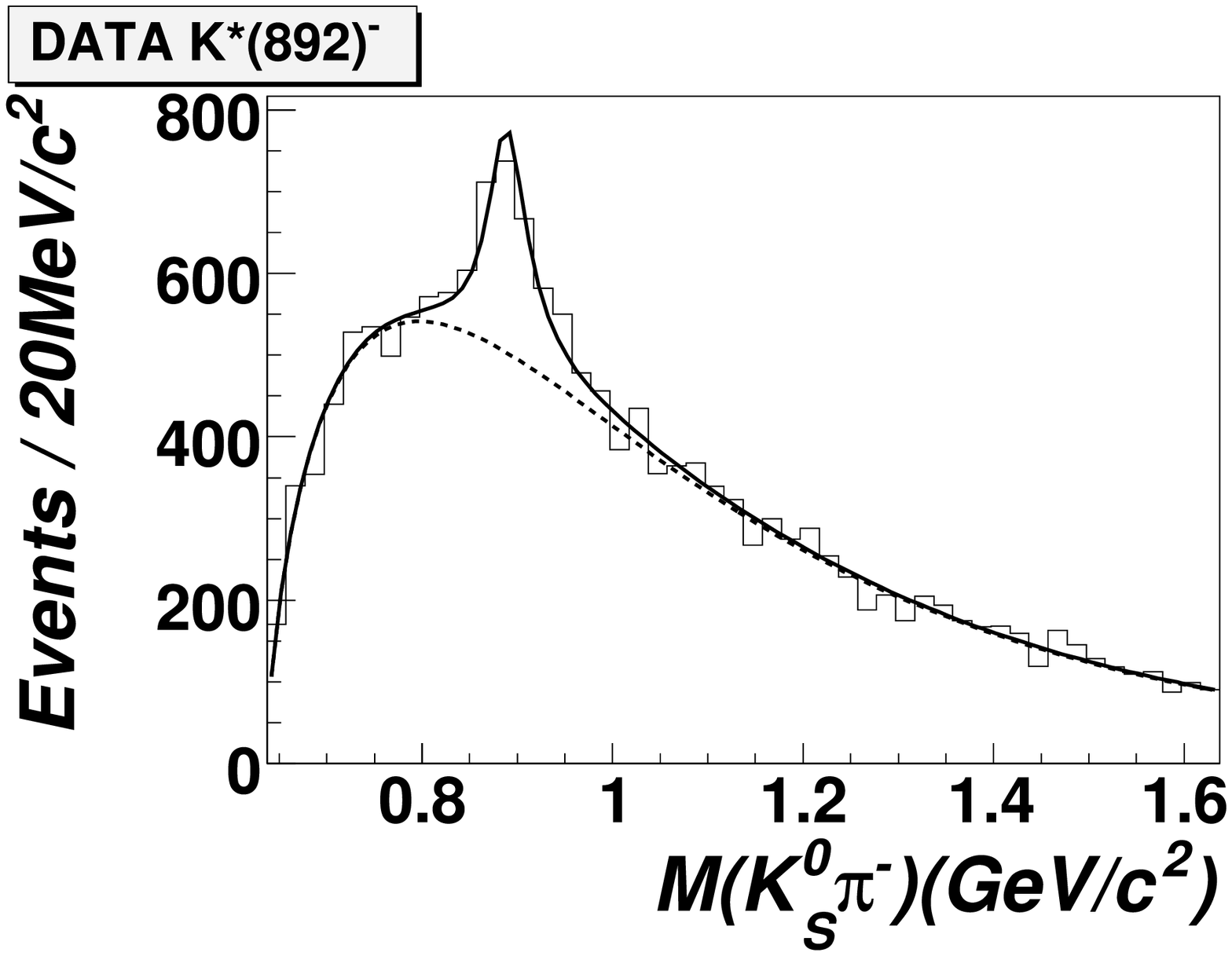,width=0.45\linewidth}}\\
\end{tabular}
\protect\caption{\label{fig:kstar_inv_cc} \it 
$\rm \ko + \mbox{positively charged track}$ (top) and 
$\rm \ko + \mbox{negatively charged track}$ (bottom) invariant mass 
distributions for both MC (left) and data (right) for the \numuCC sample.
The MC plots show the expected signal peaks. Solid line: the result of the fit 
with signal and background, dashed line: only background.}
\end{center}
\end{figure}

\begin{figure}[htb]
\begin{center}
\begin{tabular}{cc}
\mbox{\epsfig{file=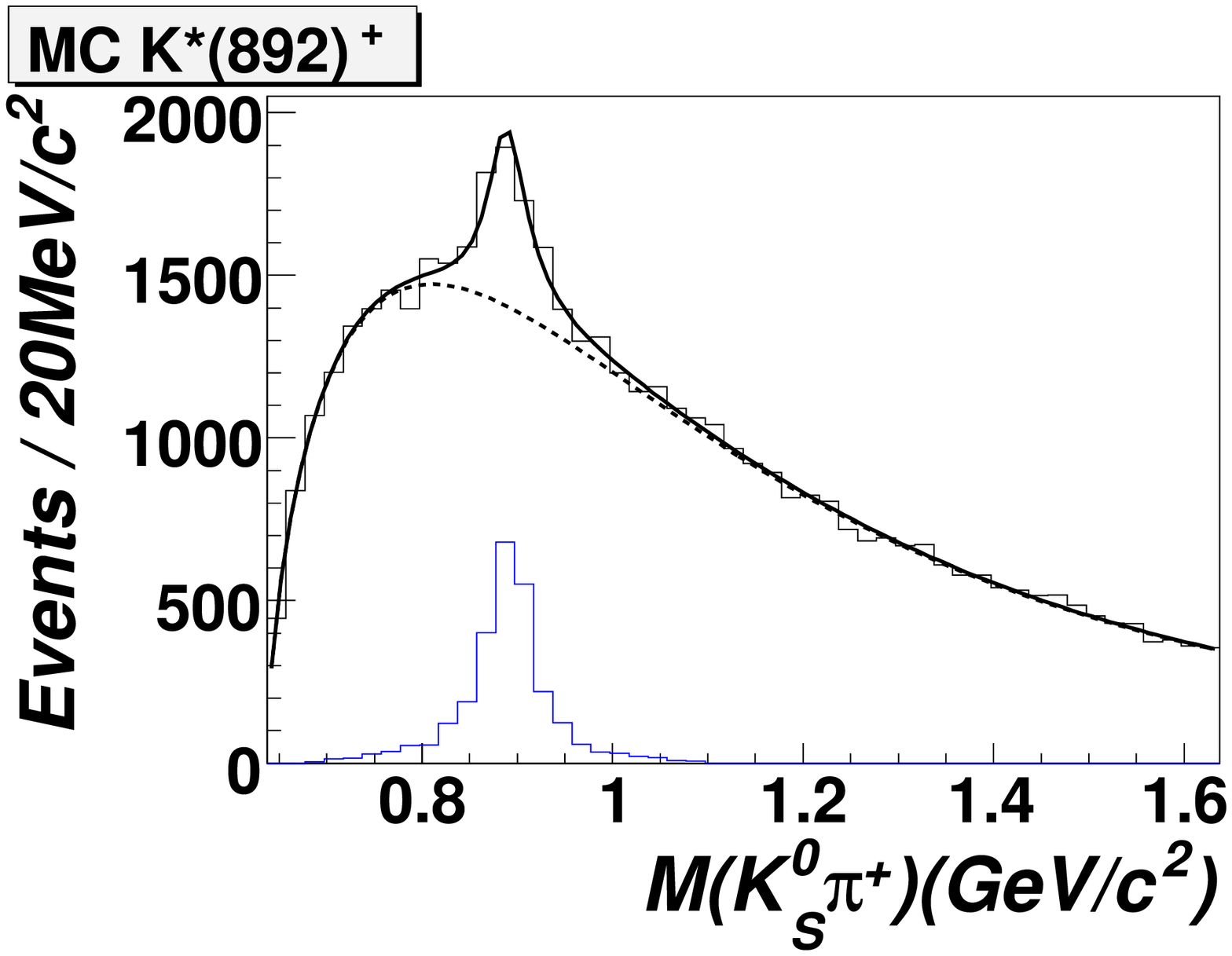,width=0.45\linewidth}} &
\mbox{\epsfig{file=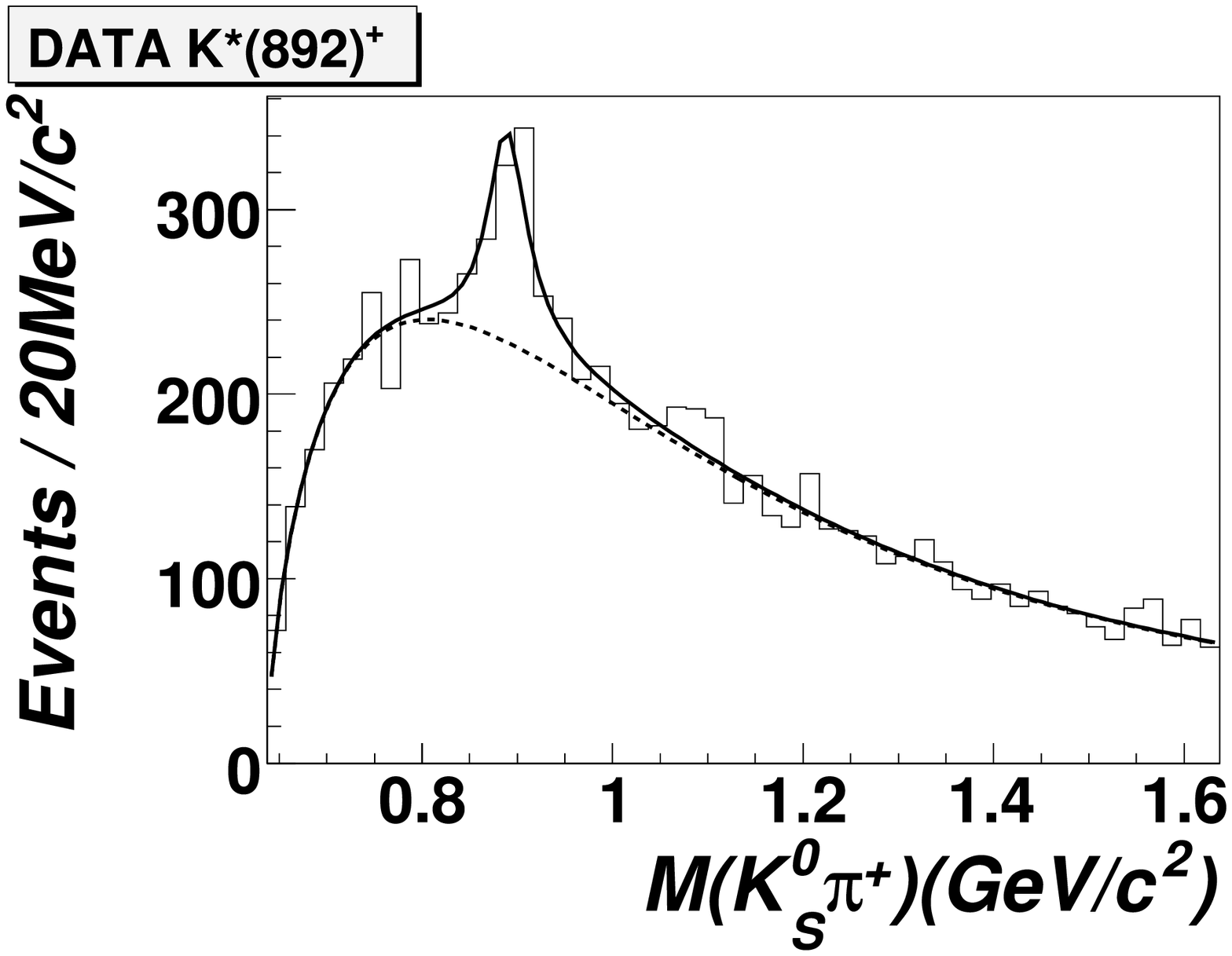,width=0.45\linewidth}}\\
\mbox{\epsfig{file=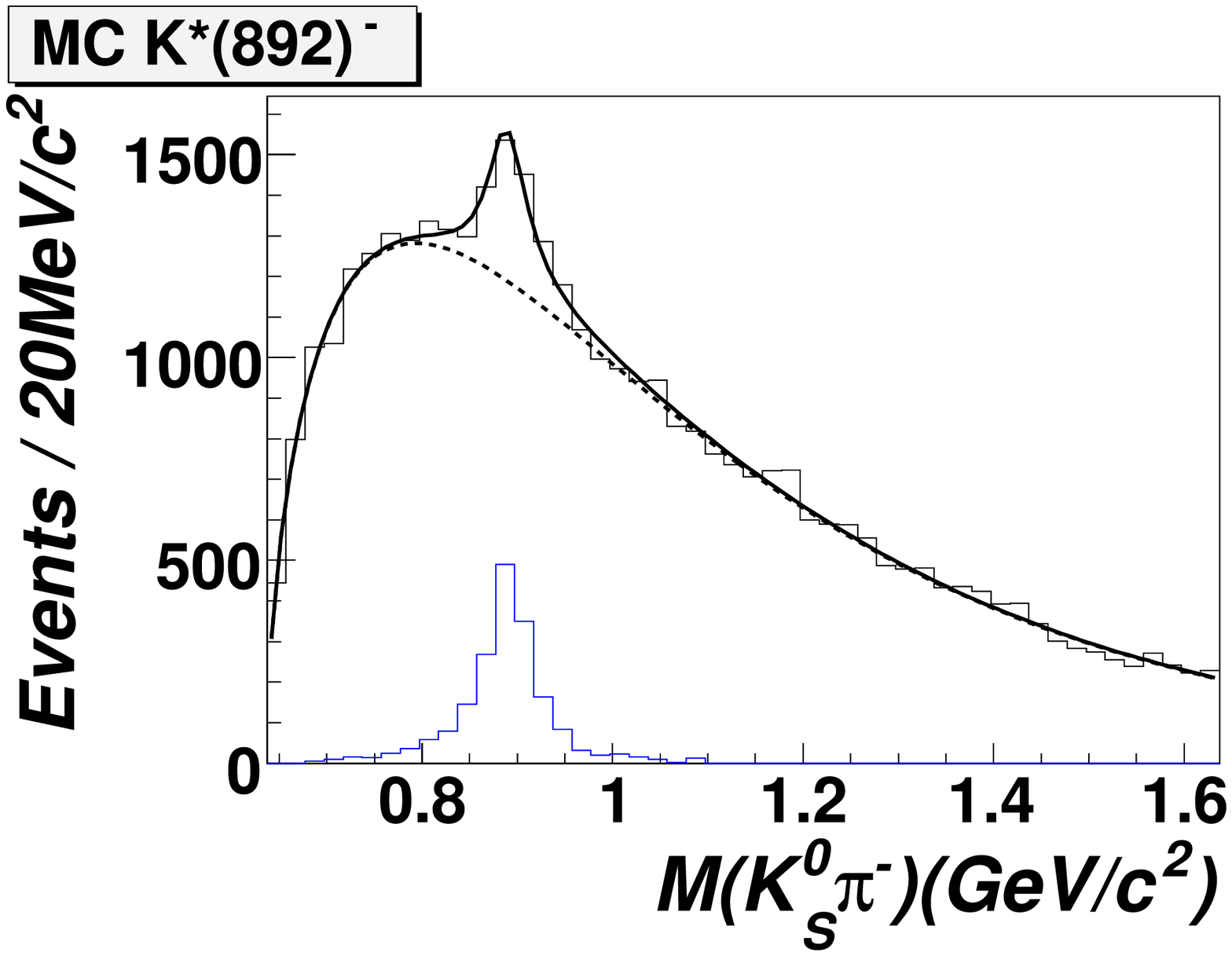,width=0.45\linewidth}} &
\mbox{\epsfig{file=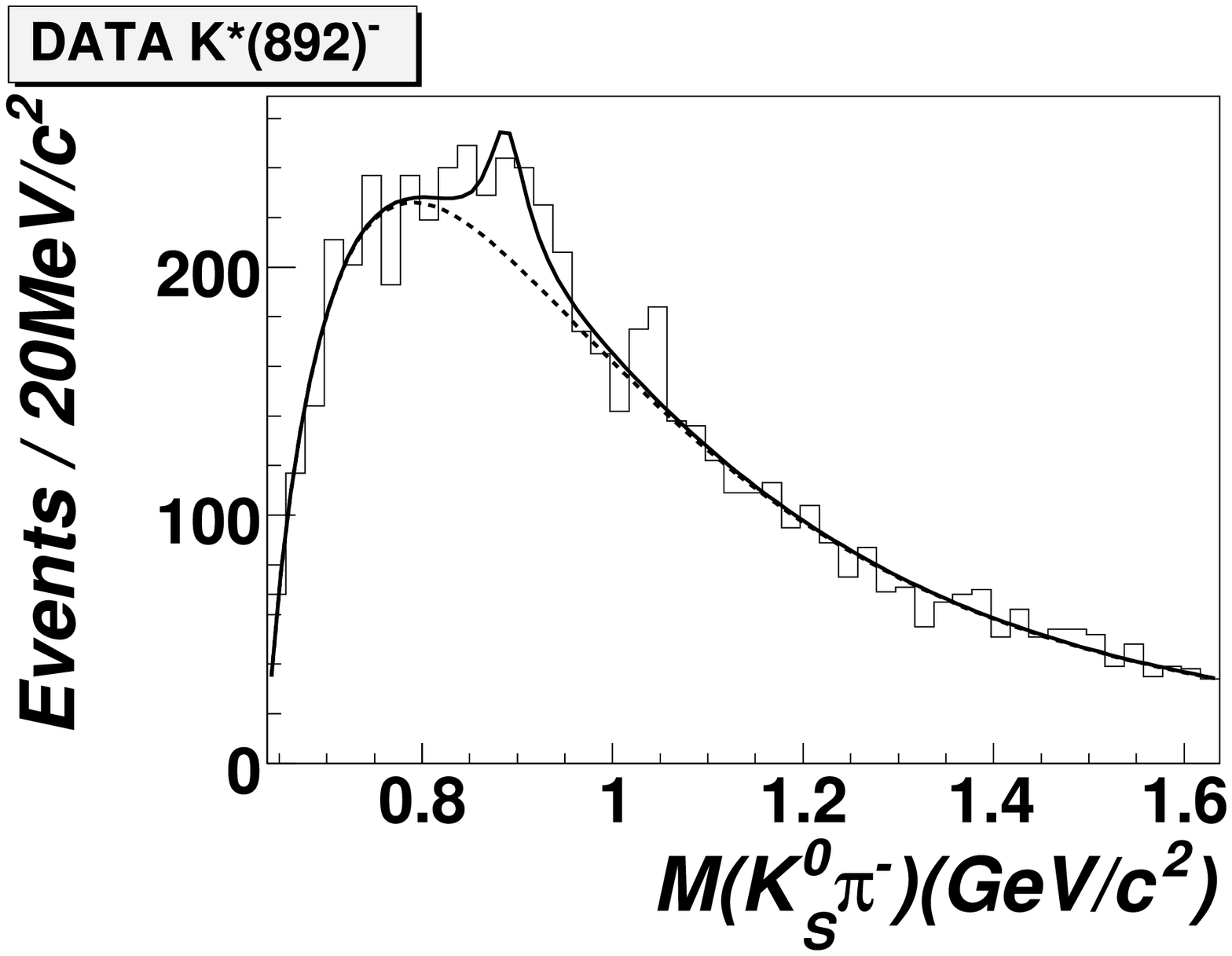,width=0.45\linewidth}}\\
\end{tabular}
\protect\caption{\label{fig:kstar_inv_nc} \it 
$\rm \ko + \mbox{positively charged track}$ (top) and 
$\rm \ko + \mbox{negatively charged track}$ (bottom) invariant mass 
distributions for both MC (left) and data (right) for the \nuNC sample.
The MC plots show the expected signal peaks. Solid line: the result of the fit 
with signal and background, dashed line: only background.}
\end{center}
\end{figure}

In Figs.~\ref{fig:kstar_inv_cc} and~\ref{fig:kstar_inv_nc} we present the 
results of the signal extraction for both data and MC, for the selected 
\numuCC and \nuNC events respectively. Detailed information about extracted 
numbers of $\Kstar^\pm$ mesons in both the tuned MC and data for the \numuCC 
and \nuNC samples can be found in Tables~\ref{tab:Kstar_stat_cc} 
and~\ref{tab:Kstar_stat_nc}, where MC(true) is the total number of 
$\Kstar \to K_S^0 \pi$ decays in the fiducial volume in the MC sample, 
MC(rec) is the number of reconstructed $\Kstar \to K_S^0 \pi$ decays and 
MC(meas) is the number of decays, extracted from the fit. The ratio of 
MC(rec) to MC(true) determines the reconstruction efficiency $\epsilon_r$, 
while the ratio MC(meas) to MC(rec) determines the signal extraction 
efficiency $\epsilon_s$, which takes into account smearing effects due to
momentum resolution. Both $\epsilon_s$ and $\epsilon_r$ were found very 
stable with respect to changes in MC tuning parameters 
(see paragraph~\ref{sec:nomad}).

\begin{table}[htb]
\begin{center}
\caption{\it 
$\Kstar^+ \to \ko \pi^+$ and $\Kstar^- \to \ko \pi^-$ summary for \numuCC \ 
interactions. The number of corresponding decays in the MC is normalized to 
the same number of $\nu_\mu$ CC events as in the real data sample.}
\begin{tabular}{ccc}
\hline
 & $N(\Kstar^+ \to K_S^0 \pi^+)$ & $N(\Kstar^- \to K_S^0 \pi^-)$ \\
\hline
DATA      & $1803 \pm 121$ & $1060 \pm  89$ \\
MC(meas)  & $1846 \pm  80$ & $1066 \pm  61$ \\
MC(rec)   &  2150 &  1374 \\
MC(true)  &  9366 &  5612 \\
\hline
$\epsilon_r$ & $0.23 \pm 0.01$ & $0.24 \pm 0.01$ \\
$\epsilon_s$ & $0.86 \pm 0.04$ & $0.78 \pm 0.05$ \\
\hline
\end{tabular}
\label{tab:Kstar_stat_cc}
\end{center}
\end{table}

\begin{table}[htb]
\begin{center}
\caption{\it 
$\Kstar^+ \to \ko \pi^+$ and $\Kstar^- \to \ko \pi^-$ summary for \nuNC \ 
interactions. The number of corresponding decays in the MC is normalized to 
the same number of \nuNC events as in the real data sample.}
\begin{tabular}{ccc}
\hline
 & $N(\Kstar^+ \to K_S^0 \pi^+)$ & $N(\Kstar^- \to K_S^0 \pi^-)$ \\
\hline
DATA      & $ 443 \pm  60$ & $ 197 \pm  53$ \\
MC(meas)  & $ 385 \pm  26$ & $ 263 \pm  24$ \\
MC(rec)   & 489  & 339  \\
MC(true)  & 2689 & 1718 \\
\hline
$\epsilon_r$ & $0.18 \pm 0.01$ & $0.20 \pm 0.01$ \\
$\epsilon_s$ & $0.79 \pm 0.06$ & $0.78 \pm 0.08$ \\
\hline
\end{tabular}
\label{tab:Kstar_stat_nc}
\end{center}
\end{table}

\clearpage

\subsection{Measurements of the $K^\star$ yields and determination 
of the $\rho_{00}$ parameter}

The measured yield per $\nu_\mu$ CC (or NC) interaction for each \Kstar \ type 
that decays into $K^0 \pi$ is defined as:
\begin{subequations}
\begin{align}
\label{kstar_yields}
T_{K^\star} = \frac{N_{K^\star}^{obs}}{Br(K^0) \cdot \epsilon_{K^\star}}\cdot 
\frac{\epsilon_{\nu_\mu CC\; (NC)}}{N_{\nu_\mu CC\; (NC)}},
\intertext{while the true number of $\Kstar \to K^0 \pi$ decays is given by:}
\label{kstar_number}
N_{K^\star}^{true} = \frac{N_{K^\star}^{obs}}{Br(K^0)\cdot\epsilon_{K^\star}},
\end{align}
\end{subequations}
where 
\begin{itemize}
\item $N_{K^\star}^{obs}$ is the number of $\Kstar$ mesons obtained from the 
fit; 
\item $N_{\nu_\mu CC\; (NC)}$ is the number of reconstructed $\nu_\mu$ CC 
(NC) events;
\item $\epsilon_{\nu_\mu CC}$ ($\epsilon_{\nu_\mu NC}$) is the 
reconstruction and identification efficiency in the fiducial volume for 
$\nu_\mu$ CC (NC) events; 
\item $Br(K^0)=0.686/2$ is the branching ratio of 
\kodecay, where the factor of 2 reflects the observation of \ko \ component 
only. 
\end{itemize}
The combined efficiency $\epsilon_{K^\star}$ is given as a product of 
reconstruction efficiency ($\epsilon_r$) and signal extraction efficiency 
($\epsilon_s$), see Tables~\ref{tab:Kstar_stat_cc} 
and~\ref{tab:Kstar_stat_nc}. The systematic uncertainties on these values 
with respect to changes in the selection criteria, are studied in the next 
subsection.

Note that for the neutral current interactions, formulae~(\ref{kstar_yields}) 
and~(\ref{kstar_number}) should be modified because of $\sim$8\% 
contamination from CC events in the \nuNC sample~\cite{nc_event}. This 
contamination is taken into account in the calculation of the $\Kstar$ yields 
in the \nuNC sample. 

\begin{figure}[htb]
\begin{center}
\begin{tabular}{ccc}
\mbox{\epsfig{file=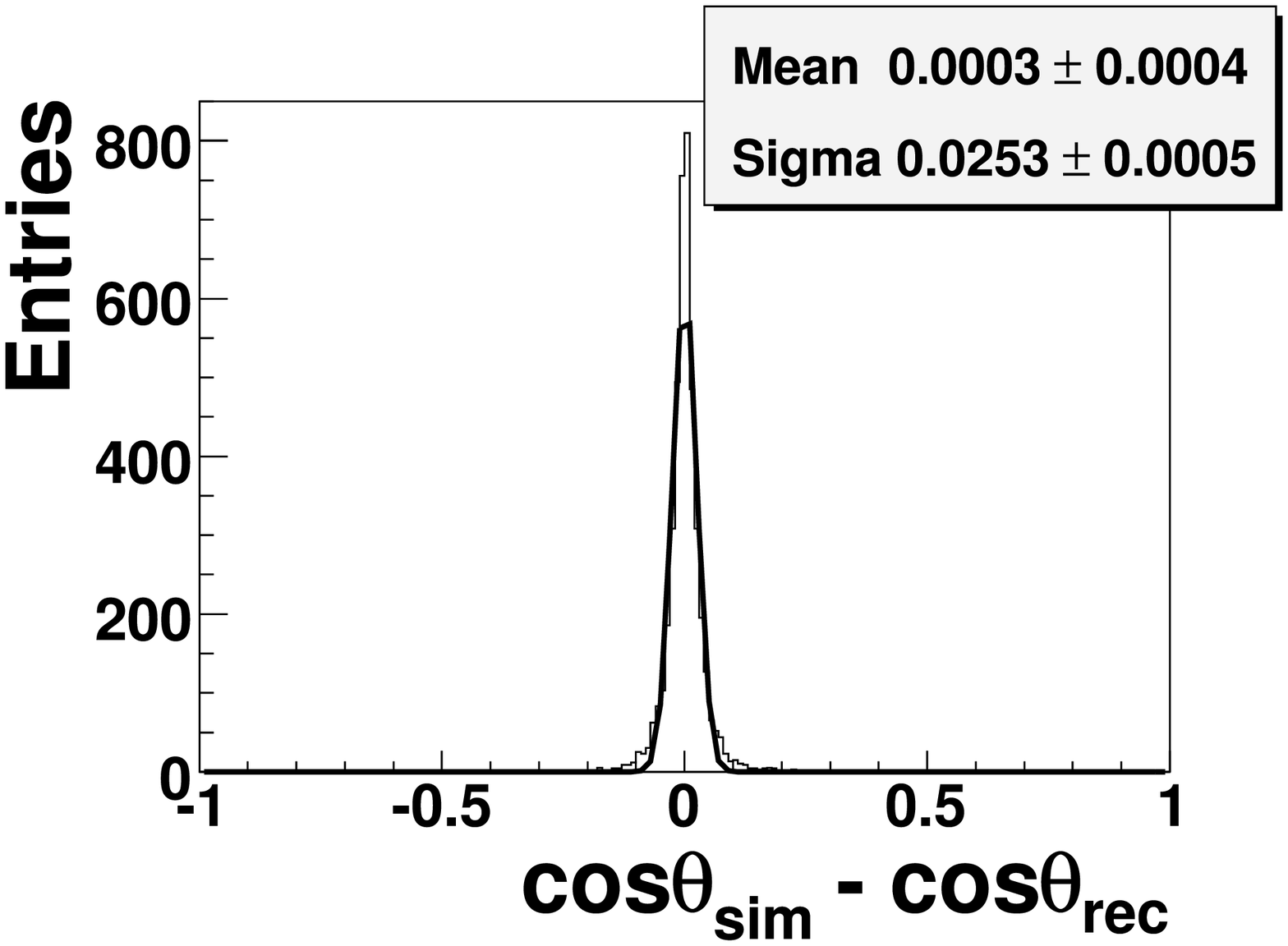,width=0.32\linewidth}}  &
\mbox{\epsfig{file=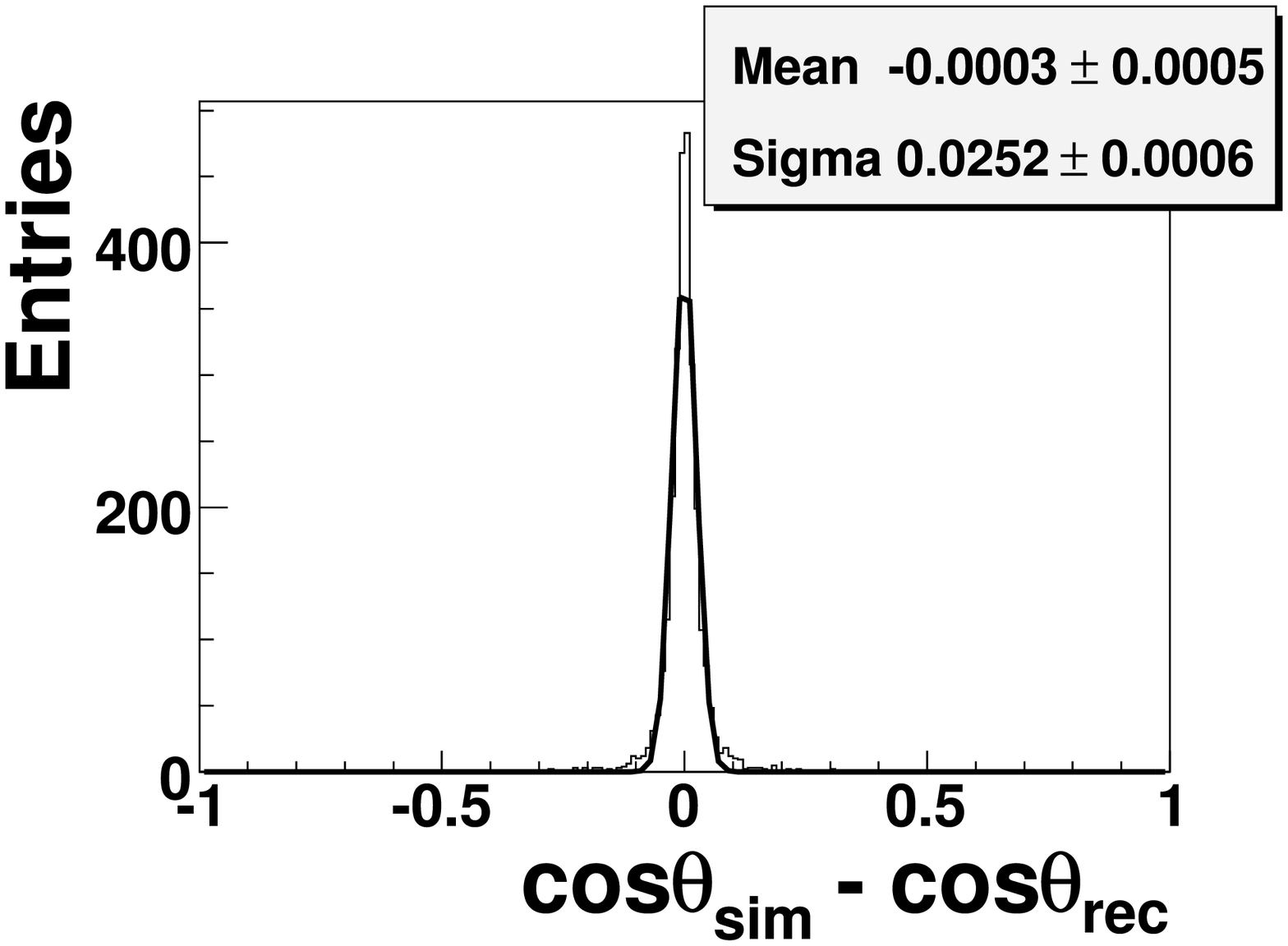,width=0.32\linewidth}} &
\mbox{\epsfig{file=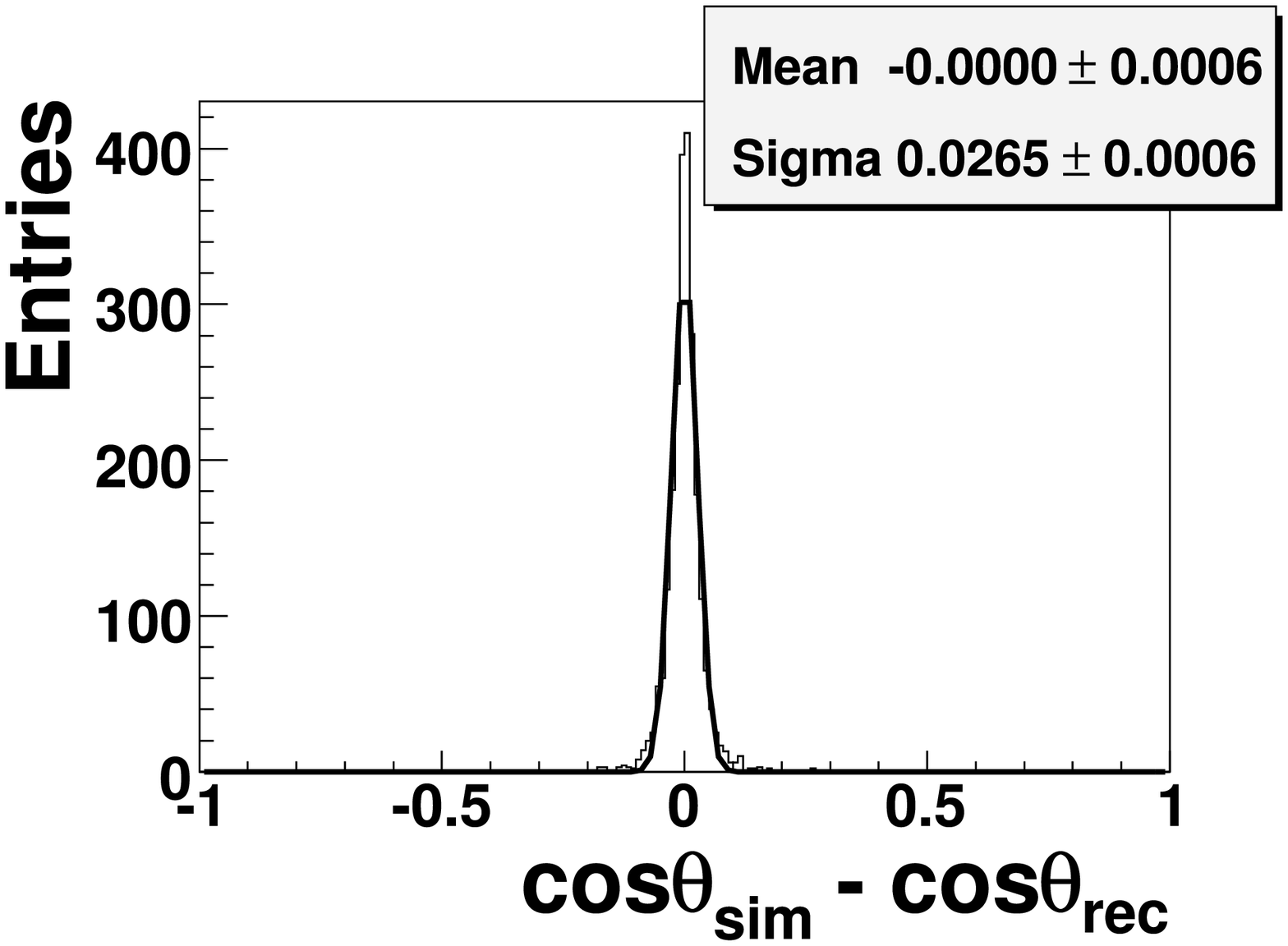,width=0.32\linewidth}}  \\
\mbox{\epsfig{file=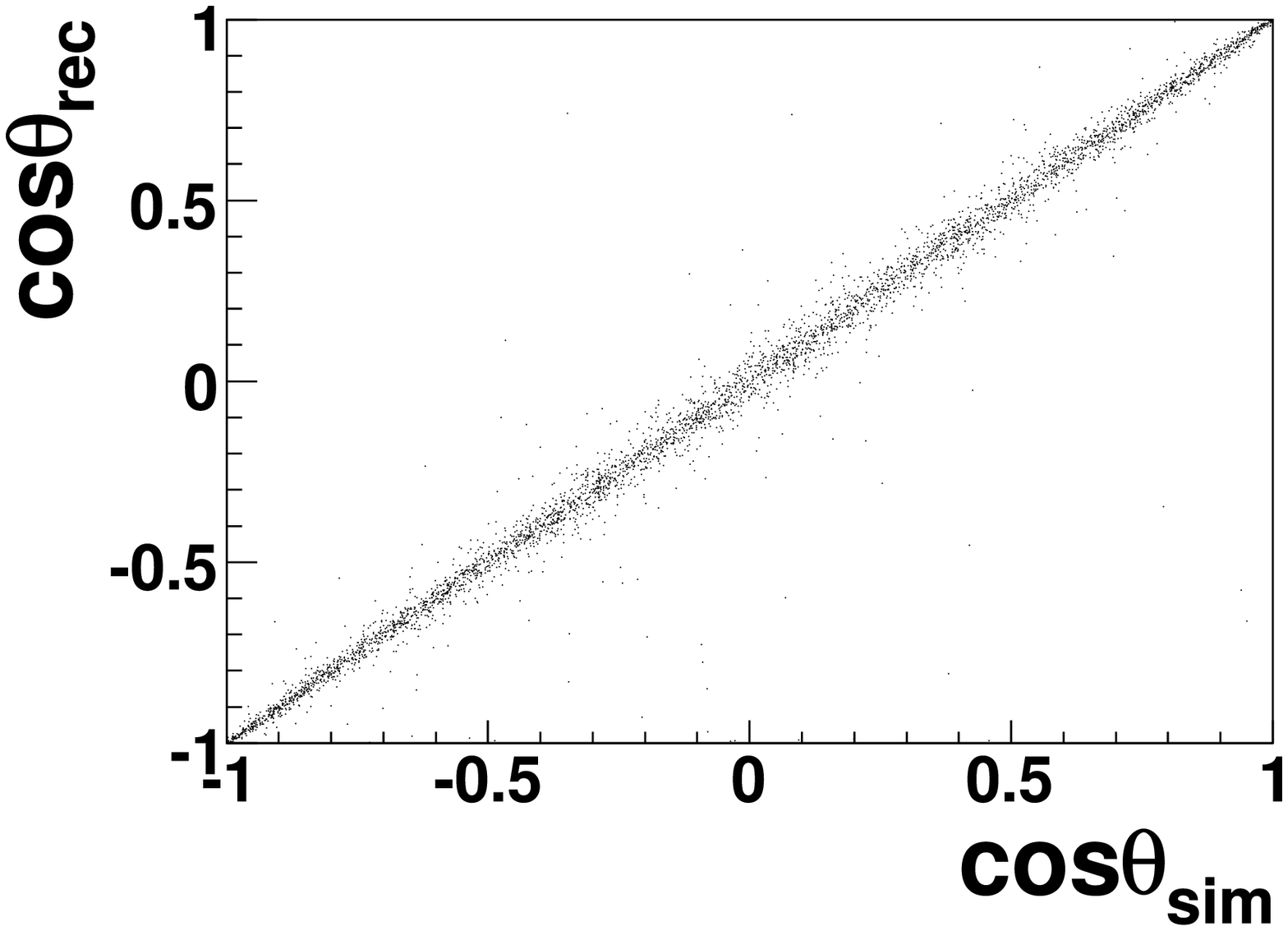,width=0.32\linewidth}} &
\mbox{\epsfig{file=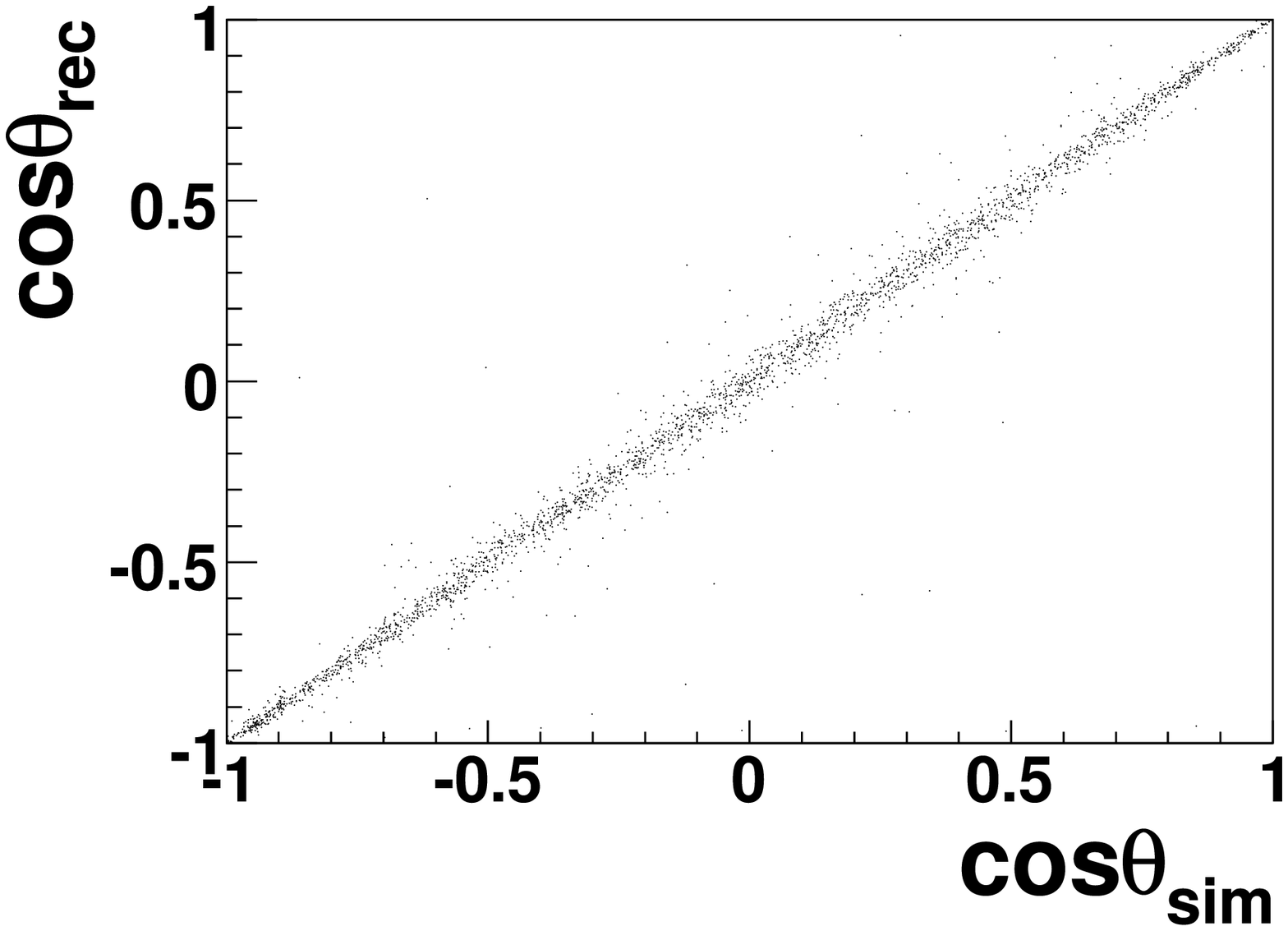,width=0.32\linewidth}}&
\mbox{\epsfig{file=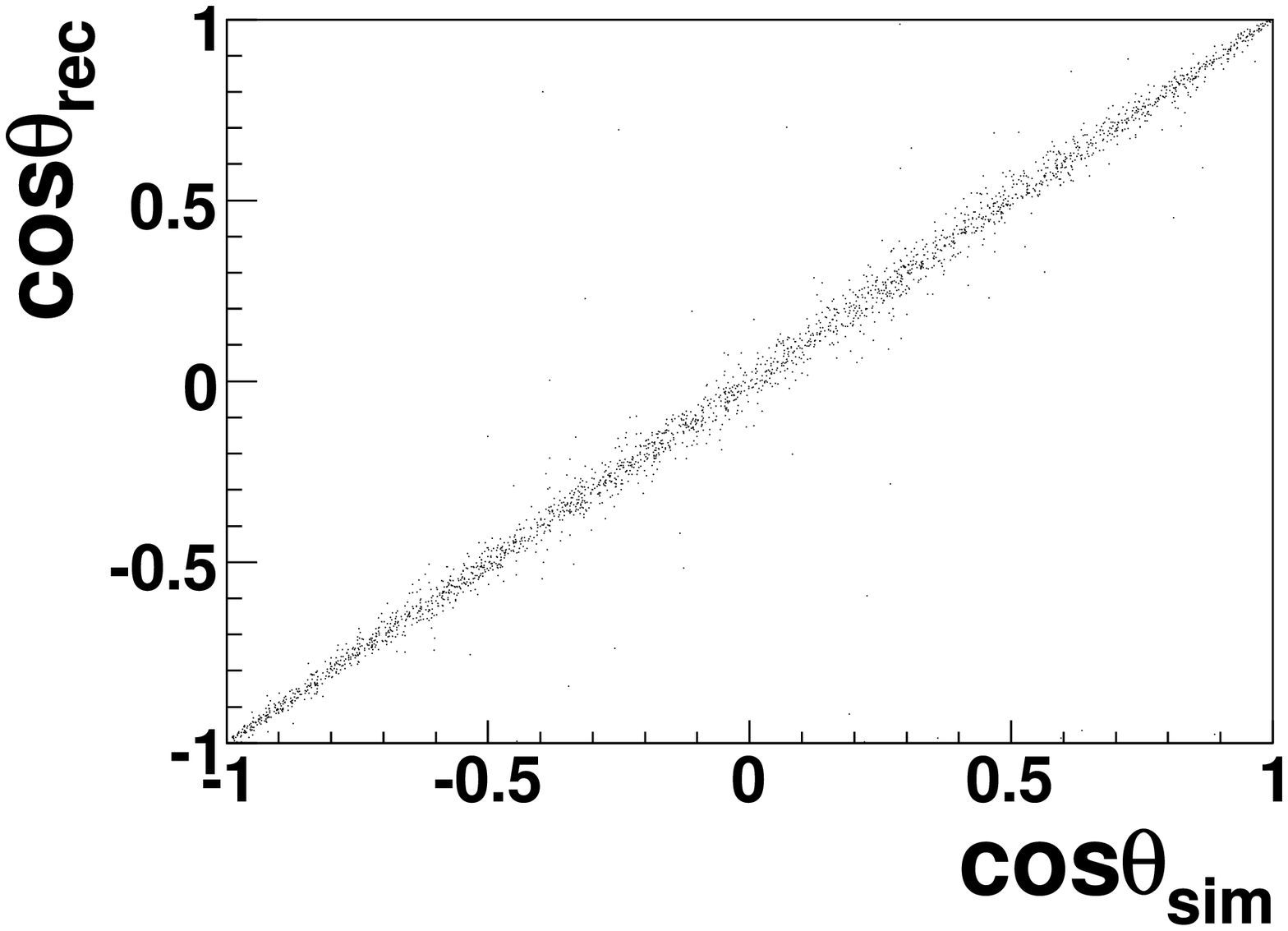,width=0.32\linewidth}}\\
\end{tabular}
\protect\caption{\label{fig:cosine_corr} \it
Correlation between the simulated and reconstructed $\cos \theta$ variables  
for $\Kstar^+$ (left), $\Kstar^-$ (middle) produced 
in \numuCC \ and for $\Kstar^+$ (right) produced in \nuNC \ interactions.}
\end{center}
\end{figure}

\begin{figure}[htb]
\begin{center}
\begin{tabular}{ccc}
\mbox{\epsfig{file=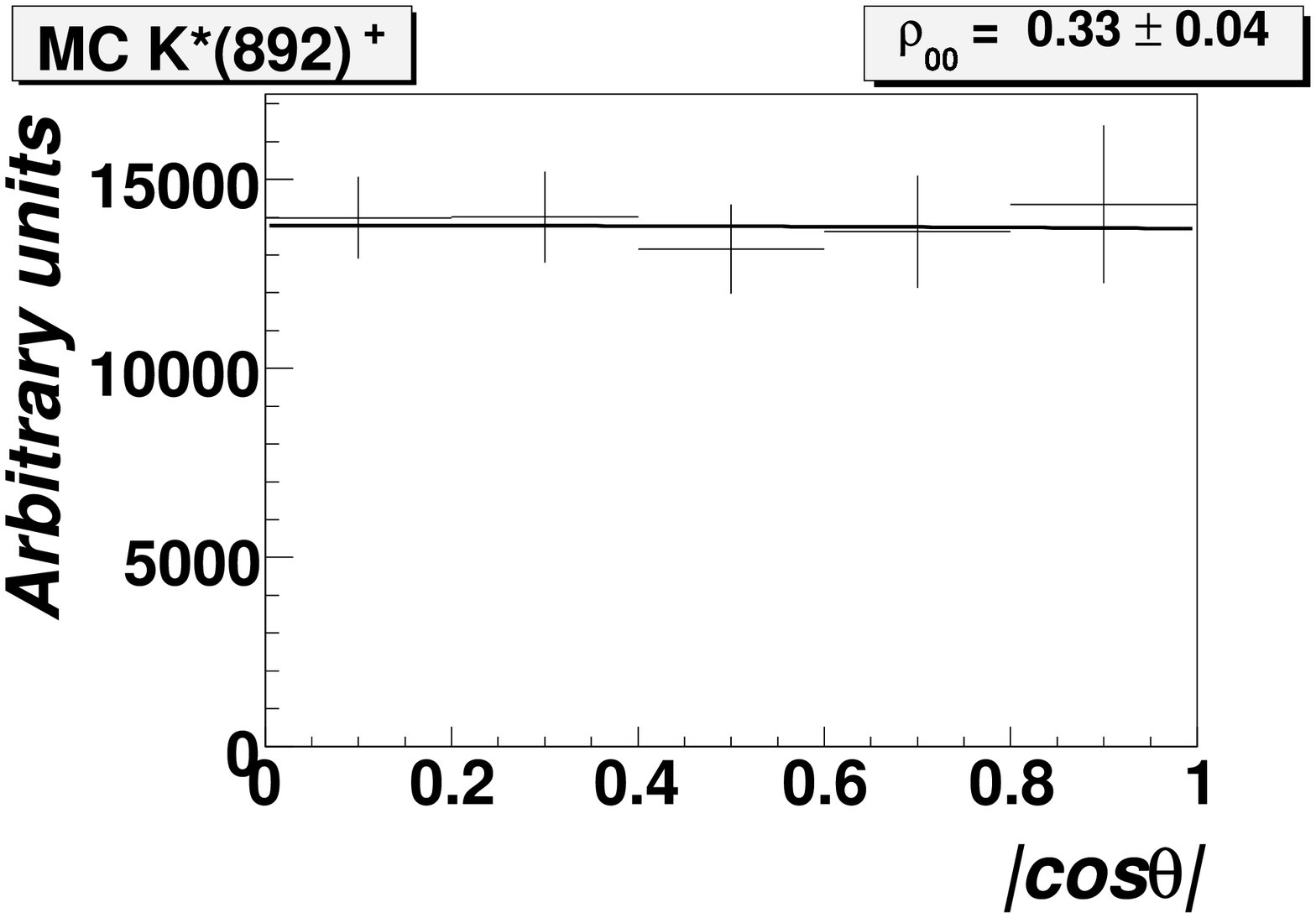,width=0.32\linewidth}} &
\mbox{\epsfig{file=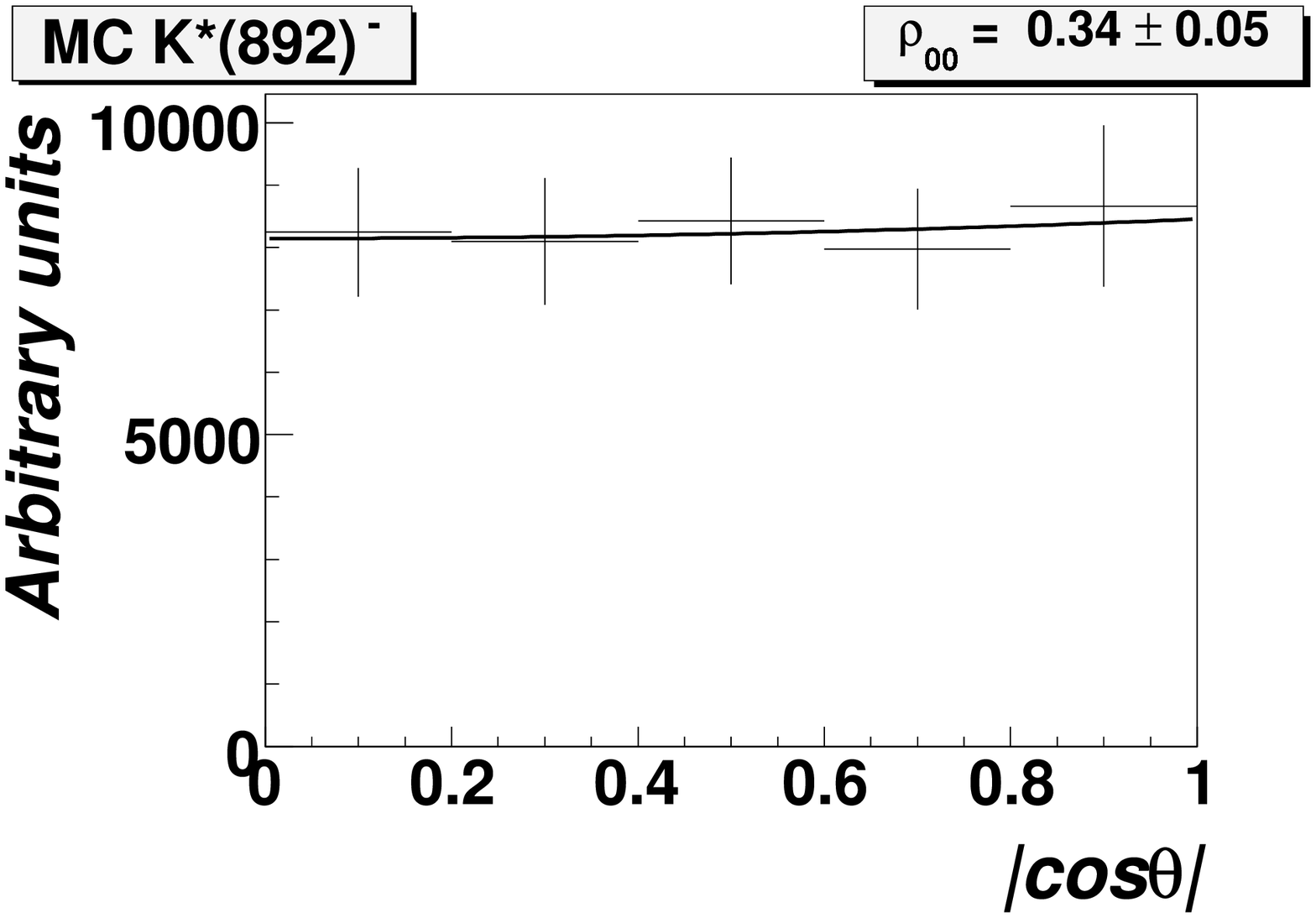,width=0.32\linewidth}} &
\mbox{\epsfig{file=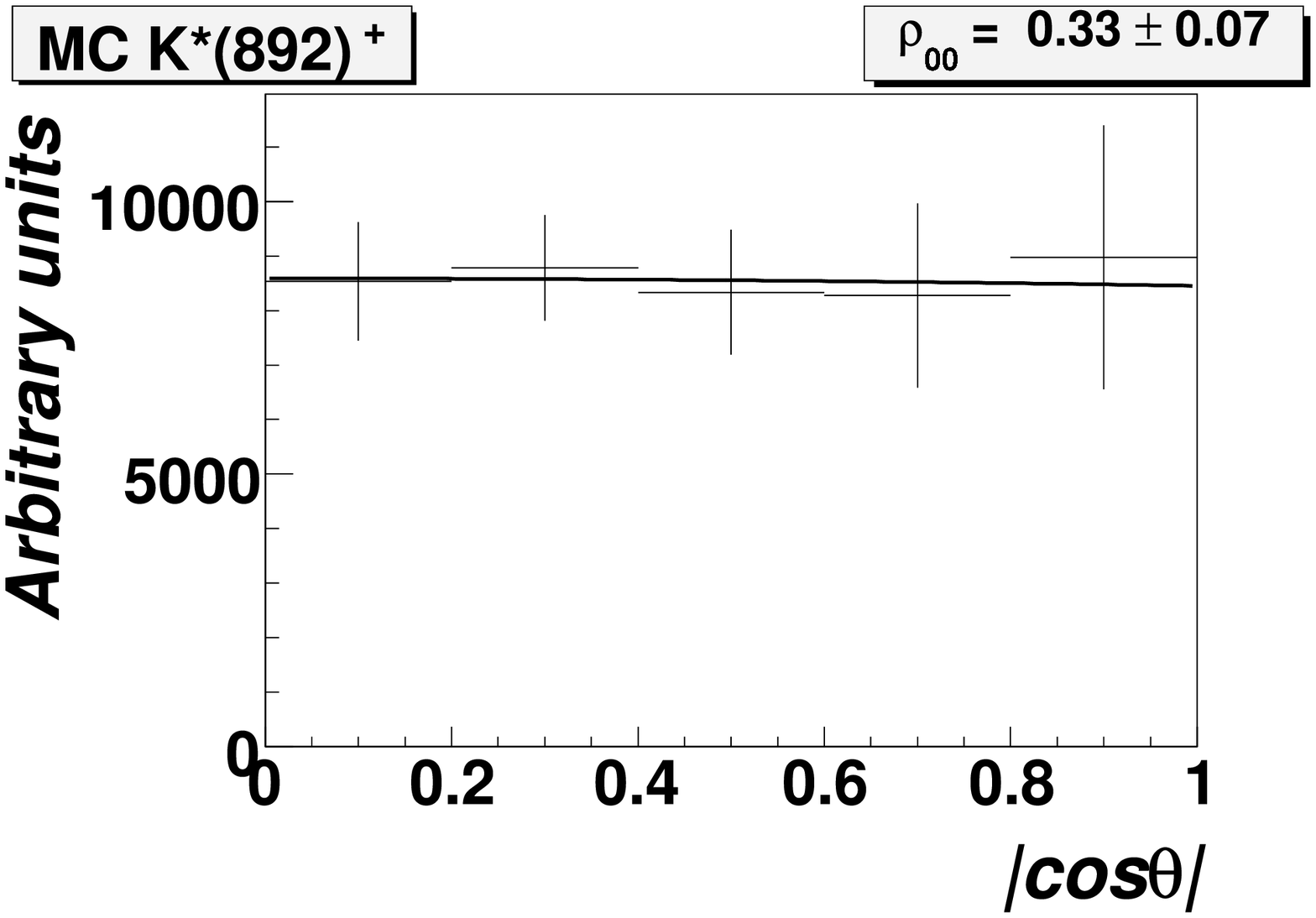,width=0.32\linewidth}} \\
\mbox{\epsfig{file=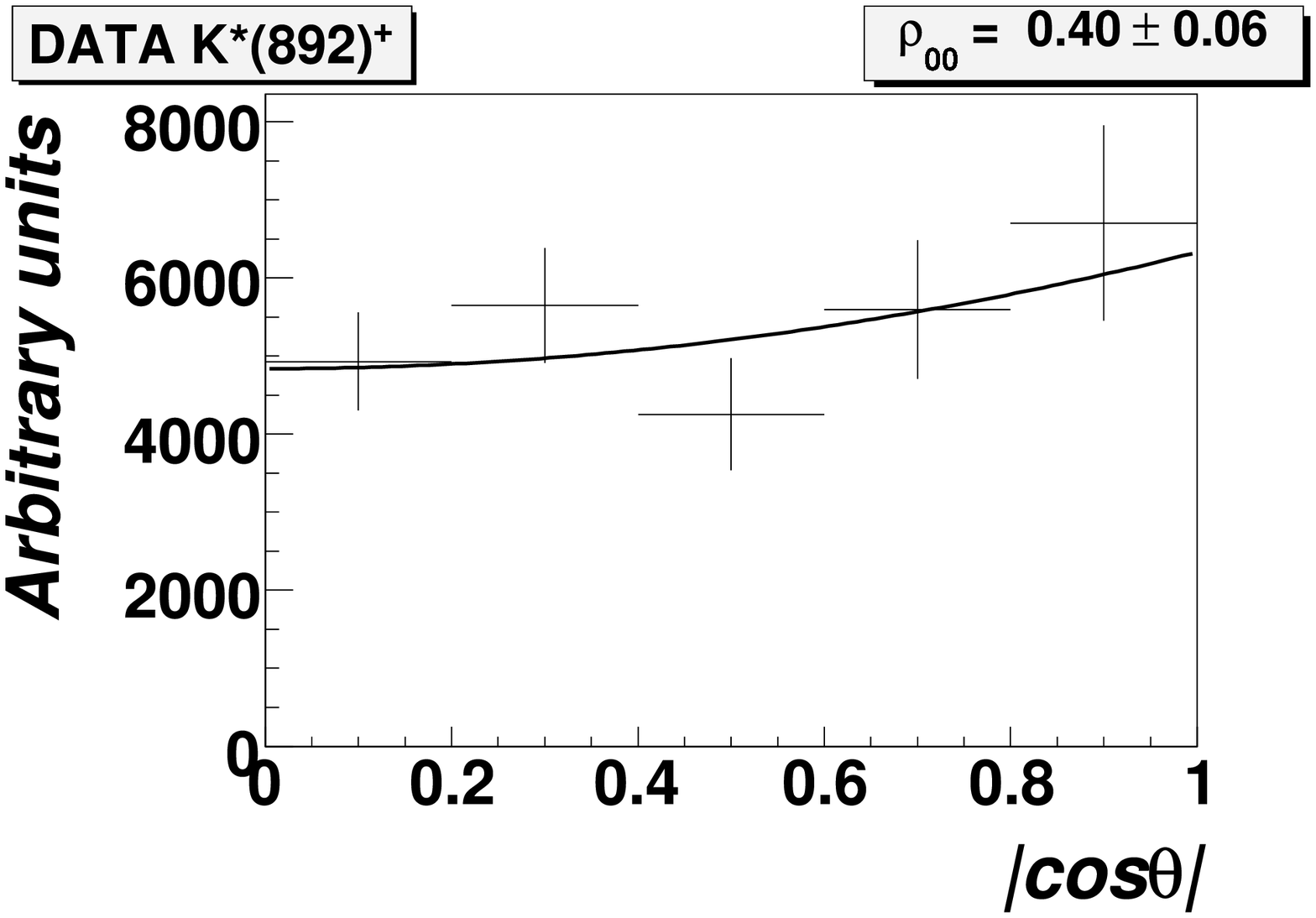,width=0.32\linewidth}} &
\mbox{\epsfig{file=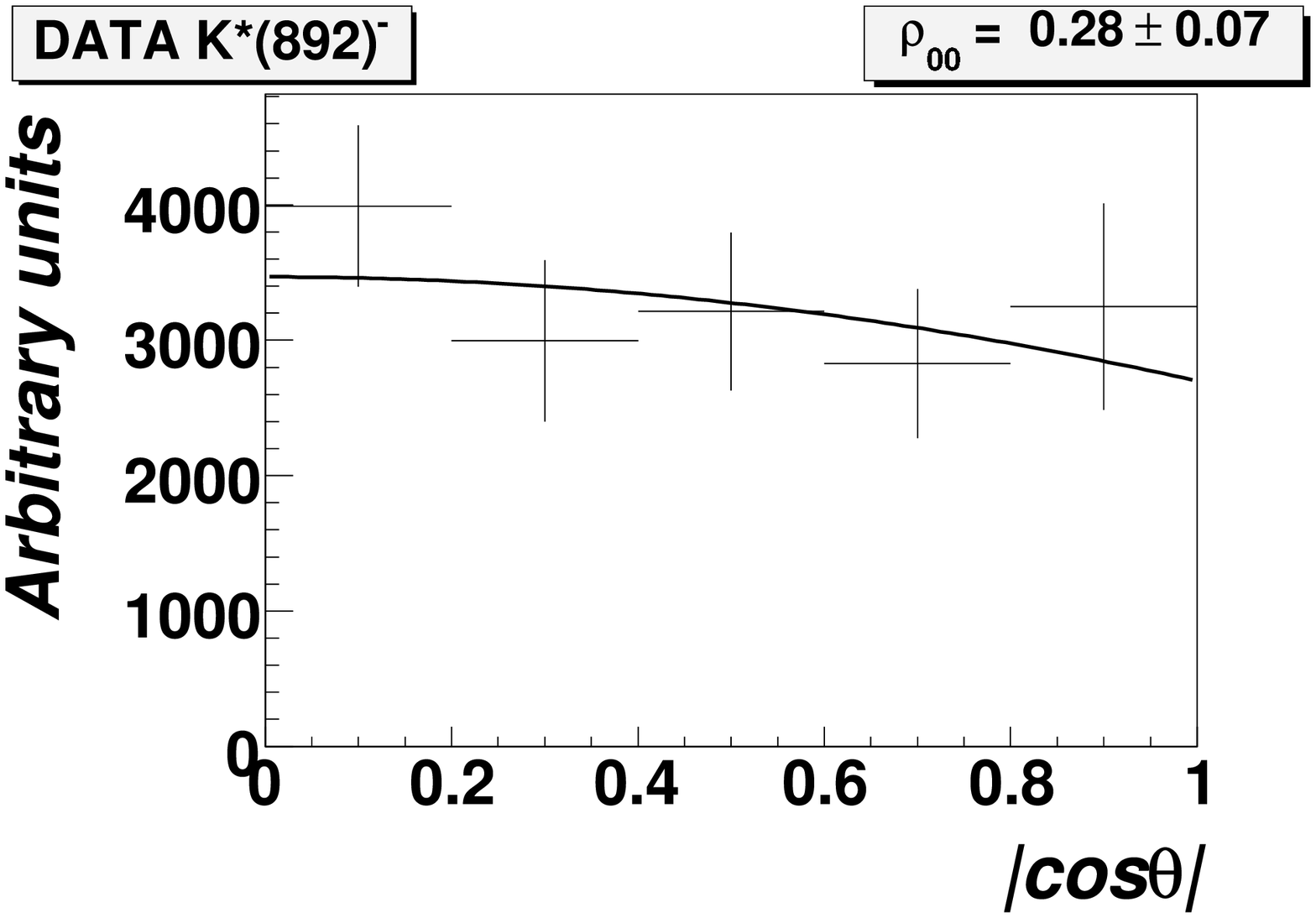,width=0.32\linewidth}} &
\mbox{\epsfig{file=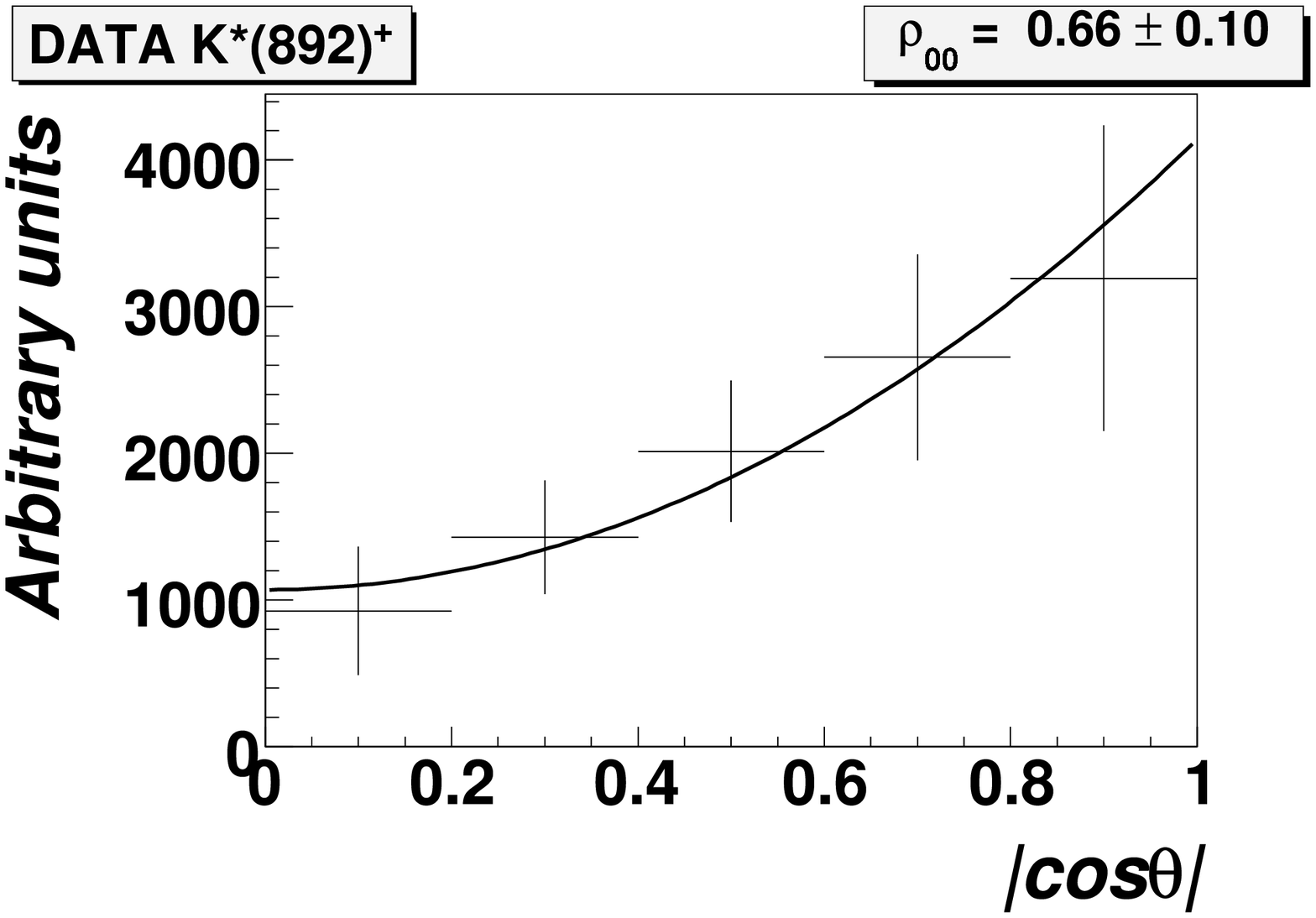,width=0.32\linewidth}} \\
\end{tabular}
\protect\caption{\label{fig:cosine_distr} \it
Angular distributions of the pions from the decay of $\Kstar(892)^+$ 
(left), $\Kstar(892)^-$ (middle) for the \numuCC sample and $\Kstar(892)^+$ 
for the \nuNC sample (right) for both MC (top) and data (bottom). Only 
statistical errors are shown.}
\end{center}
\end{figure}

We have studied the \Kstar \ production properties in bins of several deep 
inelastic and fragmentation kinematic variables. The selection efficiency in 
each bin takes into account the migration of events accross these bins.

The $\rho_{00}$ parameter is determined from the fit of the $\cos \theta$ ($\theta$ is the angle
between $K^*$ direction of flight and the direction of decay $\pi$ in the $K^*$ rest frame)
distribution using the functional form given in equation~(\ref{rho00_distr}).

In Fig.~\ref{fig:cosine_corr} we present the correlation between the simulated 
and reconstructed $\cos \theta$ variables for $\Kstar^+$ and $\Kstar^-$ 
produced in \numuCC \ MC as well as for $\Kstar^+$ produced in \nuNC \ MC 
interactions. The resolution in $\cos \theta$ is found to be better than 
$0.03$ and does not depend significantly on the value of $\cos \theta$.

The $\cos \theta$ distributions for both data and MC for the \numuCC and 
\nuNC samples are presented in Fig.~\ref{fig:cosine_distr}. The MC plots (left)
confirm that the analysis procedure is self-consistent since we do not 
observe spin alignment in the MC ($\rho_{00} = 1/3$).

\subsection{Systematic uncertainties} \label{sec:systematic}

\begin{figure}[htb]
\begin{center}
\begin{tabular}{cc}
\mbox{\epsfig{file=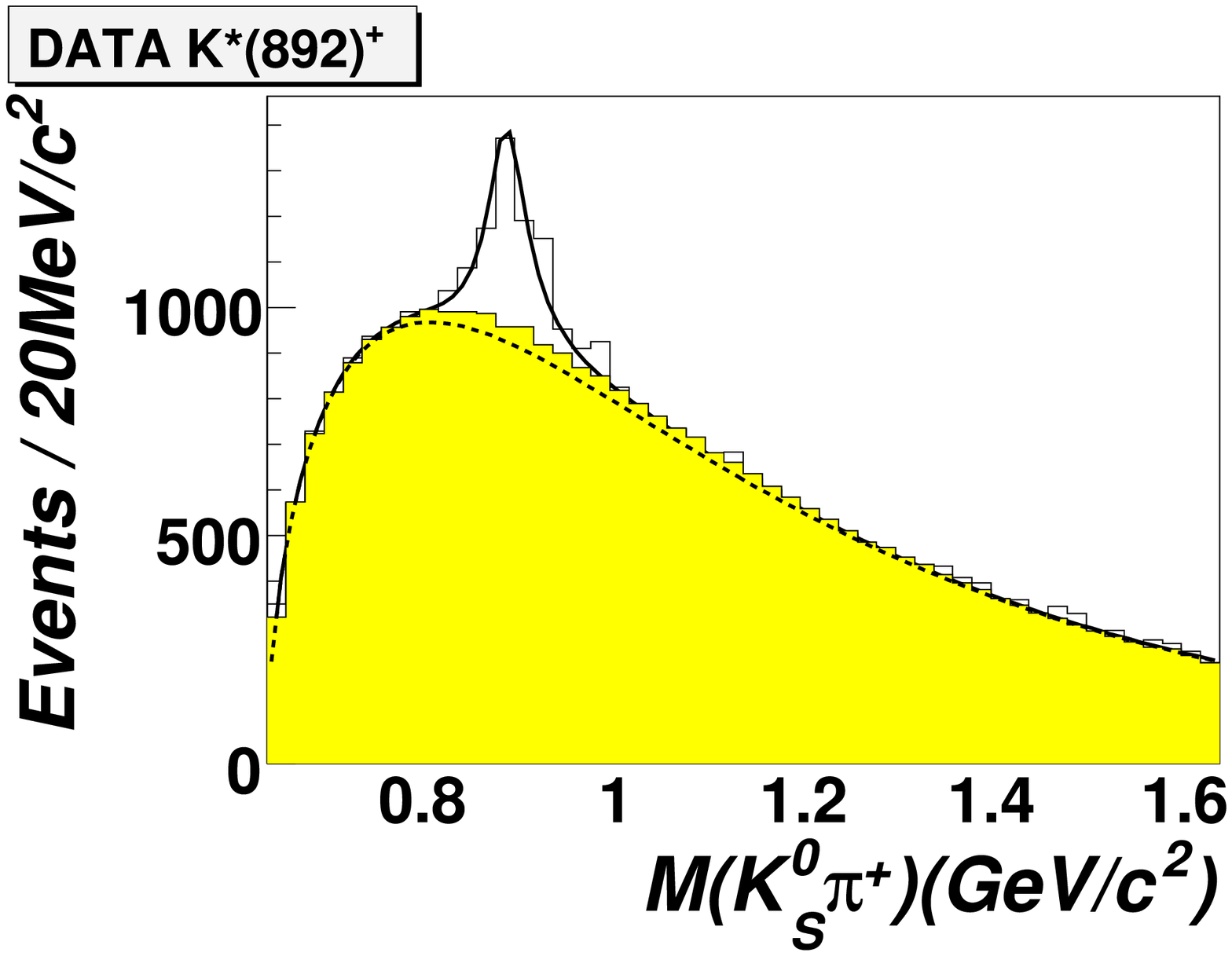,width=0.45\linewidth}} &
\mbox{\epsfig{file=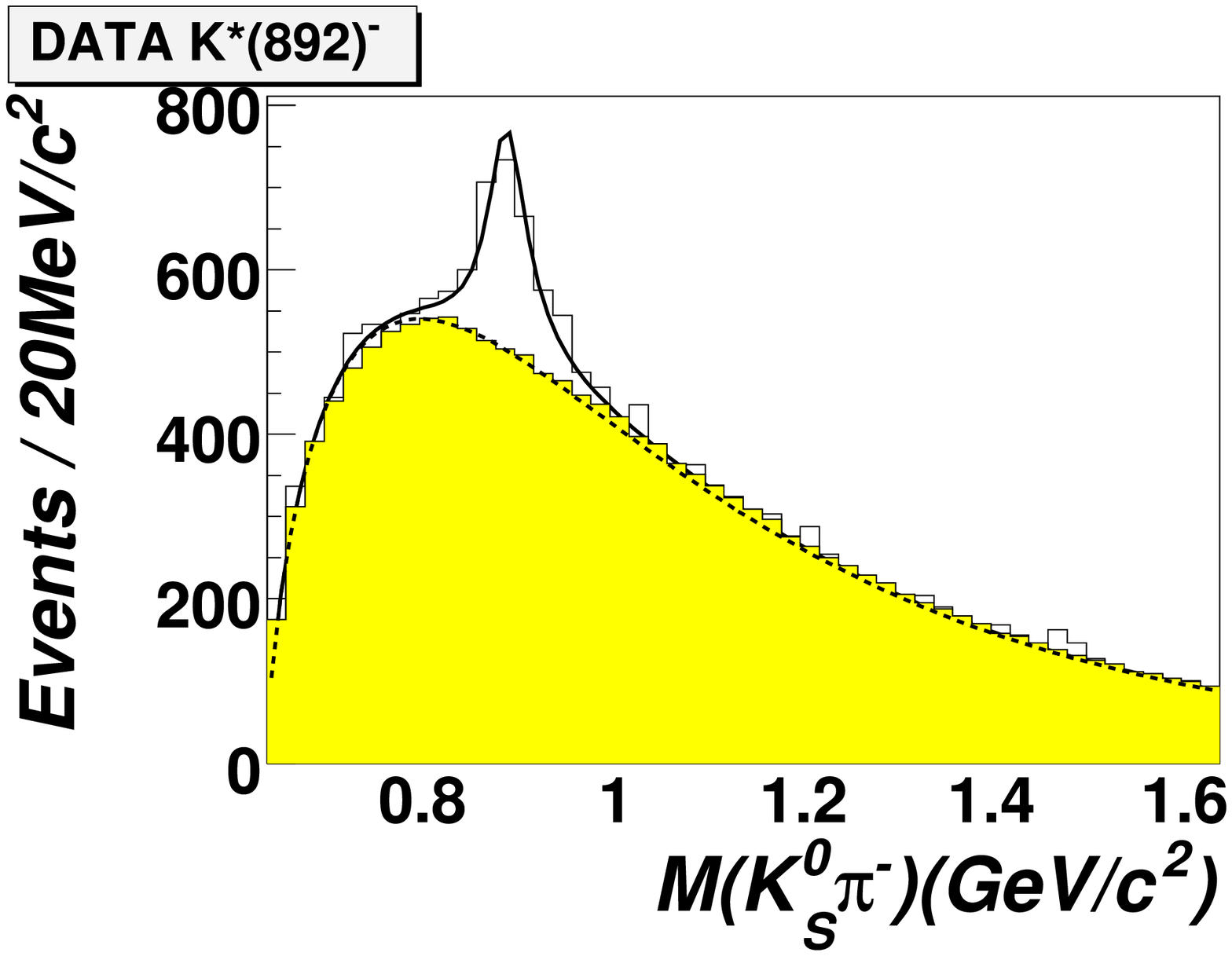,width=0.45\linewidth}}\\
\end{tabular}
\protect\caption{\label{fig:fake_inv} \it 
$\rm \ko + \mbox{positive charged track}$ (left) and 
$\rm \ko + \mbox{negative charged track}$ 
(right) invariant mass distributions for the \numuCC data sample. 
Solid line: the result of the fit with signal and background, dashed line: 
only background. Filled area shows the estimated background distribution when 
$\ko$ and {\it charged track} are taken from different events.}
\end{center}
\end{figure}

We have studied different sources of systematic uncertainties.

To investigate the dependence of the results on the $\ko$ selection 
criteria~\cite{lam_polar} we varied them within the following ranges 
(variations of these cuts correspond to changes of up to 6\% in the 
statistics of the $\ko$ sample): 

\begin{itemize}
\item cut on the transverse momentum of the pion from $\ko$ decay from 0.01 
      to 0.03 GeV/c (this cut can affect the contamination from ``fake'' 
      $\ko$'s in the data). The default value is 0.02 GeV/c;
\item cut on the $\ko$ momentum component perpendicular to the line connecting 
      the primary and \vo \ vertices from 0.09 to 0.115 GeV/c (this cut 
      affects mainly the contamination from secondary interactions). The 
      default value is 0.1 GeV/c;
\item cut on the $\chi^2$ probability of the $V^0$ vertex reconstruction 
      changed from 0.005 to 0.035. The default value is 0.01;
\item cut on the measured decay path of $\ko$ mesons varied from 12 to 30 cm.
      The default value is 16 cm.
\end{itemize}

We have also investigated the influence of the number of bins used in the 
invariant mass fit. The number of bins was changed to 40 and 70 (the default 
value is 50).

For the NC sample we varied the likelihood selection 
criteria~\cite{nc_event} from 0 to 1 (the default value is 0.5).

We estimate the total systematic uncertainty as the sum in quadrature of the 
largest deviation with respect to the reference results in each of the above 
tests (neglecting possible correlations between different cuts).

In order to check for possible enhancements in the \Kstar \ signal region due 
to combinatorial effects as well as to validate the shape of the background 
distribution, we built the invariant mass of the 
$\ko+\mbox{\it charged track}$ system where the $\ko$ candidate and the 
{\it charged track} were taken from different events, rotated such that the 
reconstructed hadronic momentum vectors coincided. Then the background was 
normalized to the invariant mass region above 1.1~GeV. The results are shown 
in Fig.~\ref{fig:fake_inv}. The shape of the background distribution is 
reasonably well described by the fitting procedure presented 
in Sec.~\ref{sec:kstar_signal}.

\section{Results} \label{sec:resuts}

In this section we present the results for $\Kstar(892)^\pm$ that decay 
into $K^0 \pi^\pm$ modes.

Tables~\ref{tab:results_yields} and~\ref{tab:results_rho} summarize results 
for the total numbers, absolute yields, relative yields and the $\rho_{00}$ 
parameters for $\Kstar(892)^\pm$ mesons in \numuCC and \nuNC interactions. 
As expected, the yields of $\Kstar^+$ mesons are larger than the yields of 
$\Kstar^-$ mesons in both \numuCC and \nuNC interactions. From 
table~\ref{tab:results_rho} one can conclude that the $\rho_{00}$ parameters 
for $\Kstar^\pm$ mesons produced in \numuCC interactions are in agreement 
within statistical errors with the value of 1/3 which corresponds to the no 
spin alignment case. Also we observe that in \nuNC interactions $\Kstar^+$ 
mesons are produced preferentially in the helicity zero state 
($\rho_{00}>1/3$), but the statistical errors are too large to reach a firm 
conclusion. For $\Kstar^-$ mesons produced in \nuNC interactions the 
$\rho_{00}$ parameter could not be accurately determined because of the small 
statistics of the corresponding event sample.

\begin{table}[htb]
\begin{center}
\caption{\it The total numbers, absolute yields and relative yields of the $\Kstar(892)^\pm$ 
mesons produced in \numuCC and \nuNC interactions that decay into 
$K^0 \pi^\pm$ modes. Both statistical and systematic errors are shown.}
\vspace*{0.3cm}
\label{tab:results_yields}
\begin{tabular}{lccc}
\hline
Sample & Number of $\Kstar$ & Yields of $\Kstar$ (\%) & 
$\frac{N(K^\star \to K^0 \pi)}{N(K^0)}$ (\%) \\
\hline
$\Kstar^+$ CC & $26676 \pm 1784 \pm 1863$ & $2.6 \pm 0.2 \pm 0.2$ & 
                $15.3 \pm 1.0 \pm 1.0$ \\
$\Kstar^-$ CC & $16278 \pm 1372 \pm  500$ & $1.6 \pm 0.1 \pm 0.1$ & 
                $9.4 \pm 0.8 \pm 0.3$ \\
$\Kstar^+$ NC & $ 9024 \pm 1216 \pm  984$ & $2.5 \pm 0.3 \pm 0.3$ & 
                $14.8 \pm 2.0 \pm 1.6$ \\
$\Kstar^-$ NC & $ 3750 \pm 1012 \pm  762$ & $1.0 \pm 0.3 \pm 0.2$ & 
                $6.1 \pm 1.7 \pm 1.2$ \\
\hline
\end{tabular}
\end{center}
\end{table}

\begin{table}[htb]
\begin{center}
\caption{\it The $\rho_{00}$ parameter for $\Kstar(892)^\pm$ mesons produced 
in \numuCC and \nuNC interactions that decay into $K^0 \pi^\pm$ modes. 
Both statistical and systematic errors are shown.}
\vspace*{0.3cm}
\label{tab:results_rho}
\begin{tabular}{lc}
\hline
Sample & $\rho_{00}$ \\
\hline
$\Kstar^+$ CC & $0.40 \pm 0.06 \pm 0.03$ \\
$\Kstar^-$ CC & $0.28 \pm 0.07 \pm 0.03$ \\
$\Kstar^+$ NC & $0.66 \pm 0.10 \pm 0.05$ \\
\hline
\end{tabular}
\end{center}
\end{table}

In Sec.~\ref{sec:kstar_results_production} and \ref{sec:kstar_results_rho00} we
present the dependencies of the $\Kstar(892)^\pm$ production yields and
the $\rho_{00}$ parameter for different kinematic variables in
\numuCC interactions.

\subsection{\label{sec:kstar_results_production}$\Kstar^\pm$ production yields}

\begin{figure}[htb]
\begin{tabular}{ccc}
\mbox{\epsfig{file=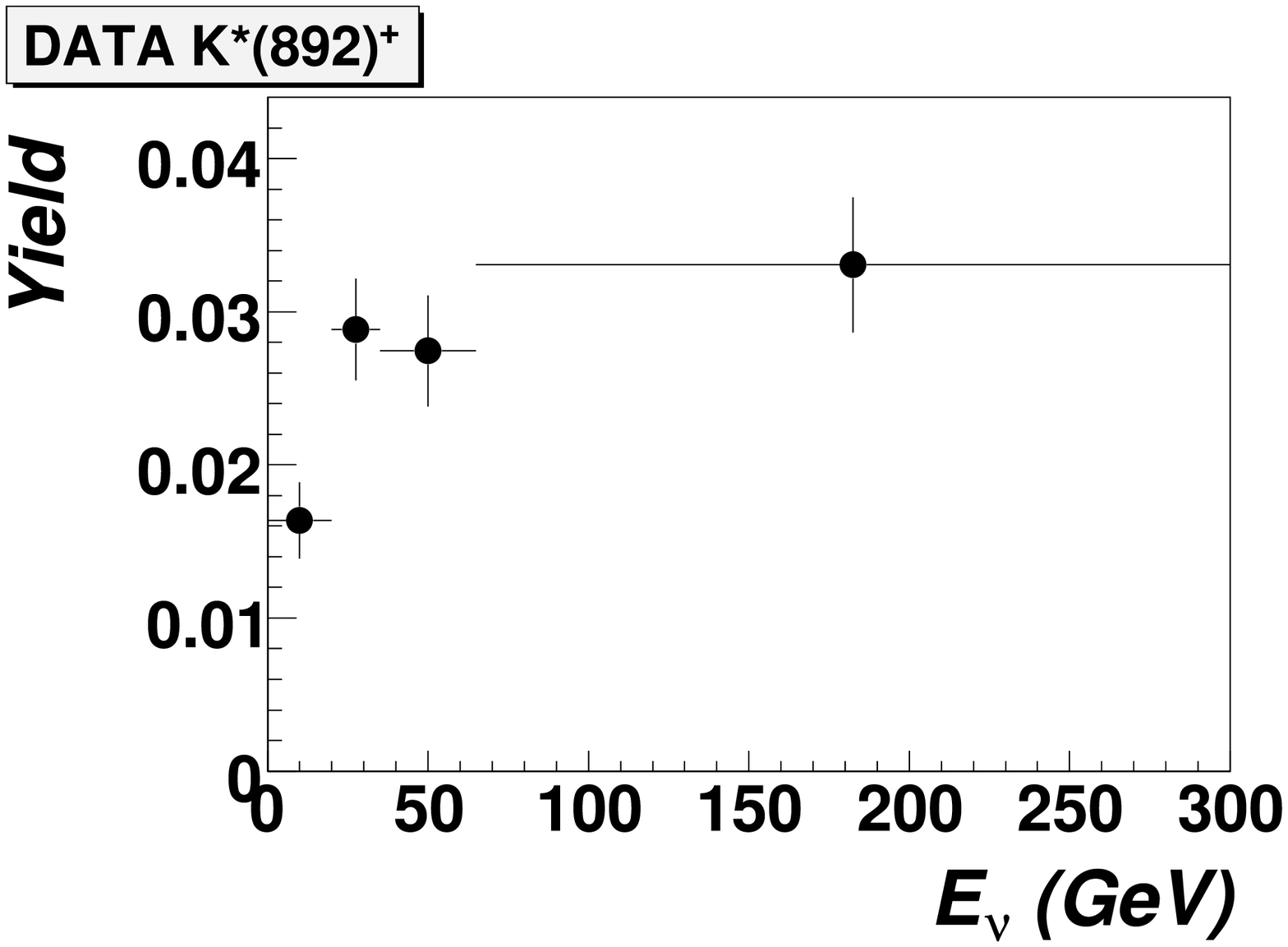,width=0.32\linewidth}} &
\mbox{\epsfig{file=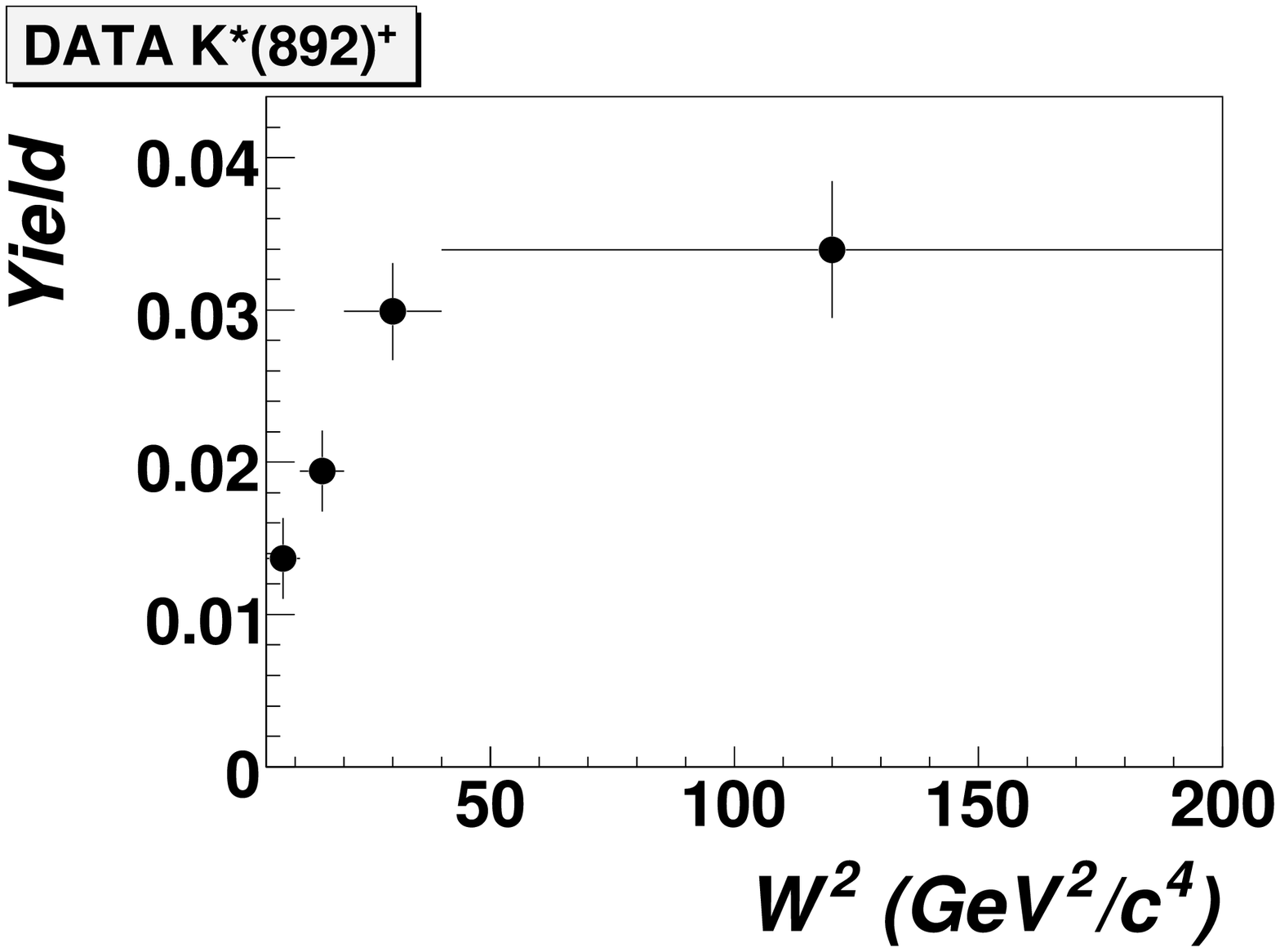,width=0.32\linewidth}} &
\mbox{\epsfig{file=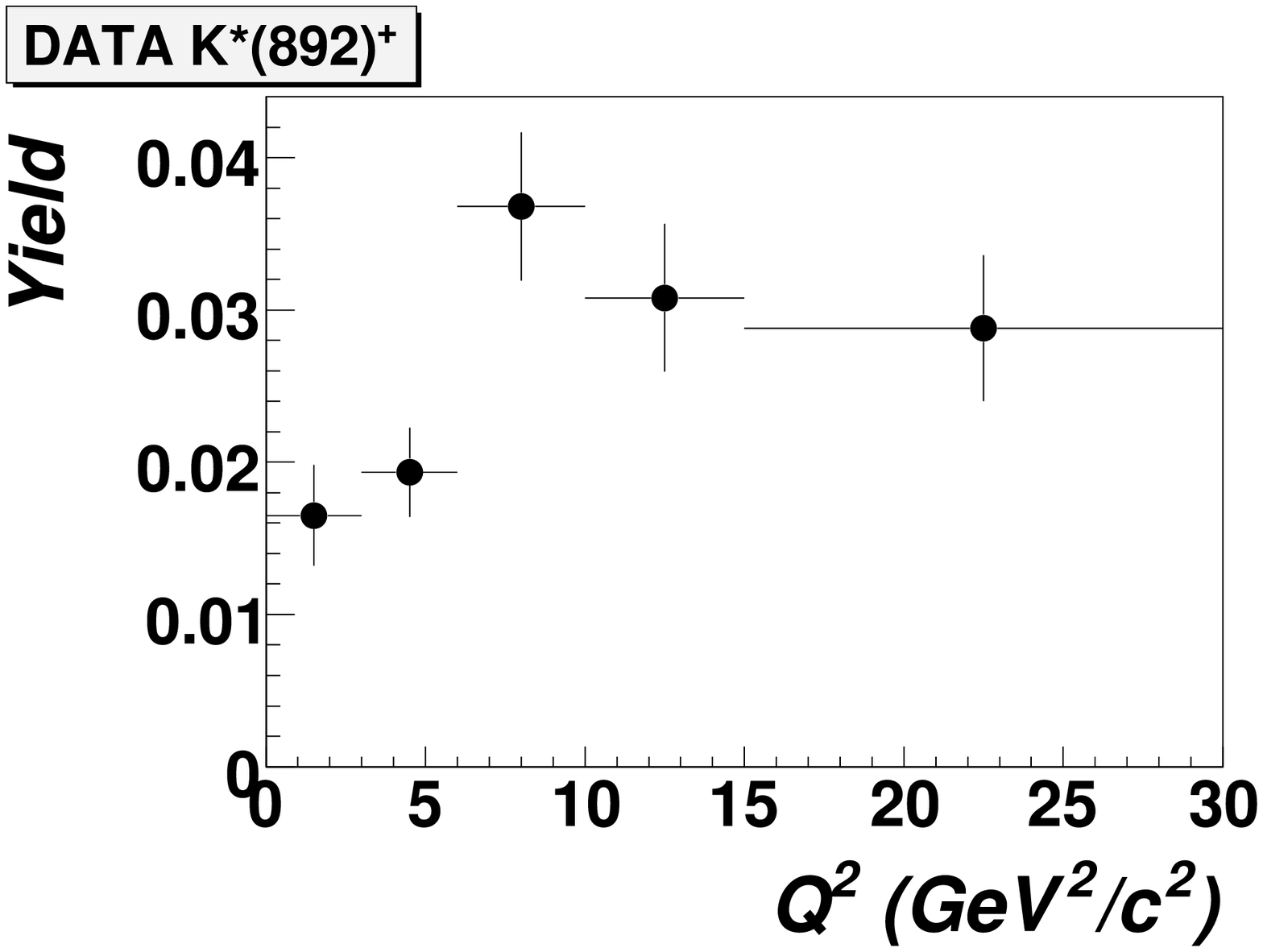,width=0.32\linewidth}} \\ 
\mbox{\epsfig{file=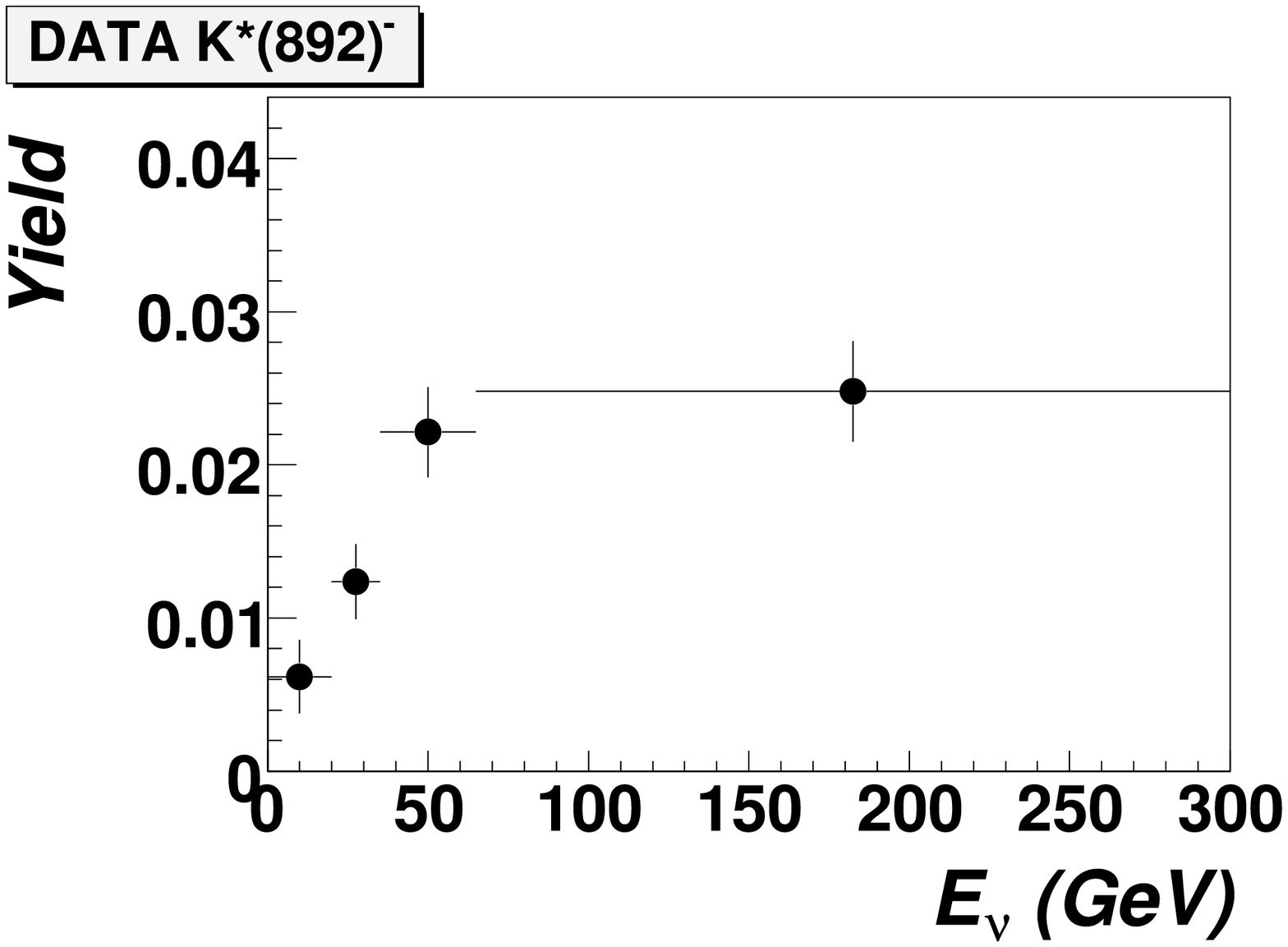,width=0.32\linewidth}} &
\mbox{\epsfig{file=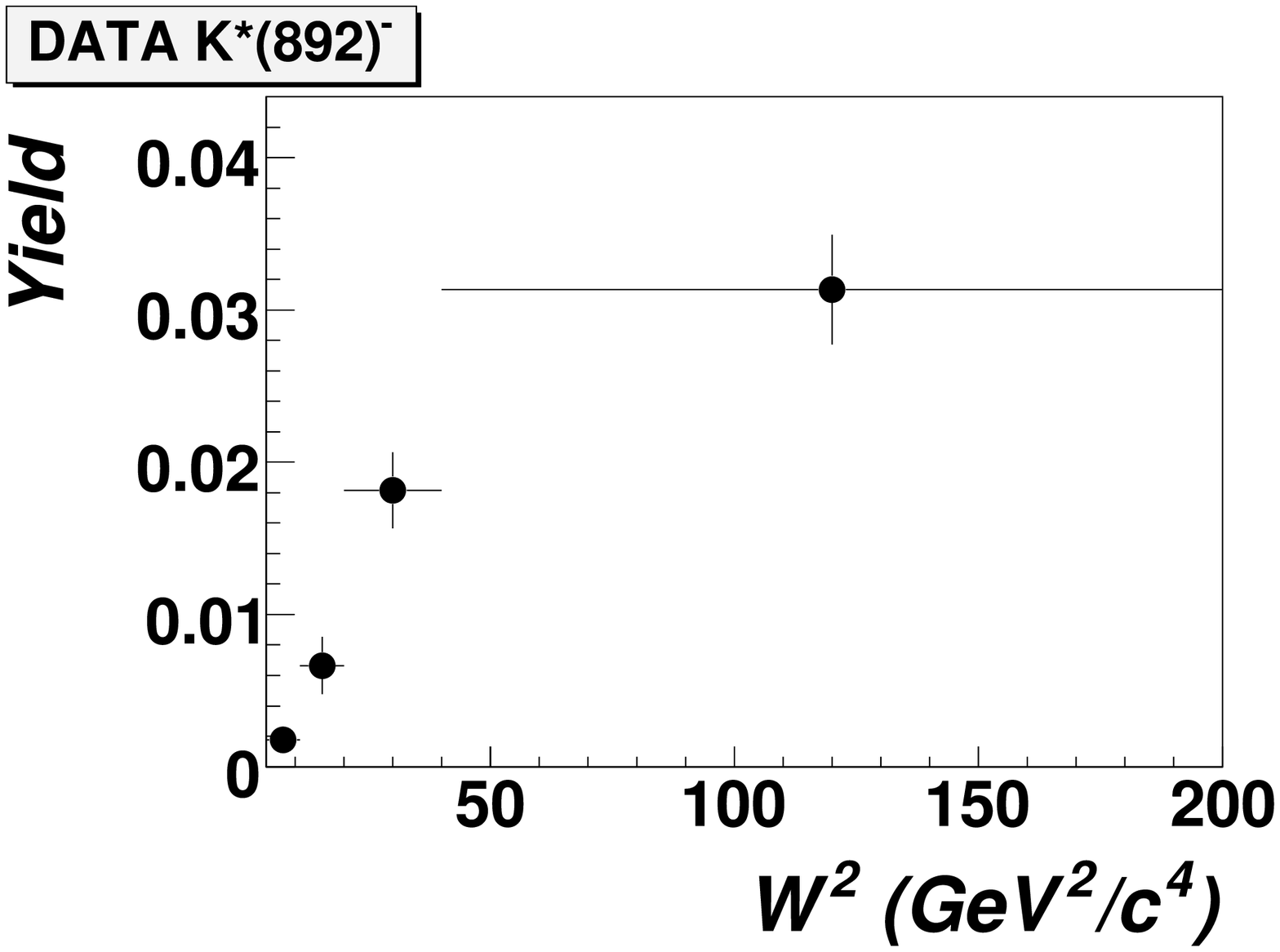,width=0.32\linewidth}} &
\mbox{\epsfig{file=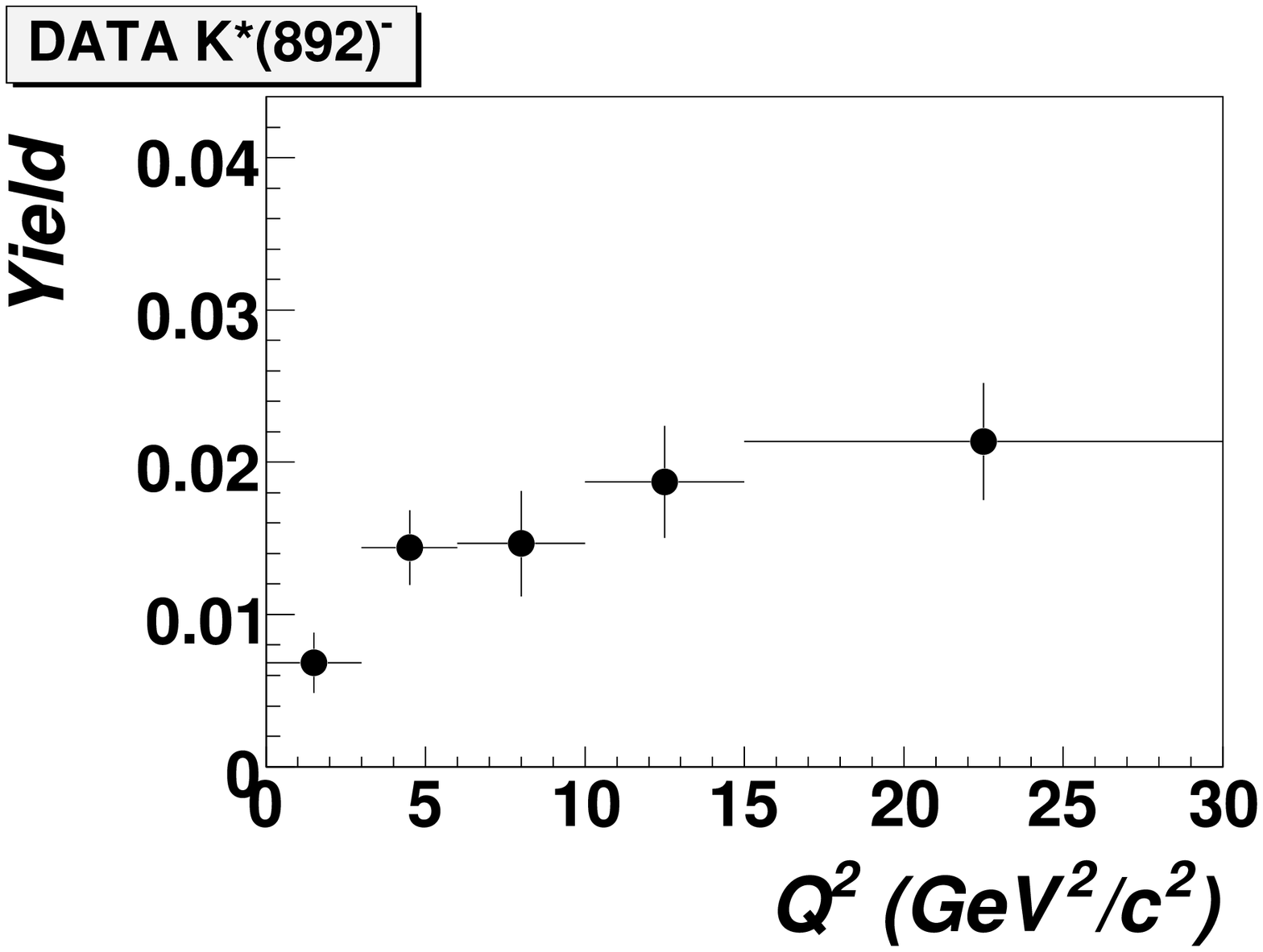,width=0.32\linewidth}} \\
\end{tabular}
\protect\caption{\label{fig:kstar_distr_1} \it 
Corrected $\Kstar^+$ (top) and $\Kstar^-$ (bottom) yields as a function of
$E_\nu,\ W^2,\ Q^2$ in \numuCC events. Only statistical errors are shown.}
\end{figure}

\begin{figure}[htb]
\begin{center}
\begin{tabular}{cc}
\mbox{\epsfig{file=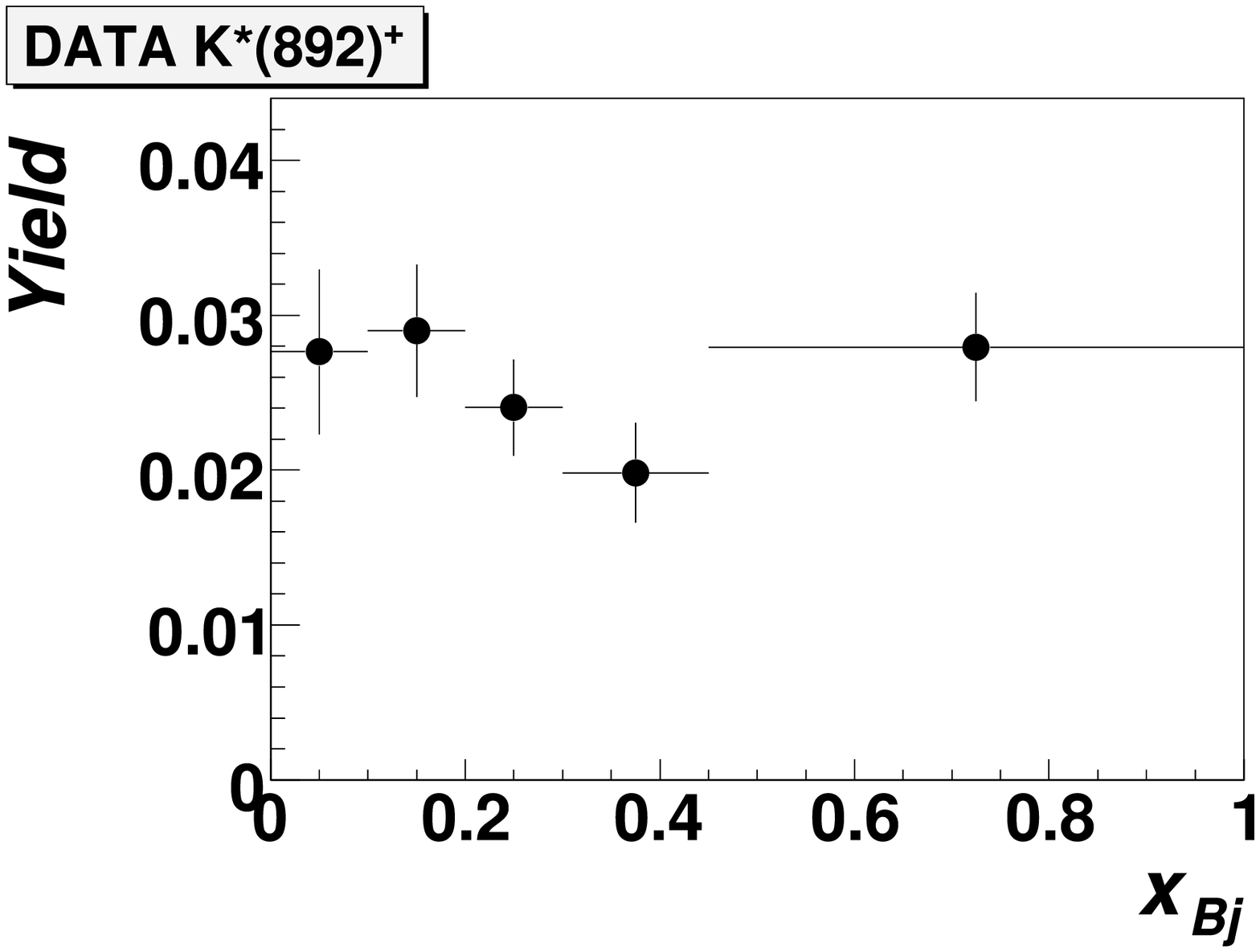,width=0.4\linewidth}} &
\mbox{\epsfig{file=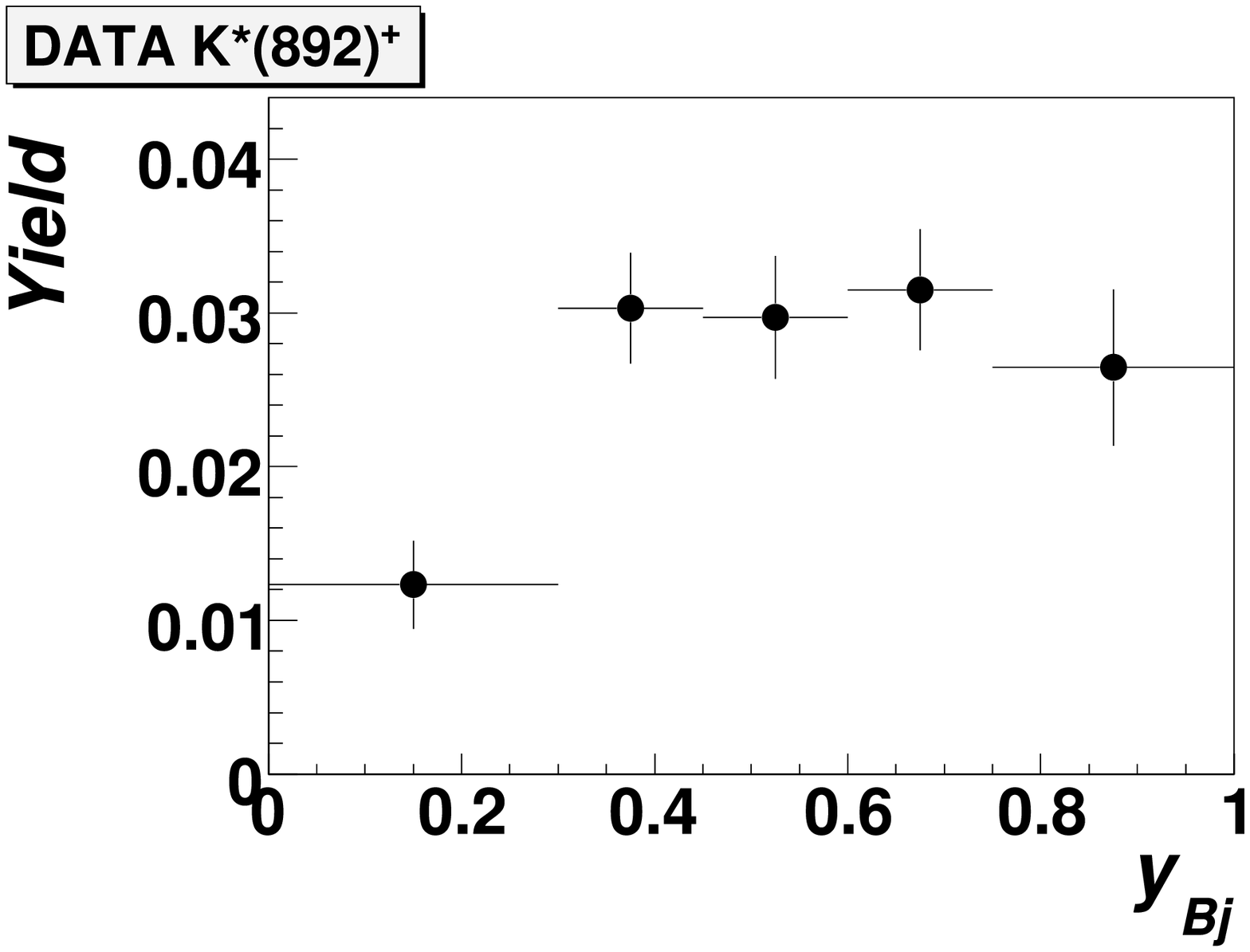,width=0.4\linewidth}} \\
\mbox{\epsfig{file=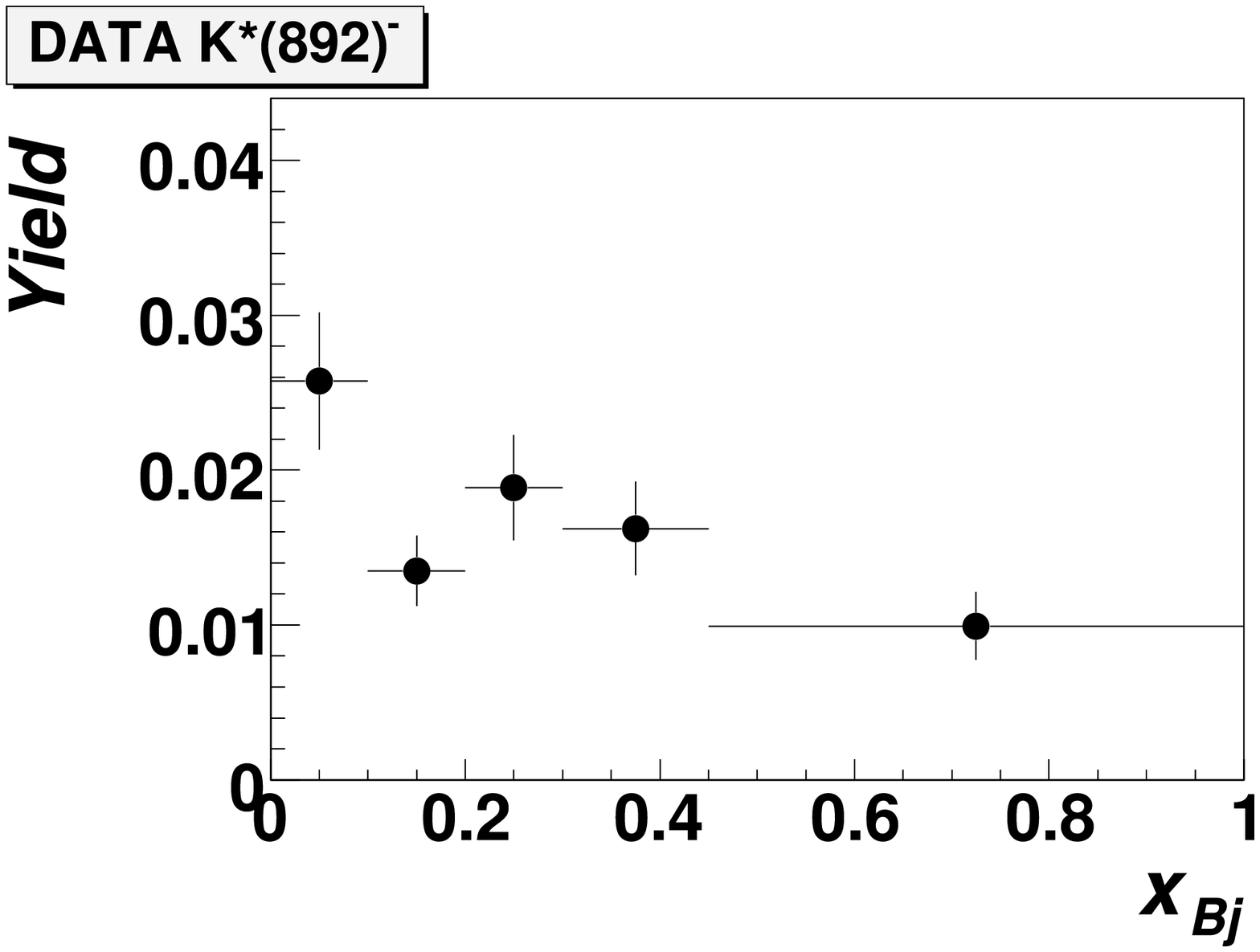,width=0.4\linewidth}} &
\mbox{\epsfig{file=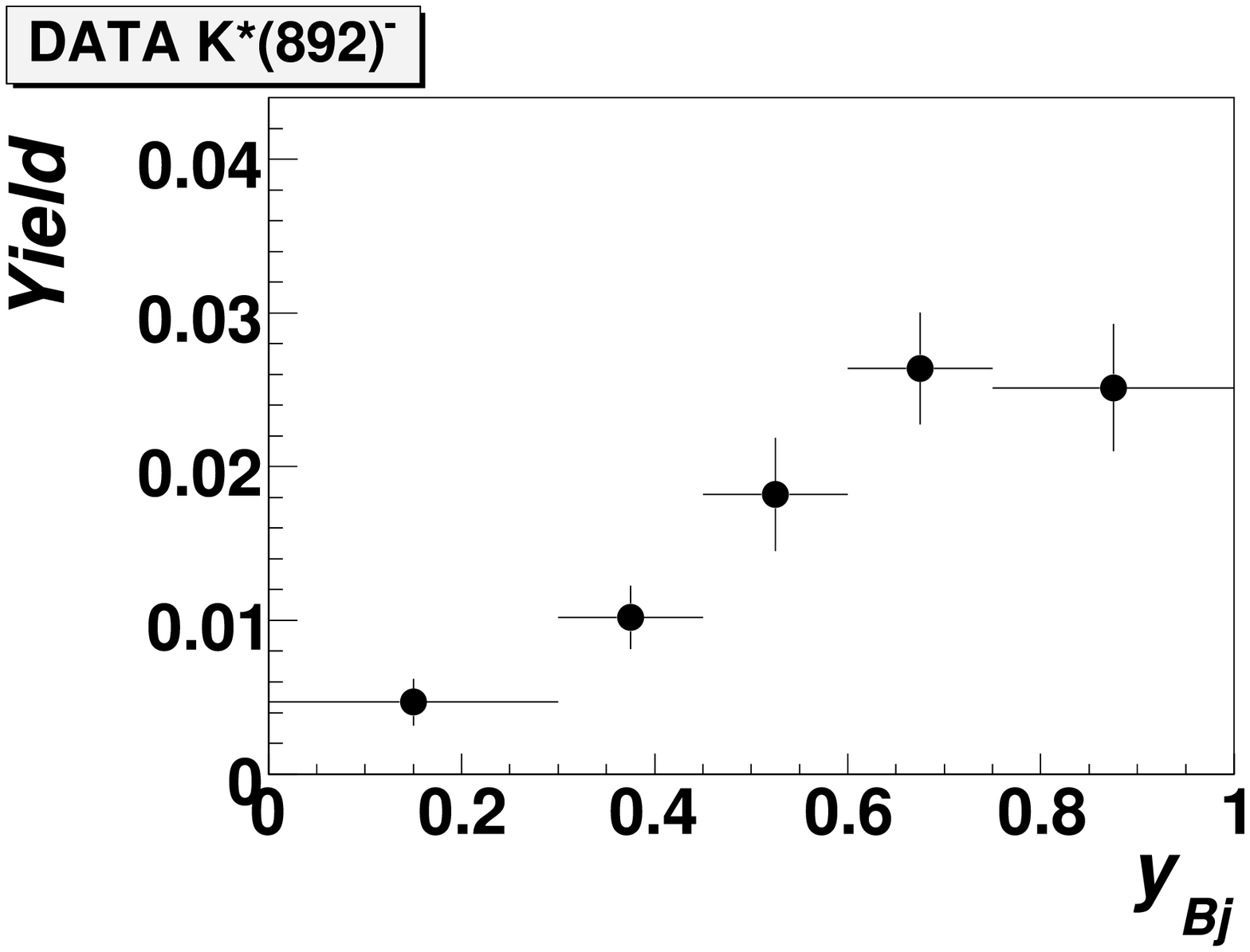,width=0.4\linewidth}} \\
\end{tabular}
\protect\caption{\label{fig:kstar_distr_2} \it Corrected $\Kstar^+$ (top) and 
$\Kstar^-$ (bottom) yields as a function of $x_{Bj},\ y_{Bj}$ in \numuCC 
events. Only statistical errors are shown.}
\end{center}
\end{figure}

In Figs.~\ref{fig:kstar_distr_1} and~\ref{fig:kstar_distr_2} we show corrected
$\Kstar^\pm$ yields in the data as a function of kinematic 
variables\footnote{$x_{Bj}$ and $y_{Bj}$ are the standard Bjorken scaling 
variables.} $E_\nu$, $W^2$, $Q^2$, $x_{Bj}$, $y_{Bj}$ in \numuCC events. 
The $\Kstar^\pm$ yields show a monotonic rise with $E_\nu$, $W^2$ and $Q^2$. 
From Fig.~\ref{fig:kstar_distr_2} we see that the dependence of the yields on 
the $x_{Bj}$ and $y_{Bj}$ variables are different for $\Kstar^+$ and 
$\Kstar^-$ mesons. This fact could be explained by different production 
mechanisms ($\Kstar^-$ mesons are produced mainly from the string 
fragmentation processes while $\Kstar^+$ mesons are produced both from string 
fragmentation processes and from struck quark fragmentation).

\begin{figure}[htb]
\begin{center}
\begin{tabular}{ccc}
\mbox{\epsfig{file=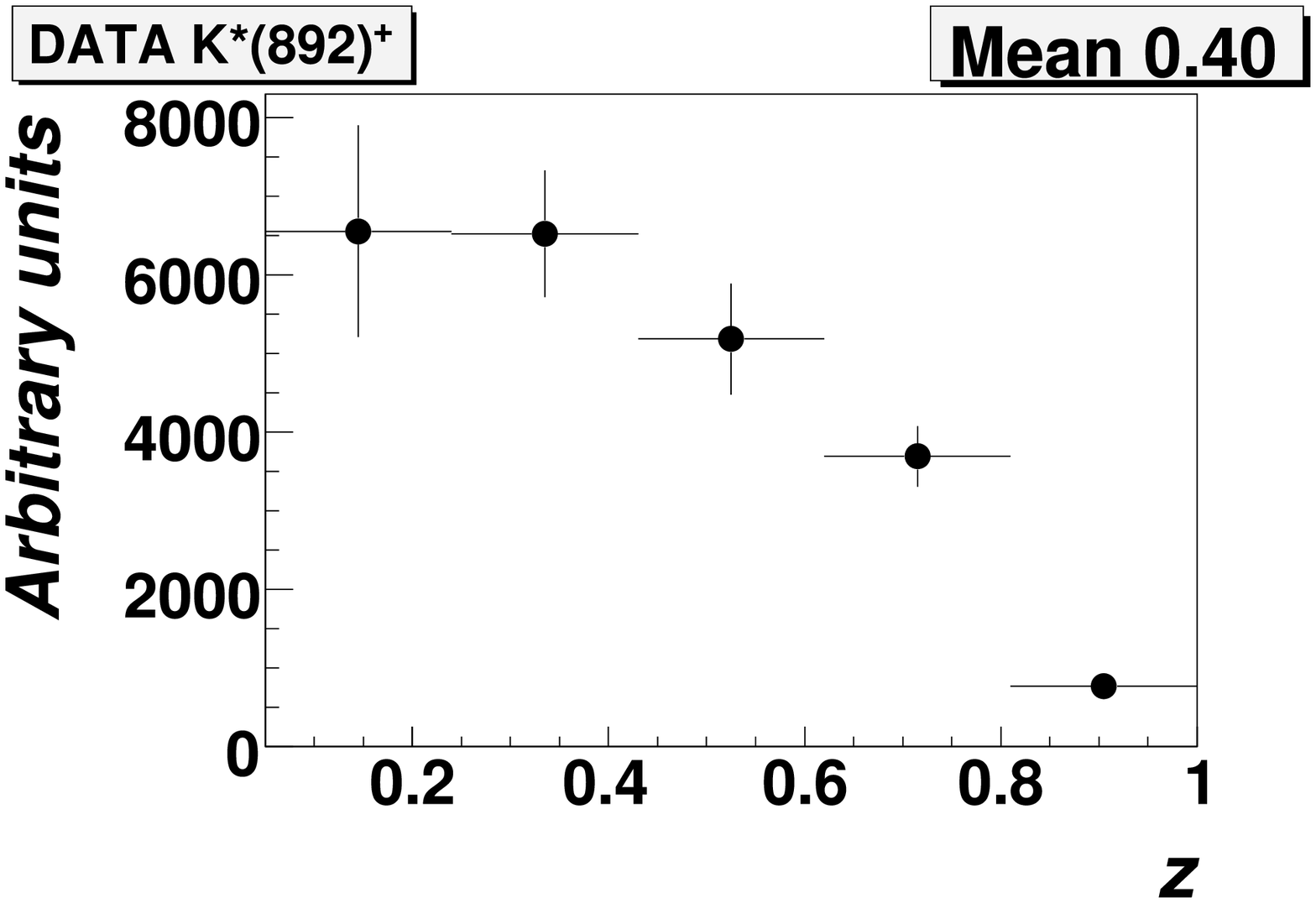,width=0.32\linewidth}}  &
\mbox{\epsfig{file=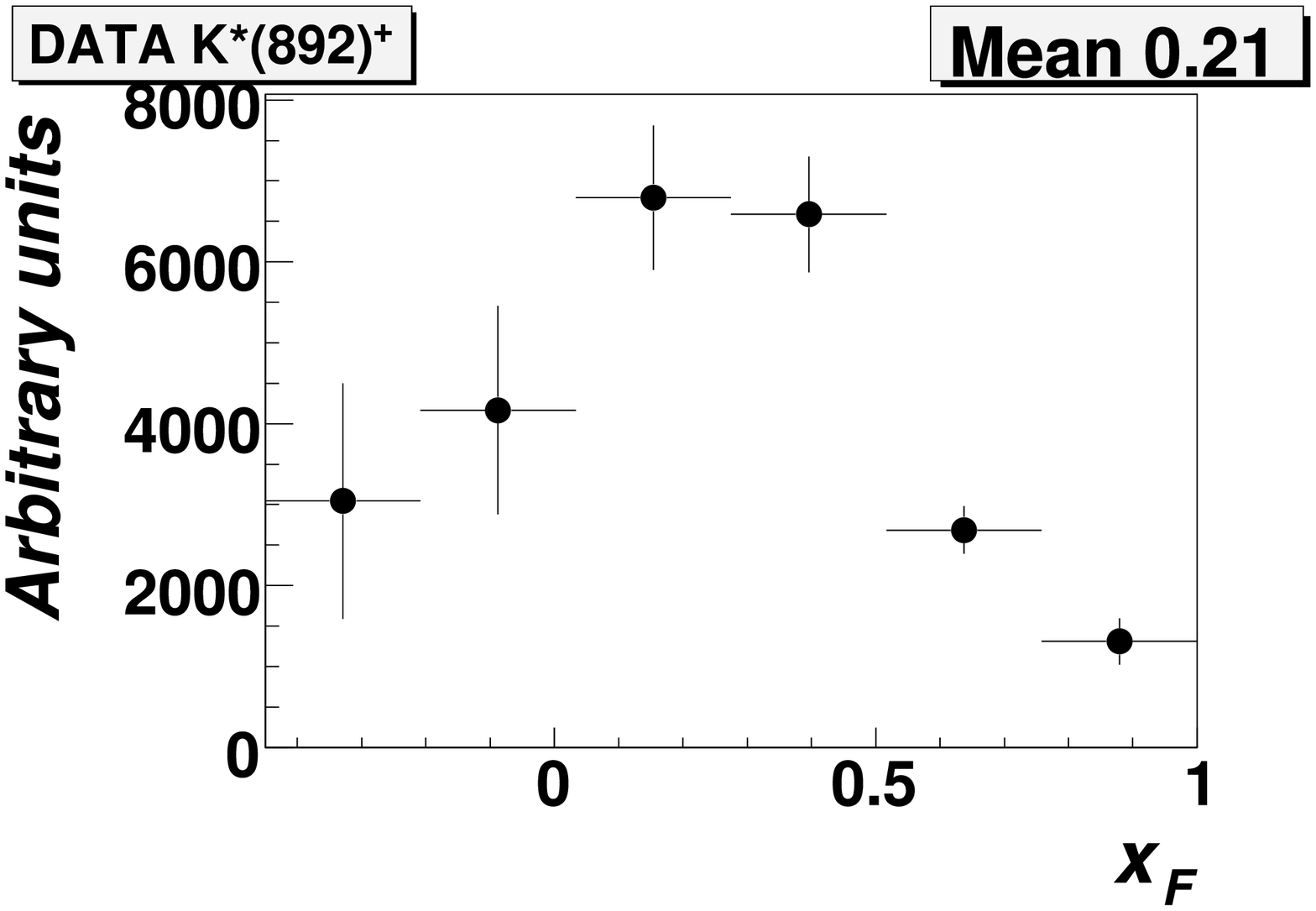,width=0.32\linewidth}} &
\mbox{\epsfig{file=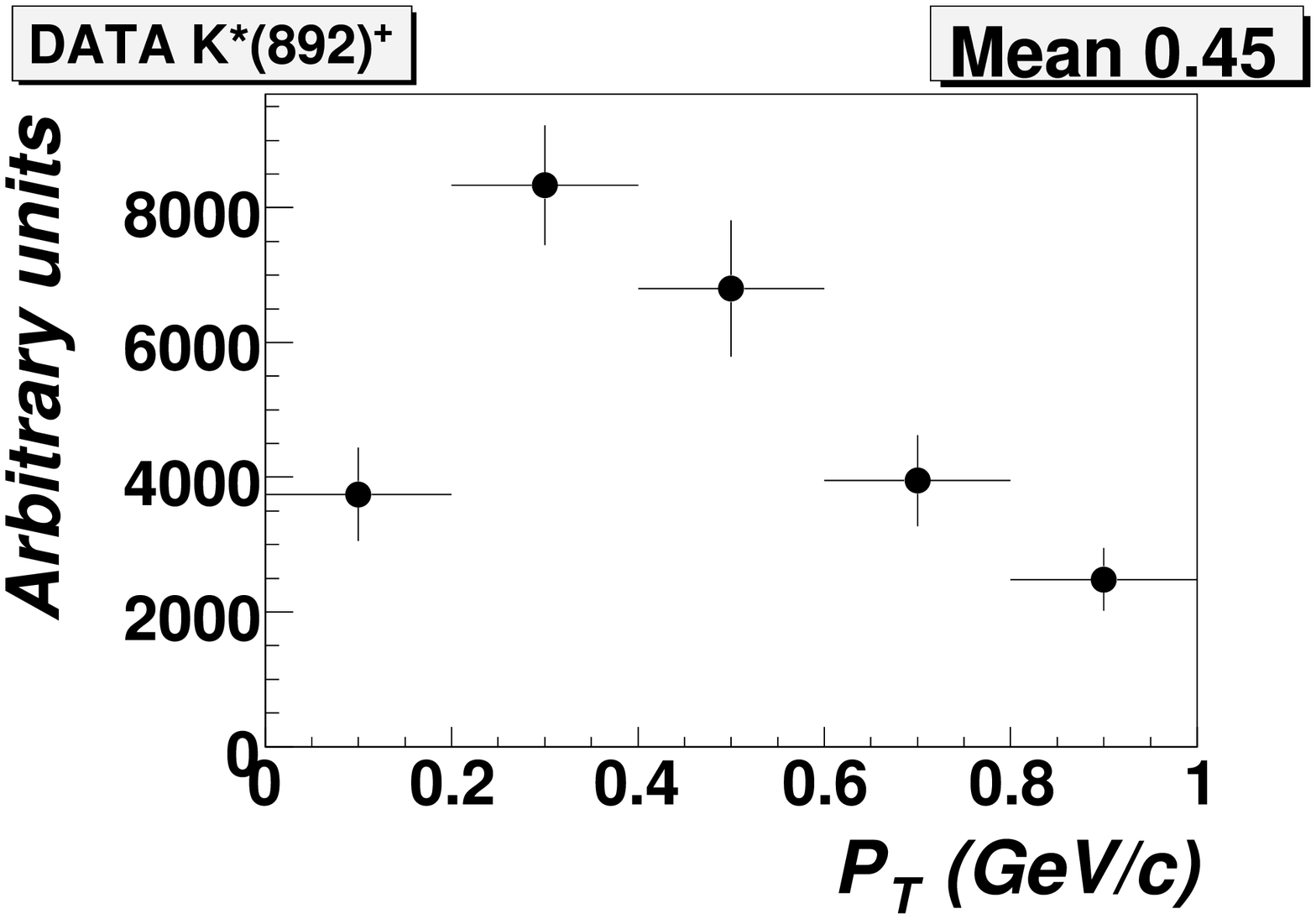,width=0.32\linewidth}} \\
\mbox{\epsfig{file=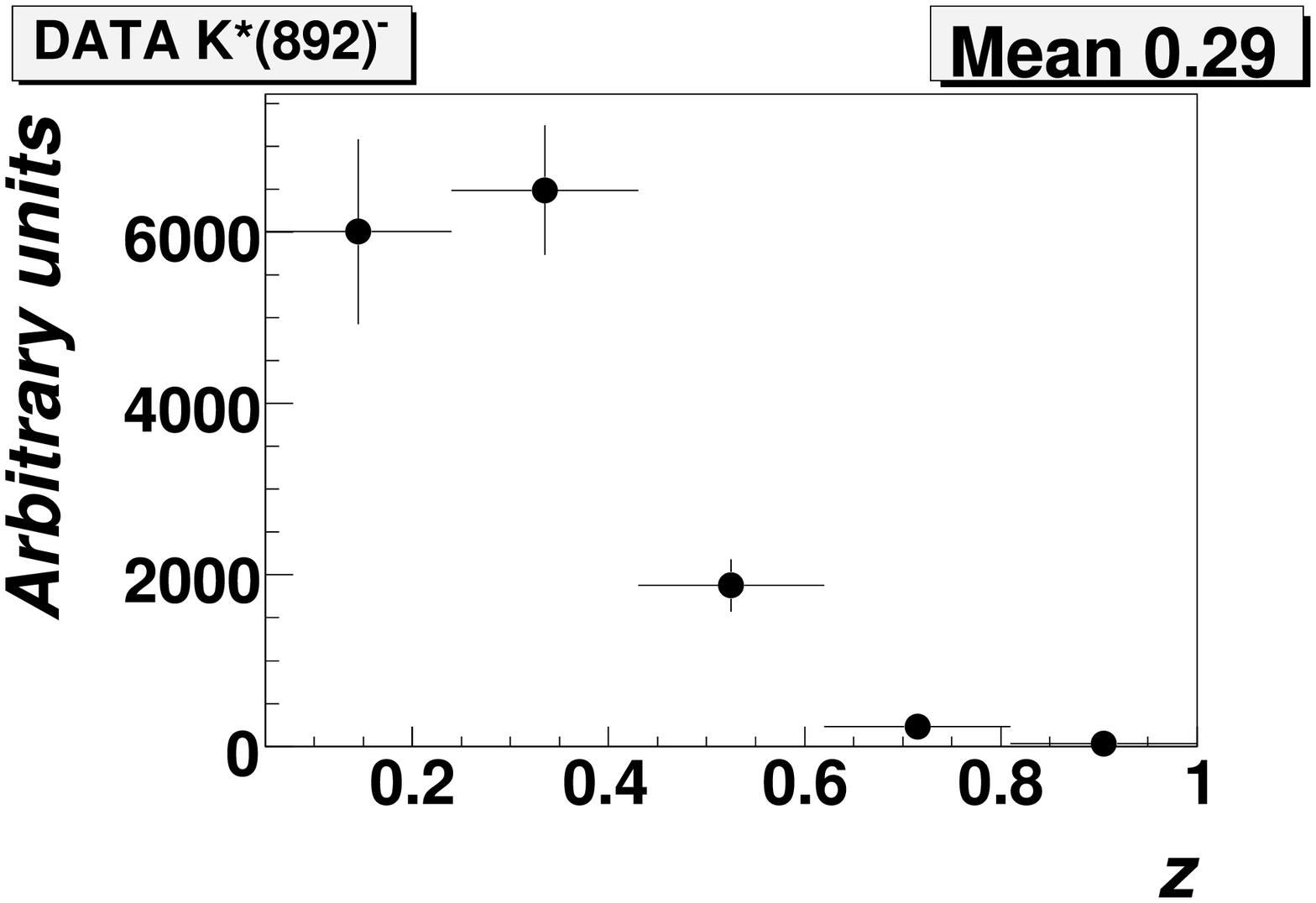,width=0.32\linewidth}} &
\mbox{\epsfig{file=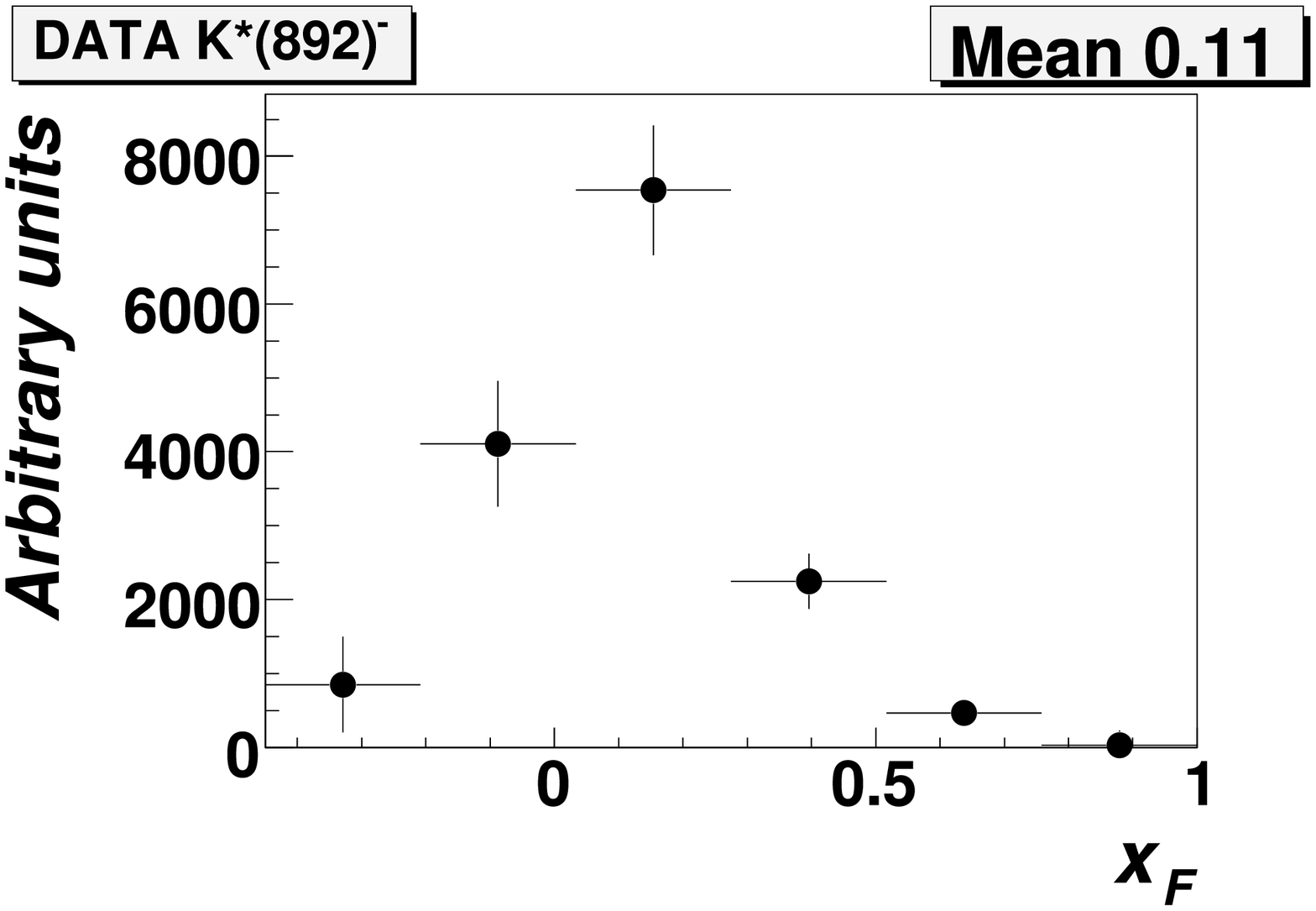,width=0.32\linewidth}}&
\mbox{\epsfig{file=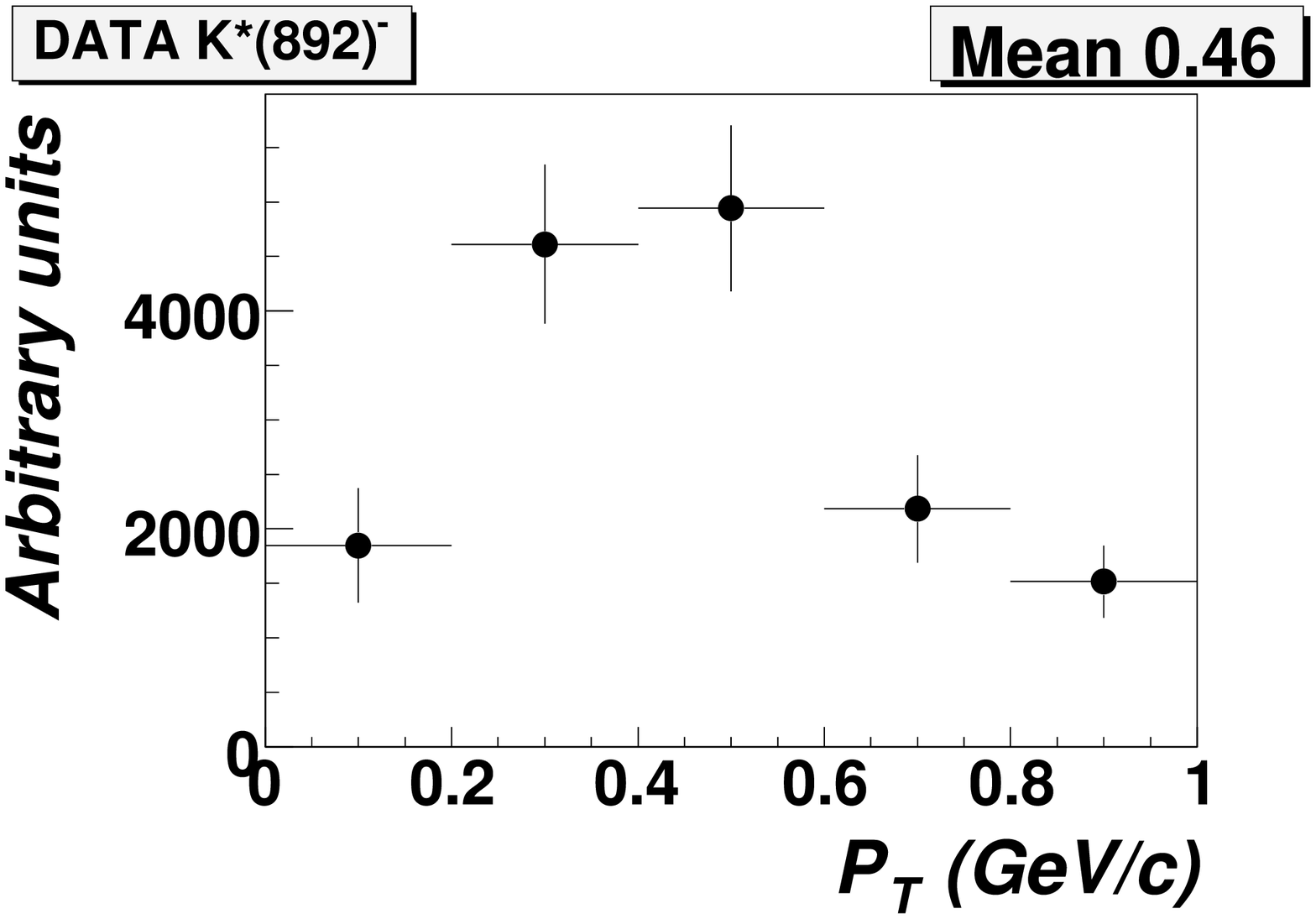,width=0.32\linewidth}}\\
\end{tabular}
\protect\caption{\label{fig:kstar_distr_3} \it Corrected $z,\ x_F,\ p_T$ 
distributions for $\Kstar^+$ (top) and $\Kstar^-$ (bottom) in \numuCC events. 
Only statistical errors are shown.}
\end{center}
\end{figure}

\begin{figure}[htb]
\begin{center}
\begin{tabular}{cc}
\mbox{\epsfig{file=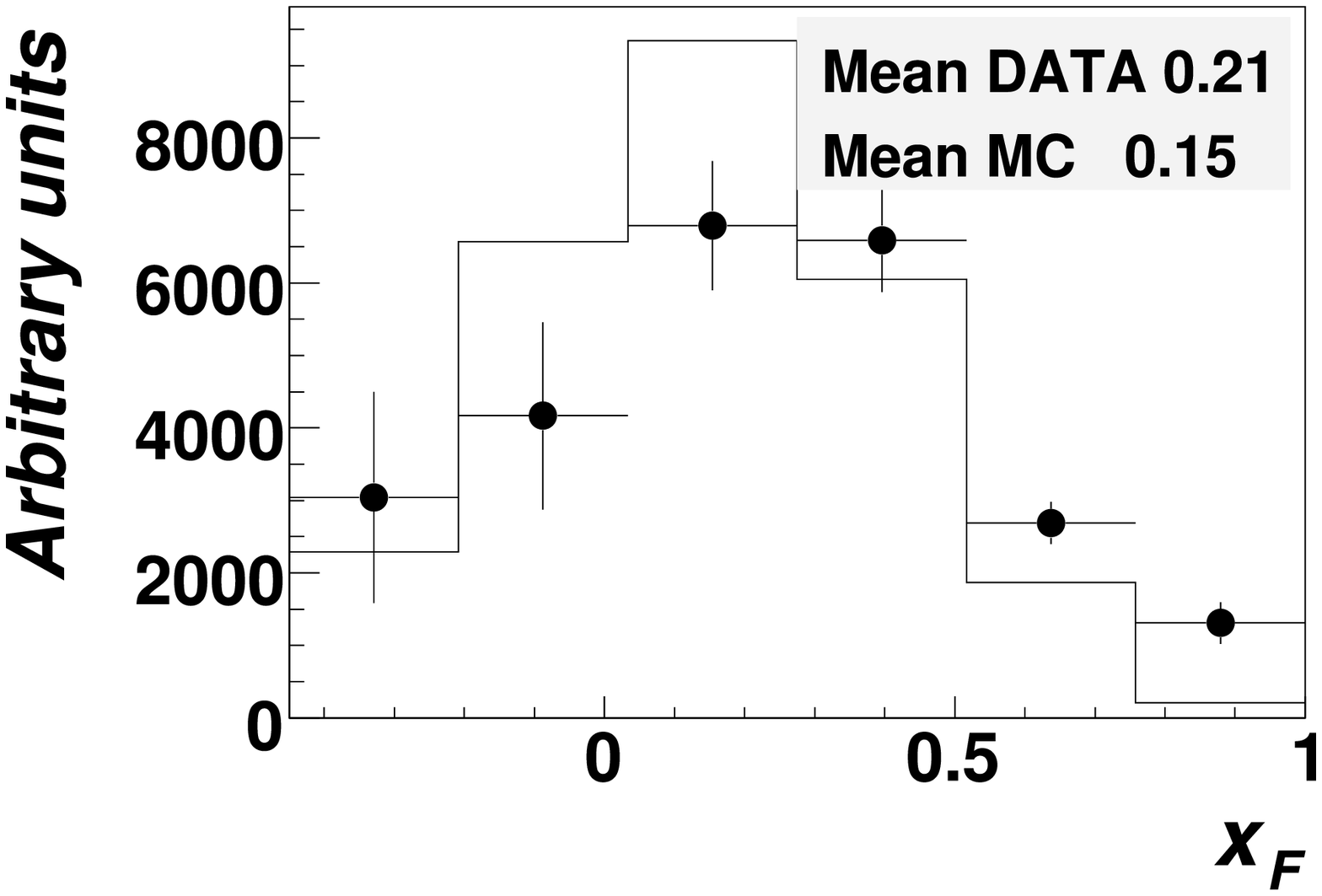,width=0.4\linewidth,height=0.3\linewidth}} &
\mbox{\epsfig{file=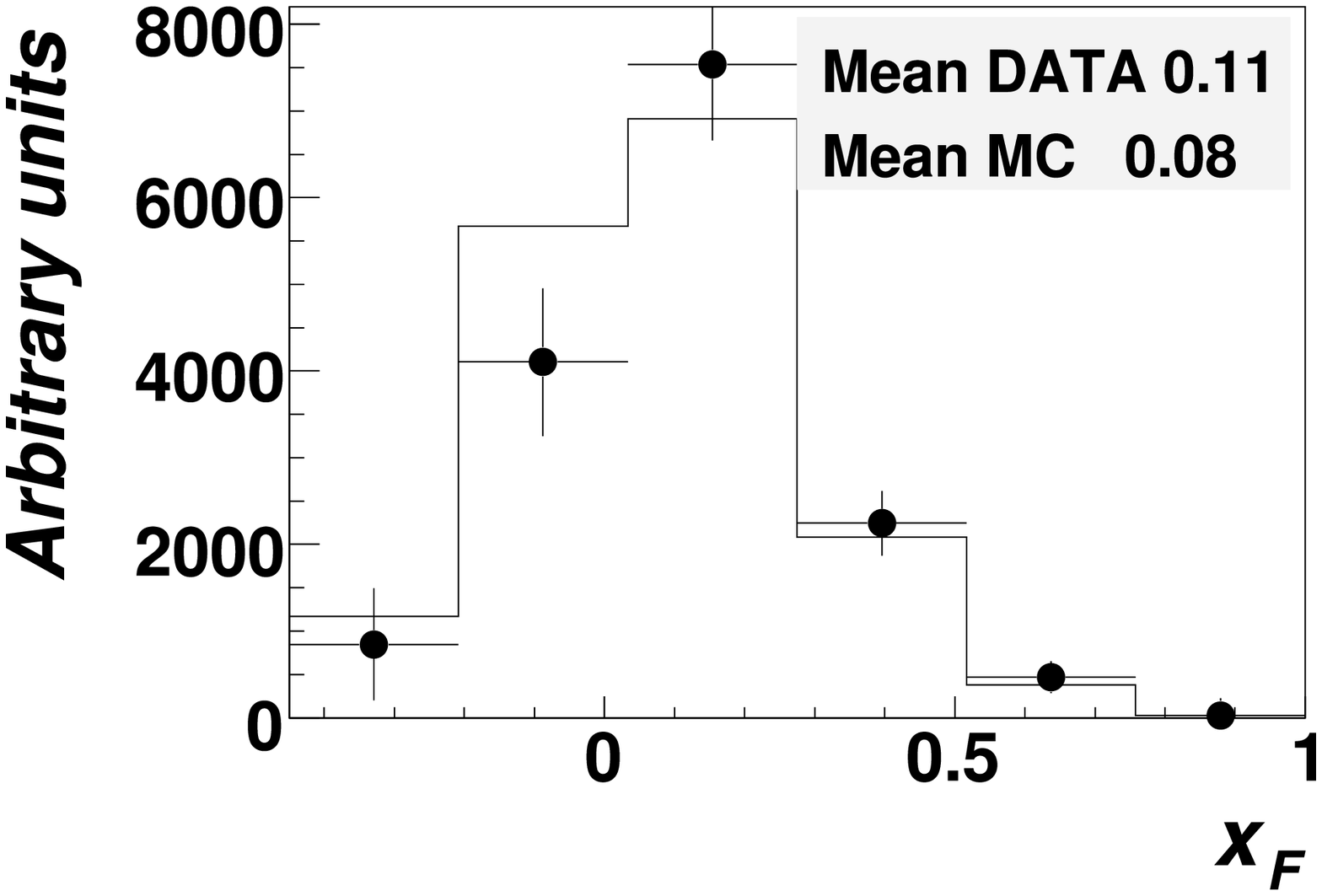,width=0.4\linewidth,height=0.3\linewidth}} \\
\end{tabular}
\protect\caption{\label{fig:kstar_distr_4} \it 
Corrected $x_F$ distribution for $\Kstar^+$ (left) and $\Kstar^-$ (right) 
in MC (histogram) and data (points with error bars) \numuCC events. 
Only statistical errors are shown.}
\end{center}
\end{figure}

$\Kstar^\pm$ distributions as a function\footnote{$z$ is the fraction of the 
total hadronic energy carried away by the $\Kstar^\pm$ in the laboratory 
system and $p_T$ is the transverse momentum the $\Kstar^\pm$ with respect to 
the hadronic jet direction.} of $z$, $x_F$, $p_T$ are shown in 
Fig.~\ref{fig:kstar_distr_3}. One can see a shift towards positive $x_F$ 
values in the distribution for $\Kstar^+$ mesons as compared to the
distribution for $\Kstar^-$ mesons. In Fig.~\ref{fig:kstar_distr_4} we present 
the comparision of $x_F$ distribution for \Kstar \ mesons in MC and data.

\subsection{\label{sec:kstar_results_rho00} The $\rho_{00}$ parameter of $\Kstar^\pm$ mesons}

\begin{figure}[htb]
\begin{center}
\begin{tabular}{ccc}
\mbox{\epsfig{file=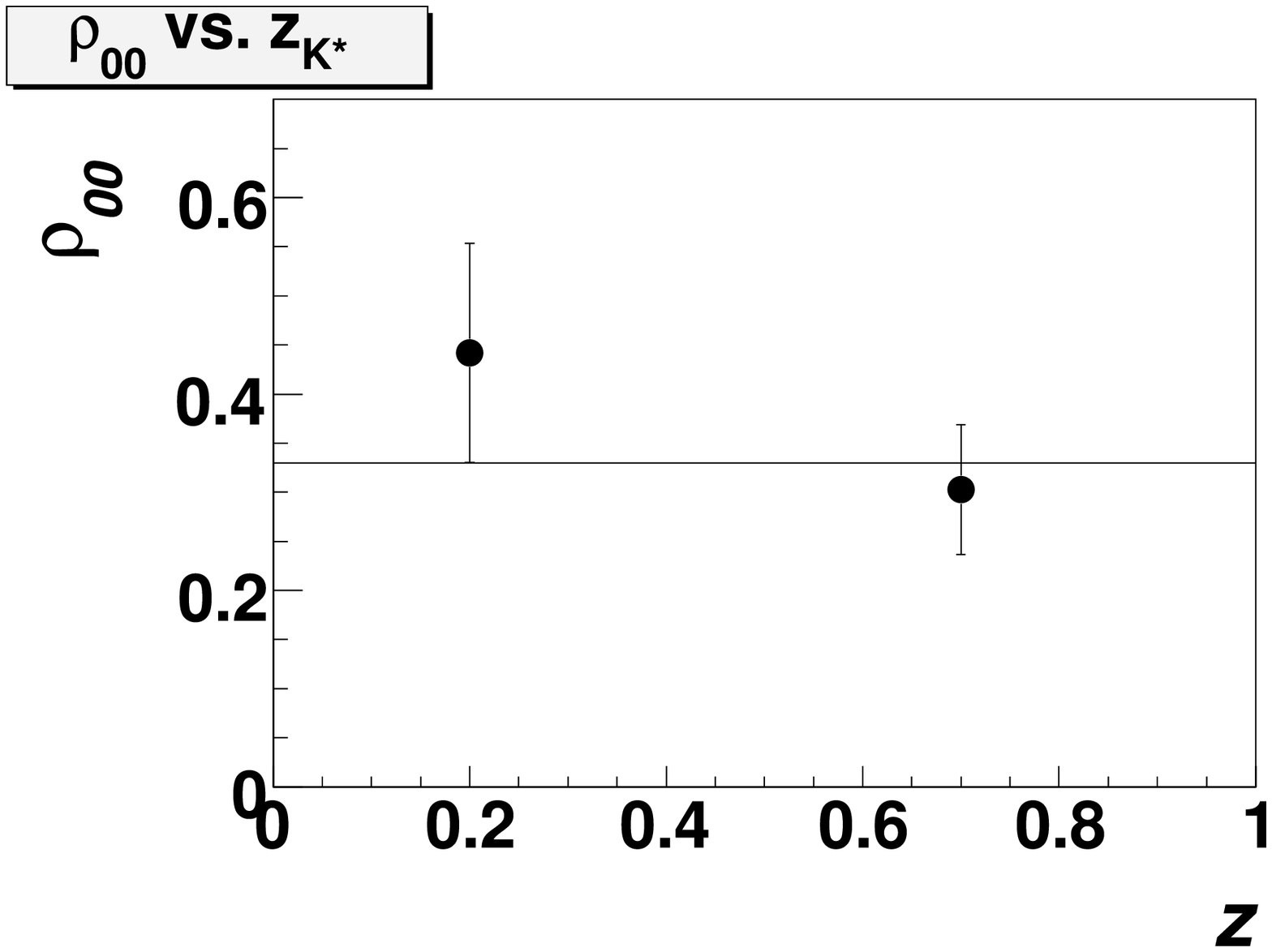,width=0.32\linewidth}} &
\mbox{\epsfig{file=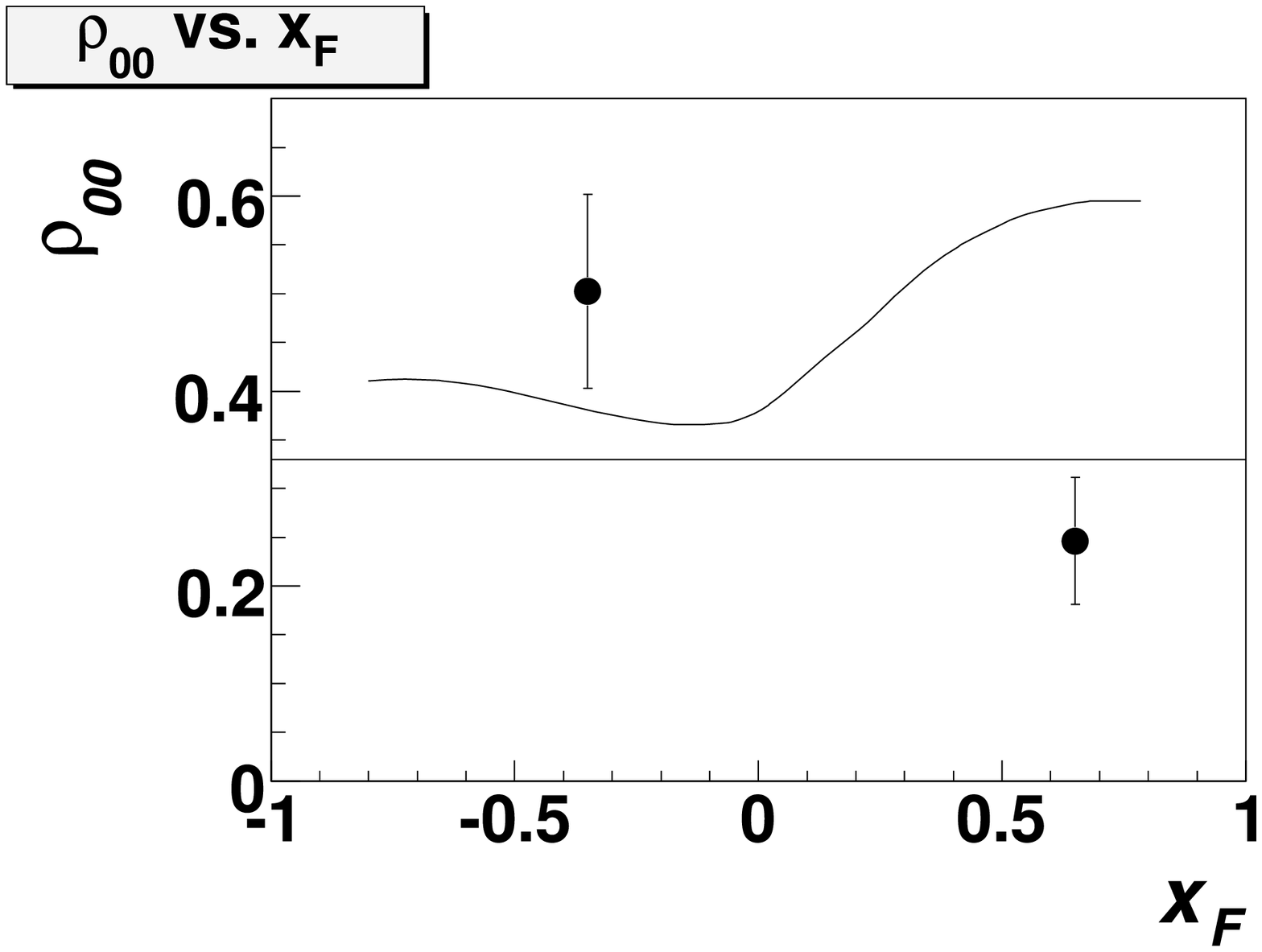,width=0.32\linewidth}} &
\mbox{\epsfig{file=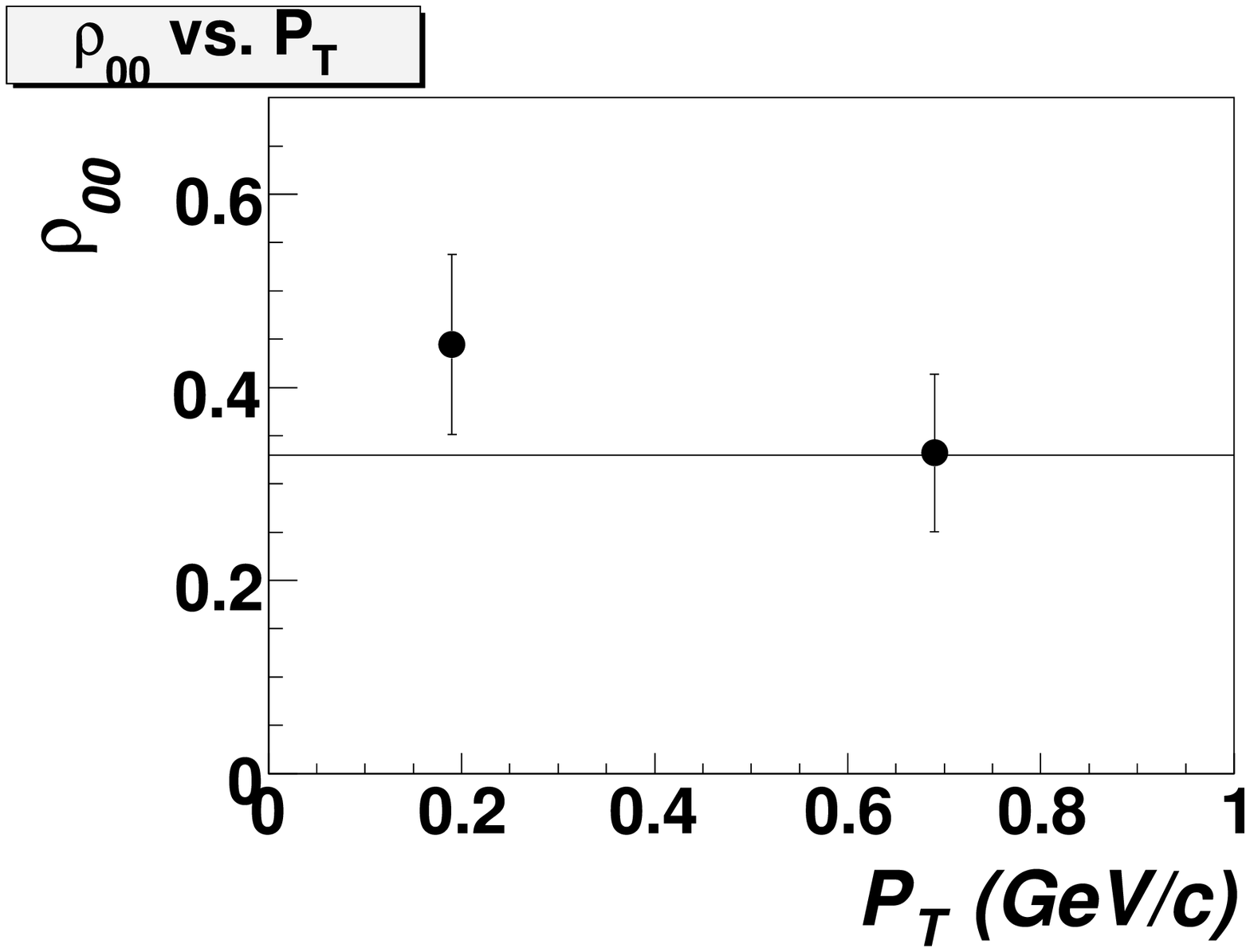,width=0.32\linewidth}} \\
\mbox{\epsfig{file=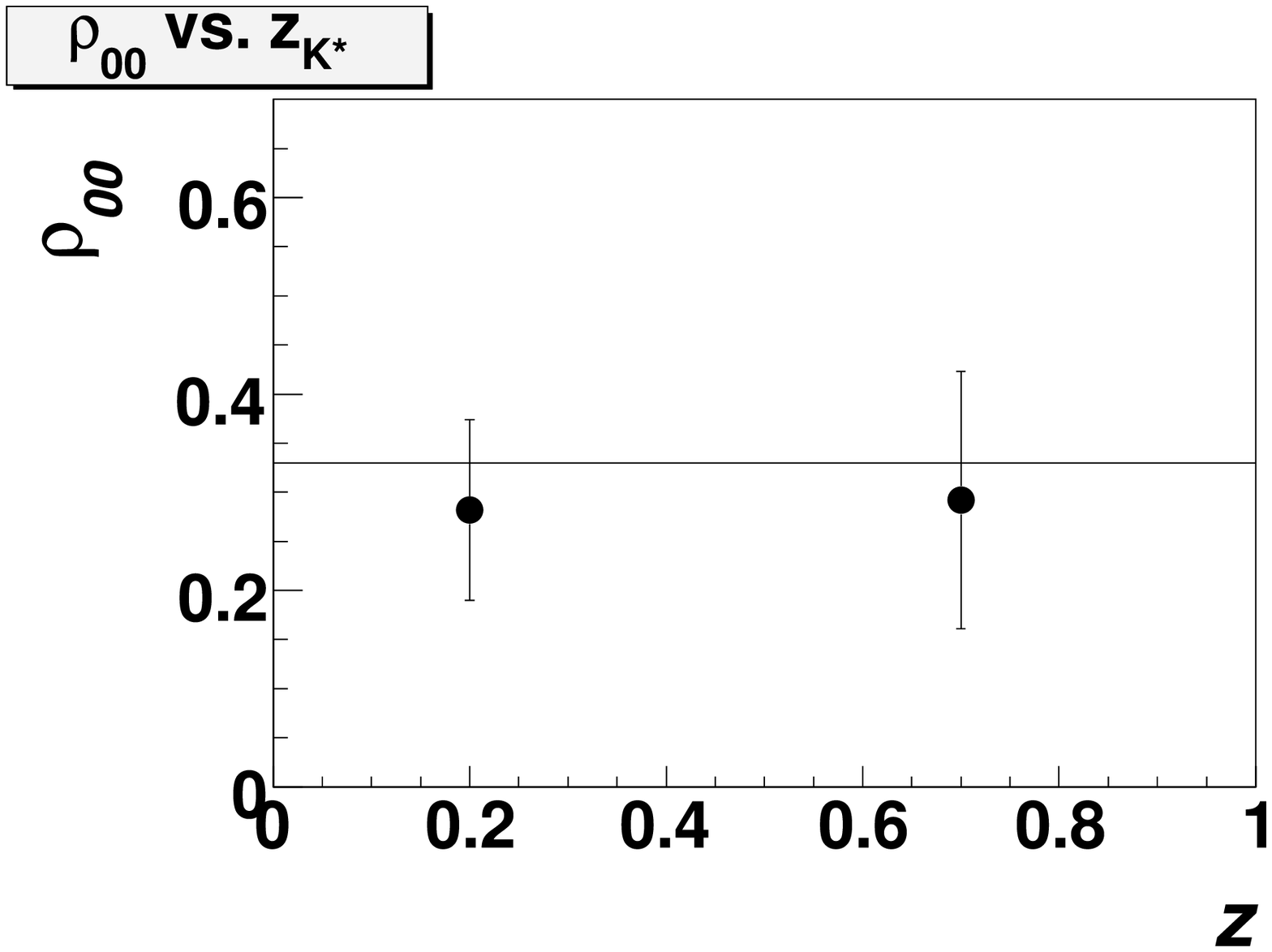,width=0.32\linewidth}} &
\mbox{\epsfig{file=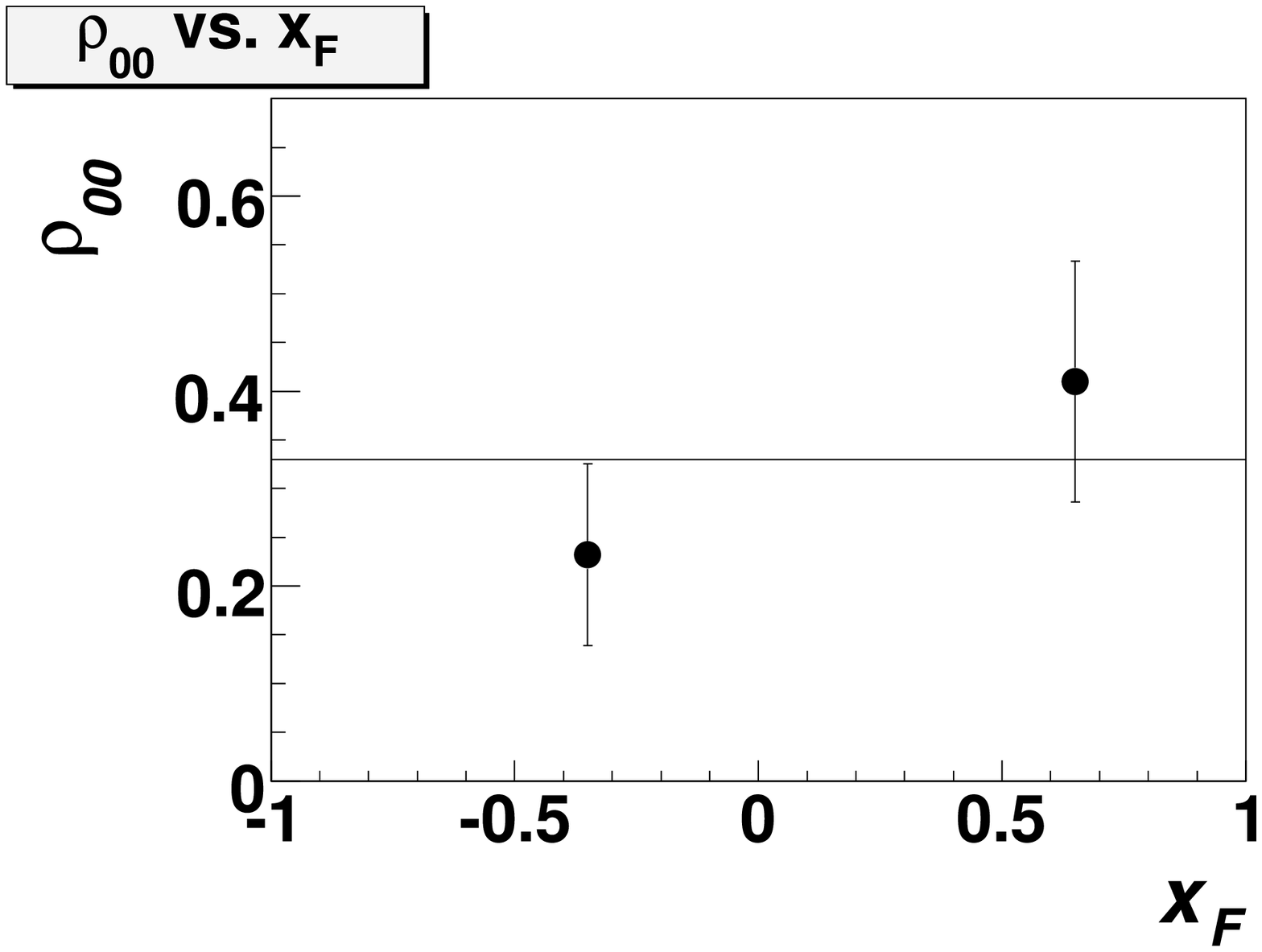,width=0.32\linewidth}} &
\mbox{\epsfig{file=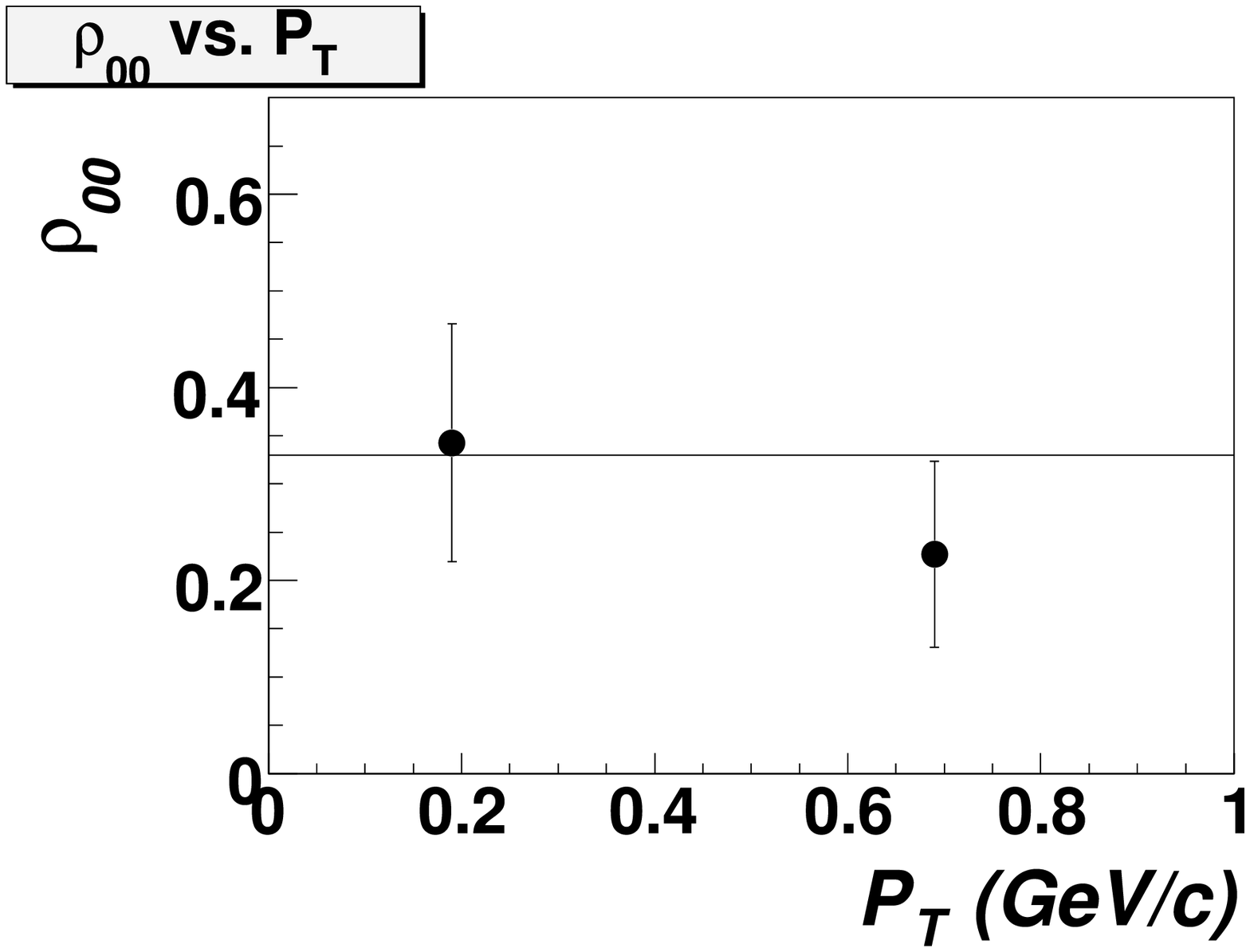,width=0.32\linewidth}}\\
\end{tabular}
\protect\caption{\label{fig:kstar_rho_jet} \it The $\rho_{00}$ parameter as a 
function of $z$, $x_F$ and $p_T$ for $\Kstar^+$ (top) and $\Kstar^-$ (bottom) 
in the \numuCC events. Only statistical errors are shown. The theoretical 
prediction~\cite{Liang1} for the dependence of the $\rho_{00}$ parameter on 
$x_F$ for $\Kstar^+$ mesons is also shown.}
\end{center}
\end{figure}

Figs.~\ref{fig:kstar_rho_jet} displays the $\rho_{00}$ parameter as a function 
of $z$, $x_F$, $p_T$ for $\Kstar^+$ (left) and $\Kstar^-$ (right).
The observed $x_F$ dependence does not seem to agree with the theoretical
predictions of Ref.~\cite{Liang,Liang1}.

The $\rho_{00}$ parameter has also been measured as a function of other 
kinematic variables ($E_\nu$, $W^2$, $Q^2$, $x_{Bj}$, $y_{Bj}$). Within the 
errors we do not observe any dependence on these variables. 

\section{Conclusion}

In this analysis we have measured the production properties and spin 
alignment of $\Kstar(892)^\pm$ vector mesons that decayed into $\ko \pi^\pm$ 
and were produced in $\nu_\mu$ CC and NC interactions in the NOMAD experiment. 

For the first time in neutrino experiments the total yields of $\Kstar^\pm$ 
vector mesons that decayed into $K^0 \pi^\pm$ modes have been measured. For 
the $\Kstar^+$ and $\Kstar^-$ mesons produced in $\nu_\mu$ CC interactions 
the following yields per event were found: 
$(2.6 \pm 0.2\, (stat.) \pm 0.2\, (syst.))\%$ and 
$(1.6 \pm 0.1\, (stat.) \pm 0.1\, (syst.))\%$ respectively, while for the 
$\Kstar^+$ and $\Kstar^-$ mesons produced in $\nu_\mu$ NC interactions the 
corresponding values are: $(2.5 \pm 0.3\, (stat.) \pm 0.3\, (syst.))\%$ and 
$(1.0 \pm 0.3\, (stat.)\pm 0.2\, (syst.))\%$. The yields of $\Kstar^\pm$ 
produced in $\nu_\mu$ CC interactions show a monotonic rise with the 
kinematic variables $E_\nu,\ W^2$ and $Q^2$.

The $\Kstar^\pm$ mesons $\rho_{00}$ parameters have been measured for the 
first time in neutrino experiments. The results obtained for the \numuCC 
sample are in agreement within errors with the $\rho_{00}=1/3$, which 
corresponds to no spin alignment for these mesons: 
$\rho_{00} = 0.40 \pm 0.06\, (stat.) \pm 0.03\, (syst.)$ for $K^\star(892)^+$ 
and $\rho_{00} = 0.28 \pm 0.07\, (stat.) \pm 0.04\, (syst.)$ for 
$K^\star(892)^-$. For $K^\star(892)^+$ mesons produced in \nuNC interactios 
we observed an indication for preferential production in the helicity zero 
state ($\rho_{00} > 1/3$): 
$\rho_{00} = 0.66 \pm 0.10\, (stat.) \pm 0.05\, (syst.)$, but the statistical 
errors do not allow us to reach a firm conclusion.

{\large \bf Acknowledgements}

We gratefully acknowledge  the CERN SPS accelerator and beam-line staff
for the magnificent performance of the neutrino beam. The experiment was 
supported by the following funding agencies: Australian Research Council 
(ARC) and Department of Education, Science, and Training (DEST), Australia;
Institut National de Physique Nucl\'eaire et Physique des Particules (IN2P3), 
Commissariat \`a l'Energie Atomique (CEA), France; Bundesministerium 
f\"ur Bildung und Forschung (BMBF, contract 05 6DO52), Germany; Istituto 
Nazionale di Fisica Nucleare (INFN), Italy; Joint Institute for Nuclear 
Research and Institute for Nuclear Research of the Russian Academy of 
Sciences, Russia; Fonds National Suisse de la Recherche Scientifique, 
Switzerland; Department of Energy, National Science Foundation 
(grant PHY-9526278), the Sloan and the Cottrell Foundations, USA. 

We also thank Liang Zuo-tang and Oleg Teryaev for valuable discussions.


\begin{thebibliography}{99}
\bibitem{nomad-strange-cc} P.~Astier {\it et al.}, [NOMAD  Collaboration],  
                    Nucl. Phys. B \textbf{621} (2001) 3.
\bibitem{lam_polar} P.~Astier {\it et al.}, [NOMAD Collaboration], 
                    Nucl. Phys. B \textbf{588} (2000) 3.
\bibitem{alam_polar} P.~Astier {\it et al.}, [NOMAD Collaboration], 
                    Nucl. Phys. B \textbf{605} (2001) 3.
\bibitem{nc_event}  D.~Naumov {\it et al.}, [NOMAD Collaboration],
                    Nucl. Phys. B \textbf{700} (2004) 51.
\bibitem{LUND}      B.~Andersson,~G.~Gustafson,~G.~Ingelman and 
                    T.~Sj\"ostrand,~Phys.~Rep.~\textbf{97} (1983) 31;\\
                    T.~Sj\"ostrand et al., Int. J. Mod. Phys A \textbf{3} 
                    (1988) 751.
\bibitem{alignMore} C.~Bourrely, E.~Leader and J.~Soffer, 
                    Phys. Rep. \textbf{59} (1980) 95. 
\bibitem{donoghue78} J.~F.~Donoghue, Phys.\ Rev.\ D \textbf{17} (1978) 2922.
\bibitem{ALEPH}     D.~Buskulic {\it et al.}, [ALEPH Collaboration],
                    Z. Phys. C \textbf{69} (1995) 393. 
\bibitem{DELPHI}    P.~Abreu {\it et al.}, [DELPHI Collaboration],
                    Phys. Lett. B \textbf{406} (1997) 271; \\
                    P.~Abreu {\it et al.}, [DELPHI Collaboration],
                    Z. Phys. C \textbf{68} (1995) 353.
\bibitem{OPAL}      K.~Ackerstaff {\it et al.}, [OPAL Collaboration],
                    Phys. Lett. B \textbf{412} (1997) 210; \\
                    G.~Abbiendi {\it et al.}, [OPAL Collaboration],
                    Eur. Phys. J. C \textbf{16} (2000) 61; \\
                    K.~Ackerstaff {\it et al.}, [OPAL Collaboration],
                    Z. Phys. C \textbf{74} (1997) 437.
\bibitem{EXCHARM}   A.N.~Aleev {\it et al.}, [EXCHARM Collaboration],
                    JINR preprint, \textbf{E1-99-178} (1999).
\bibitem{BEBC_align} W.~Wittek {\it et al.}, [BEBC WA59 Collaboration],
                     Phys. Lett. B \textbf{187} (1987) 179.
\bibitem{Donoghue:1978yb} J.~F.~Donoghue, Phys.\ Rev.\ D \textbf{19} 
                    (1979) 2806.
\bibitem{EfremovTeryaev} A.V.~Efremov and O.V.~Teryaev,
                    Sov.\ J.\ Nucl.\ Phys.\  \textbf{36} (1982) 557; \\
                    A.V.~Efremov and O.V.~Teryaev,
                    JINR preprint, \textbf{P2-82-832} (1982).
\bibitem{Liang}     Xu~Qing-hua, Liu~Chun-xiu and Liang~Zuo-tang, Phys. Rev. 
                    D {\bf 63} (2001) 111301.
\bibitem{Liang1}    Xu~Qing-hua and Liang~Zuo-tang, hep-ph/0205291; \\
                    Xu~Qing-hua and Liang~Zuo-tang, Phys. Rev. \textbf{D66}
                    (2002) 017301.
\bibitem{NOMAD_NIM} J.~Altegoer {\it et al.}, [NOMAD Collaboration],
                    Nucl. Instr. and Meth. A \textbf{404} (1998) 96.
\bibitem{NOMAD_OSC} J.~Altegoer {\it et al.}, [NOMAD Collaboration],
                    Phys. Lett. B \textbf{431} (1998) 219; \\
                    P.~Astier {\it et al.}, [NOMAD Collaboration],
                    Phys. Lett. B \textbf{453} (1999) 169; \\
                    P.~Astier {\it et al.}, [NOMAD Collaboration],
                    Phys. Lett. B \textbf{483} (2000) 387; \\
                    P.~Astier {\it et al.}, [NOMAD Collaboration],
                    Nucl. Phys. B \textbf{611} (2001) 3.
\bibitem{NOMAD_NUE} P.~Astier {\it et al.}  [NOMAD Collaboration],
                    Phys.\ Lett.\ B \textbf{570} (2003) 19.
\bibitem{LEPTO}     G.~Ingelman, LEPTO version 6.1, ``The Lund Monte Carlo for 
                    Deep Inelastic Lepton-Nucleon Scattering'', 
                    TSL-ISV-92-0065 (1992); \\
                    G.~Ingelman, A.~Edin, J.~Rathsman, LEPTO version 6.5,
                    Comp. Phys. Comm. \textbf{101} (1997) 108, 
                    [hep-ph/9605286].
\bibitem{JETSET}    T.~Sj\"ostrand, ``PYTHIA 5.7 and JETSET 7.4: physics and 
                    manual'', LU-TP-95-20 (1995), [hep-ph/9508391]; \\
                    T.~Sj\"ostrand, Comp. Phys. Comm \textbf{39} (1986) 347, 
                    \textbf{43} (1987) 367.
\bibitem{GEANT}     GEANT : Detector Description and Simulation Tool, 
                    {\em CERN Programming Library Long Writeup} 
                    \textbf{W5013}, GEANT version 3.21.
\bibitem{Alekhin}   S.~Alekhin, Phys.\ Rev.\ D \textbf{68} (2003) 014002.
\bibitem{DPMJET}    J.~Ranft, Phys. Rev. D51, (1995) 64; \\
                    J.~Ranft, arXiv:hep-ph/9911213; \\
                    J.~Ranft, arXiv:hep-ph/9911232.
\bibitem{Jackson}   J.D.~Jackson, Nuovo Cimento \textbf{34} (1964) 1644.
\end{thebibliography}
\end{document}